\documentclass[fleqn,usenatbib]{mnras}

\usepackage{newtxtext,newtxmath}

\usepackage[T1]{fontenc}
\usepackage{ae,aecompl}

\usepackage{graphicx}	
\usepackage{amsmath}	
\usepackage{amssymb}	
\usepackage{appendix}

\usepackage{multirow}
\usepackage{booktabs}
\newcommand{\fe}{\ensuremath{^{60}\mathrm{Fe}}}
\newcommand{\co}{\ensuremath{^{60}\mathrm{Co}}}
\newcommand{\al}{\ensuremath{^{26}\mathrm{Al}}}
\newcommand{\ex}[1]{\ensuremath{^{#1}}}
\newcommand{\tn}[1]{\ensuremath{\textnormal{#1}}}
\newcommand{\feng}{\ensuremath{^{59}\mathrm{Fe}(n,\gamma)^{60}\mathrm{Fe}}}

\newcommand{\nean}{\ensuremath{^{22}\mathrm{Ne}(\alpha,n)^{25}\mathrm{Mg}}}
\newcommand{\msun}{\ensuremath{\mathrm{M}_{\sun}}}

\title[$^{60}$Fe in CCSNe and electromagnetic emission]{$^{60}$Fe in core-collapse supernovae and
prospects for X-ray and $\gamma$-ray detection in supernova remnants}

\author[S.~W.~Jones et al.]{
Samuel W.~Jones$^{1,2,\dagger,}$\thanks{E-mail: swjones@lanl.gov},
Heiko M\"oller$^{2,3,4,\dagger}$,
Chris L.~Fryer$^{2,\dagger}$,
Christopher J.~Fontes$^1$,
\newauthor
Reto Trappitsch$^{5,\dagger}$,
Wesley P.~Even$^{2,\dagger}$,
Aaron Couture$^{6,\dagger}$,
Matthew R.~Mumpower$^{7}$,
\newauthor
and
Samar Safi-Harb$^{8}$
\\
$^1$X Computational Physics (XCP) Division, Los Alamos National Laboratory, New Mexico 87545, USA \\
$^2$Computer, Computational  and Statistical Sciences (CCS) Division, Los Alamos National Laboratory, New Mexico 87545, USA \\
$^3$Institut f\"ur Kernphysik (Theoriezentrum), Technische Universit\"at  Darmstadt, Schlossgartenstra{\ss}e 2, 64289 Darmstadt, Germany\\
$^4$GSI Helmholtzzentrum f\"ur Schwerioneneforschung, Planckstra{\ss}e 1, 64291 Darmstadt, Germany\\
$^5$Nuclear and Chemical Sciences Division, Lawrence Livermore National Laboratory, Livermore, CA 94550, USA \\
$^6$Physics Division, Los Alamos National Laboratory, New Mexico 87545, USA \\
$^7$Theoretical Division, Los Alamos National Laboratory, New Mexico 87545, USA \\
$^8$Department of Physics and Astronomy, University of Manitoba, Winnipeg, MB R3T 2N2, Canada \\
$^\dagger$ NuGrid Collaboration \\
}

\date{Accepted 15 February 2019. Received 10 December 2018}

\pubyear{2019}

\hypersetup{
	colorlinks=true,
	linkcolor=blue,
	citecolor=blue,
	urlcolor=cyan
}
\begin{document}
\label{firstpage}
\pagerange{\pageref{firstpage}--\pageref{lastpage}}
\maketitle

\begin{abstract} 
	We investigate \fe~in massive stars and core-collapse supernovae
	focussing on uncertainties that influence its production in 15, 20 and
	25~\msun~stars at solar metallicity.  We find that the \fe~yield is a
	monotonic increasing function of the uncertain \feng~cross section and
	that a factor of 10 reduction in the reaction rate results in a factor
	8--10 reduction in the \fe~yield; while a factor of 10 increase in the
	rate increases the yield by a factor 4--7. We find that none of the 189
	simulations we have performed are consistent with a core-collapse
	supernova triggering the formation of the Solar System, and that only
	models using \feng~cross section that is less than or equal to that from
	NON-SMOKER can reproduce the observed \fe/\al~line flux ratio in the
	diffuse ISM.  We examine the prospects of detecting old core-collapse
	supernova remnants (SNRs) in the Milky Way from their $\gamma$-ray
	emission from the decay of \fe, finding that the next generation of
	gamma-ray missions could be able to discover up to $\sim100$ such old
	SNRs as well as measure the \fe~yields of a handful of known Galactic
	SNRs. We also predict the X-ray spectrum that is produced by atomic
	transitions in \co~following its ionization by internal conversion and
	give theoretical X-ray line fluxes as a function of remnant age as well
	as the Doppler and fine-structure line broadening effects.  The X-ray
	emission presents an interesting prospect for addressing the missing SNR
	problem with future X-ray missions.

\end{abstract}




\section{Introduction}

Gamma-ray detections of nuclear decay lines in the Galaxy from either point or diffuse sources
provide an opportunity to test stellar evolution theory \citep[see, e.g.][and references
therein for a review centered around the INTEGRAL mission]{Diehl2013a}.  Short-lived
radioactive (SR or SLR)\footnote{ There are conflicting definitions of what constitutes a
	short-lived radionuclide (SR or SLR) in the literature, but broadly speaking they are
	nuclides with half-lives on the order of a million years or less (some definitions
	extend up to half-lives of 50 million years). The same isotopes have also been labeled
	\emph{long-lived} radioactive isotopes in relevant works.  } isotopes synthesized in
	and ejected from stars contribute a portion of the diffuse Galactic gamma-ray
	foreground in the Milky Way. In particular, decay lines from the radionuclides
	$^{26}$Al and \fe, with half lives of $7.17\times10^5$~yr and $2.62\times10^6$~yr,
	respectively, have been measured by multiple instruments, most recently by the
	INTEGRAL/SPI mission
	\citep{Smith2004a,Wang2007a,Bouchet2011a}.

\begin{figure}
	\centering
	\caption{\feng~reaction rate for neutron densities of ($10^7$, $10^9$, $10^{10}$,
	$10^{11}$) cm$^{-3}$ and the $\beta$-decay rate of $^{59}\mathrm{Fe}$ as functions of
	temperature. Creating \fe~via the s-process requires that the $(n,\gamma)$ reaction
	competes with or is faster than the decay rate, which in turn requires neutron
	densities in excess of a few $10^{10}$~cm$^{-3}$.}
	\label{fig:rates_sproc}
	\includegraphics[width=\linewidth]{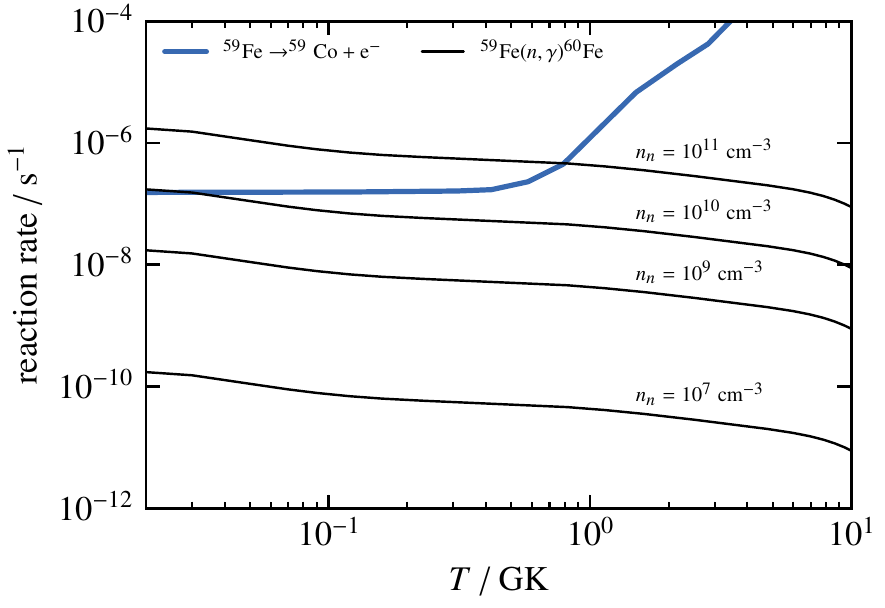}
\end{figure}

Although it is not naturally occurring, terrestrial deposits of live \fe~have in
fact been detected in layers of sea-floor sediment on Earth \citep[][see also
\citealp{Fitoussi2008a}]{Knie1999a,Knie2004a,Wallner2016a} and must have
accreted from outside of the Solar System.  The most recent measurements have
identified two such accretion events $(1.5-3.2)\times10^6$~yr and
$(6.5-8.7)\times10^6$~yr ago \citep{Wallner2016a}, consistent with the origin
being ejecta from supernovae approximately $50-120$~pc from Earth
\citep{Fields2005a}.  The Upper Centaurus Lupus and Lower Centaurus Crux stellar
subgroups have been kinematically identified as having passed through the region
currently occupied by the Local Bubble\footnote{the Local Bubble is a cavity in
	the ISM that contains the Solar System} (LB) $(10-15)\times10^6$~yr ago
	\citep{Fuchs2006a}. Supernovae exploding in those subgroups would
	consistently explain both the origin of the LB and the terrestrial
	deposits of \fe~\citep{Feige2014a,Schulreich2017a}.

The ratio of gamma-ray fluxes from \al~and \fe~in the interstellar medium (ISM)
is of particular significance. This value, since the first measurements from
HEAO-3 \citep{Mahoney1982a}, has stayed in the range $0.09 \leq F(\fe)/F(\al)
\leq 0.21$ (see \citealp{Wang2007a}, their Figure~7 and Table~2).
\citet{Bouchet2011a}, performing new analysis of the SPI/INTEGRAL data in 2011,
found a flux ratio of $\sim 0.17$ (cf.~\citeauthor{Wang2007a}'s $0.148 \pm
0.06$).  Recent measurements \citep{feige18} further determined a lower limit
for the \fe/\al\ ratio in the local interstellar medium of
$0.18^{+0.15}_{-0.08}$ by analyzing these SLRs in deep-sea sediments. This
measurement agrees well with gamma-ray flux measurements.  These values present
a constraint or calibration point for stellar evolution theory. That is, massive
stars are thought to produce the bulk of the \al~and \fe~in the Galaxy
and therefore, if accurate, models of massive stars should be able to reproduce
the observed ratio \citep{Timmes1995a,Rauscher2002,Limongi2003a,Limongi2006a}.

\citet{Timmes1995a} present galactic chemical evolution (GCE) simulations for
\al~and \fe~in the Milky Way.  Remarkably, the flux ratio their models predicted
was 0.16 -- a near-perfect match to the observational measurements.  Newer
stellar evolution and explosion calculations \citep{Rauscher2000a,Limongi2003a}
predicted ratios further from the measured value ($\sim0.9$ and $\sim0.4-0.9$,
respectively; see \citealp{Prantzos2004a}, their Figure~2). A few years later,
\citet{Limongi2006a} published a set of models that were able to reproduce the
observed flux ratio when adopting the mass loss rates of \citet{Langer1989a} as
opposed to those published a decade later by \citet{Nugis2000} for the
Wolf-Rayet phase.

The massive star and core-collapse supernova (CCSN) models used by
\citet{Timmes1995a} were from \citet{Woosley1995} and used the \feng~cross
section from \citet{Woosley1978a}. The \citet{Rauscher2002} models used the
cross section from \citet{Rauscher2000a}. While both cross sections originated
from Hauser-Feshbach calculations there were still differences
\citep{Woosley2003a}.  This could explain at least part of the discrepancy
between the \fe/\al~flux ratios predicted by \citet{Timmes1995a} and
\citet{Rauscher2002}. We note at this point that in both examples given here
(winds and cross section), adopting more modern physics has exacerbated the
tension between models and observations.

\citet{Tur2010a} showed that the yields of radioactive \fe~and \al~in massive
stars and CCSNe are sensitive to the $3\alpha$ and particularly the
$^{12}$C$(\alpha,\gamma)^{16}$O He-burning reaction rates. Varying these
reaction rates by up to twice the experimental uncertainty resulted in up to a
factor of five change in the ejected mass of \fe~and a factor of 10 change in
the \fe/\al~yield ratio.  Some of the more drastic alterations to the amount of
\fe~inside a massive star come from structural changes such as the size of a
convective region or the occurrence or not of convective shell mergers
\citep{Rauscher2002,Ritter2018a}.
Substantial structural changes can be triggered by seemingly small changes to
the reaction rates \citep{Tur2010a}, illustrating the cumulative effect of the
prevalent uncertainties in stellar evolution modelling such as rotation
\citep{Palacios2005a,Edelmann2017a}, convection and convective boundaries
\citep[e.g.][]{Meakin2007,Jones2017a,Davis2018a}, mass loss
\citep{Palacios2005a,Limongi2006a,Renzo2017a}, and opacities
\citep{Woosley2007a}.

There are other astrophysical sites that have been proposed as candidates in
which \al~and \fe~could be produced in significant quantities that we mention
here for completeness. For example, \fe~can be made in super-AGB stars
\citep{Lugaro2012a} and high-density type-Ia supernovae that include a
deflagration phase \citep{Woosley1997a}. Electron-capture supernovae (ECSNe)
have also been proposed to contribute at least $4\%-30\%$ of the live \fe~in the
Galaxy \citep{Wanajo2013a}, although the progenitor scenario remains rather
uncertain with respect to the fraction of stars that explode as ECSNe
\citep{Doherty2017a}. The nucleosynthesis is likely similar in the lowest-mass
CCSNe, however \citep{Wanajo2018a}. It is still an open question as to whether
ECSNe are indeed core-collapse supernovae \citep{Jones2016a}, and in the case
that they are thermonuclear explosions they would exhibit very large \fe~yields
and \fe/\al~ratios \citep{Jones2018a}.

\begin{figure*}
	\centering
	\caption{Top panel: \fe~mass fraction in the inner 6~\msun~of a 20~\msun~stellar model
	computed with the KEPLER stellar evolution code and post-processed with the NuGrid
	nuclear reaction network. Convective regions are hatched. Bottom panel: neutron density.}
	\label{fig:fe60-kip}
	\includegraphics[width=\textwidth]{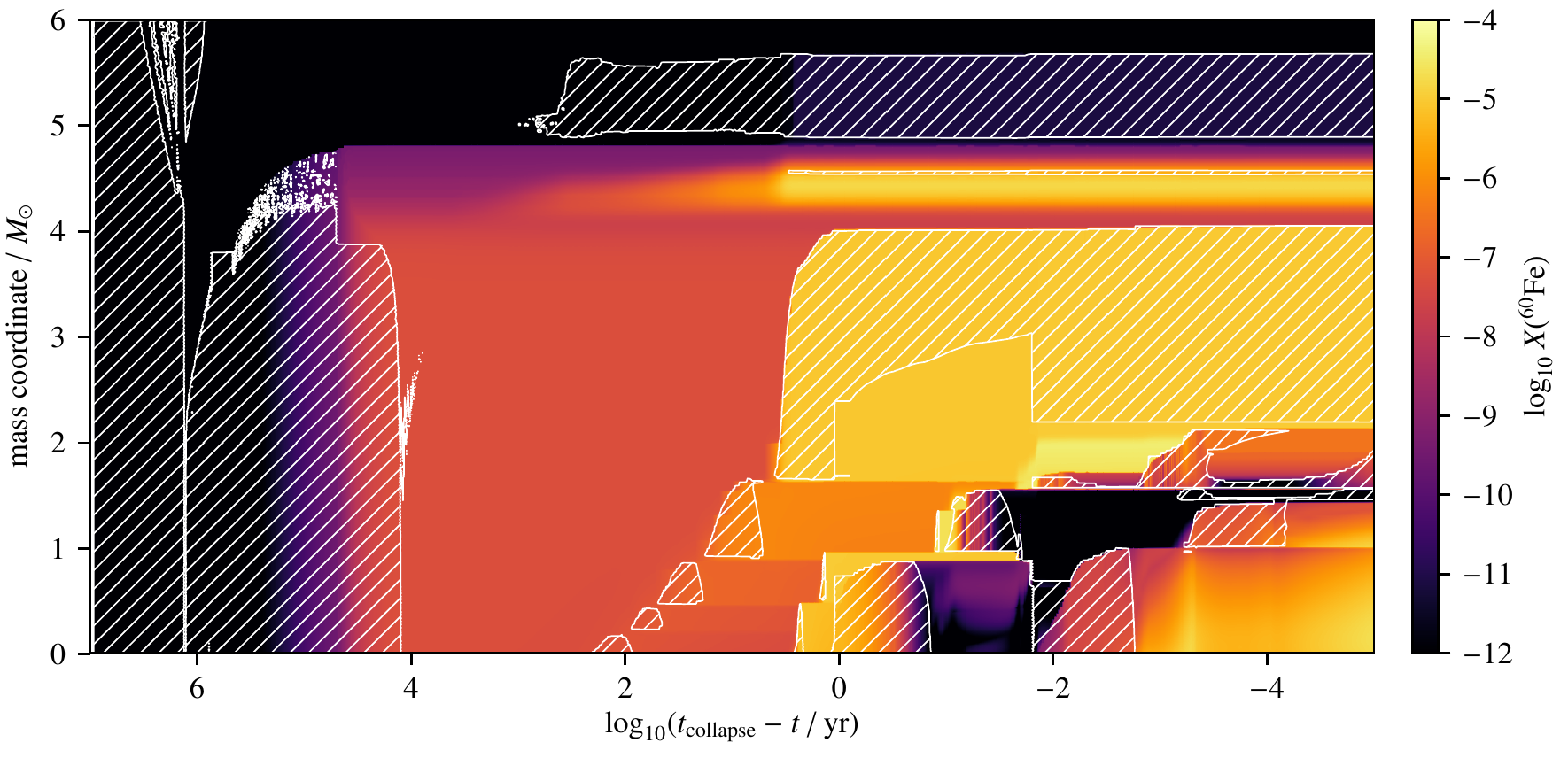} \\
	\includegraphics[width=\textwidth]{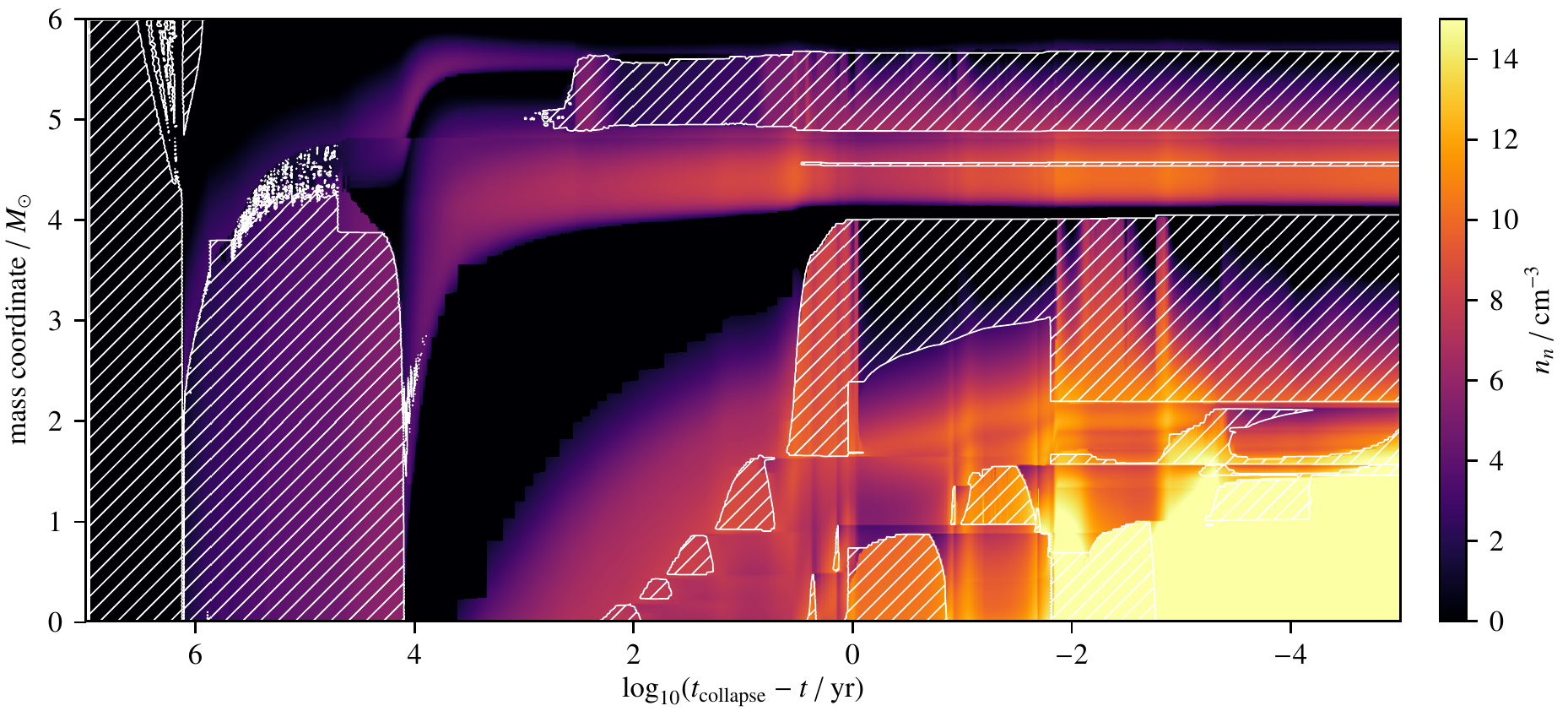}
\end{figure*}

\begin{figure*}
	\centering
	\caption{\feng~s-process impact study: \fe~mass fraction profiles in the stellar core
		at the pre-supernova stage, depending on three realizations of the
		\feng~reaction cross section, which are scalar multiples of $\langle\sigma
		v\rangle_\mathrm{NS}$ -- the \feng~reaction cross section from
		\citet{Rauscher2000a} -- as indicated in the legend.}
	\label{fig:presnfe60}
	\includegraphics[width=\textwidth]{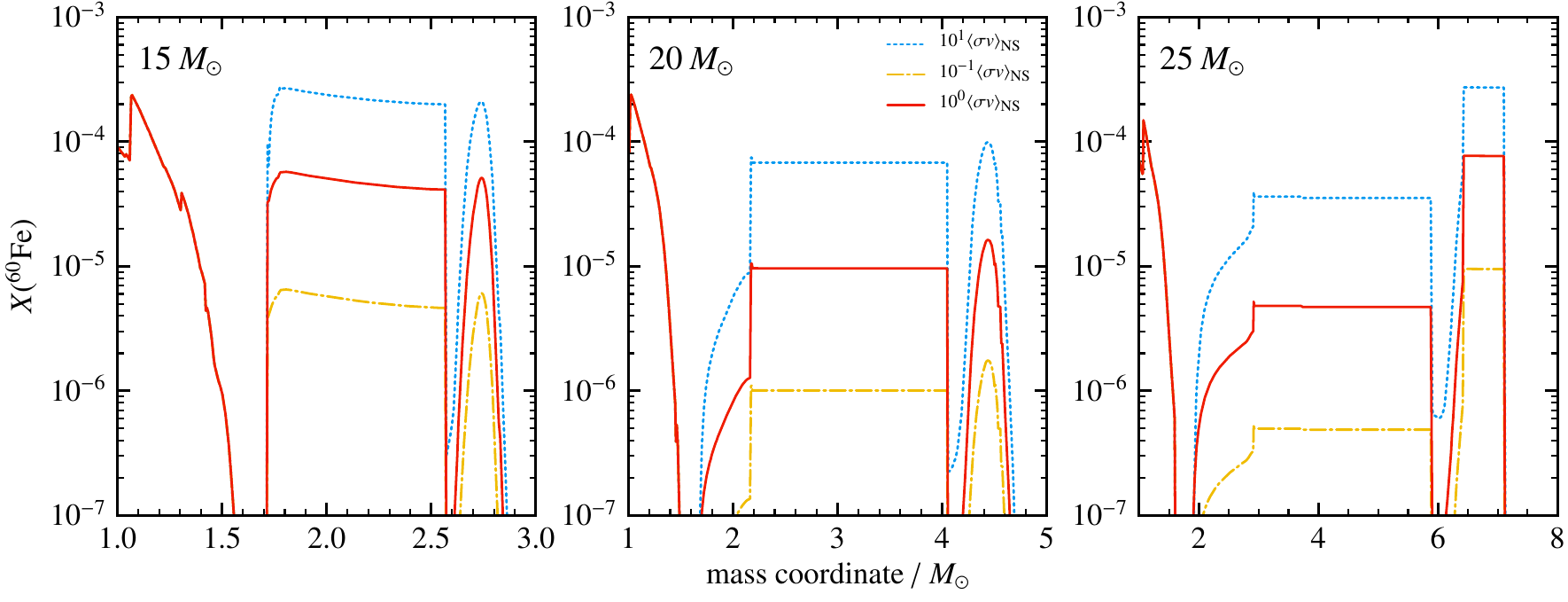}
\end{figure*}

Our work is also motivated by addressing the missing SNR problem. Radio and
high-energy observations of SNRs reveal nearly 400 SNRs in our Galaxy
\citep{Green2017, ferrand12}. However, at a Galactic supernova frequency of one
event every 30--50 years \citep{Tammann1994}, and the lifetime of an SNR being
$\sim$10$^6$ years, the total number of SNRs should be at least an order of
magnitude higher than observed. This missing SNR problem can be at least partly
attributed to a selection effect against the old, low-surface brightness SNRs.
We suggest the decay of $^{60}$Fe as a means to detect SNRs at very late times.

In this paper we focus on two  relatively unexplored -- but known --
uncertainties in the production of \fe~in CCSN models. The first is the
\feng~cross section for which there are still no direct measurements available
\citep[see][]{Woosley2003a}. The second is the way in which the 1D CCSN
simulations are performed, how they are parameterized and with what parameter
choices. A third relevant uncertainty but one that we do not address in this
work is the dependence on the dimensionality of the CCSN simulation.

\section{Methodology and Simulations}
We present spherically symmetric (1D) stellar evolution (SE) and CCSN
simulations with emphasis on their nucleosynthesis, which is
calculated in post-processing.  We used three different \feng~reaction cross
sections for the nucleosynthesis covering a range of two orders of magnitude.

\subsection{Stellar evolution models}

\label{sec:progenitors} Stellar evolution models were computed using the KEPLER
code \citep{WZW78,Rauscher2002,WH07}, ending when the collapsing iron core
achieves an infall velocity of 1000~km~s$^{-1}$. We computed models with initial
masses of 15, 20 and 25~\msun, all with initial metallicity $Z_\mathrm{ini} =
0.02$ and relative fractions of the metals after \citet{GrevesseNoels1993}.  The
modelling assumptions made for these simulations were the same as for the KEPLER
models by \citet[][]{Jones2015a} but with different rates for three of the key
reactions responsible for energy generation during H and He burning. The
$^{14}$N$(p,\gamma)^{15}$O reaction rate is now taken from \citet{Imbriani2004},
the triple-$\alpha$ reaction rate is from \citet{Fynbo_3a_2005}, and the
$^{12}$C$(\alpha,\gamma)^{16}$O reaction rate is taken from \citet{Kunz2002}.
\citet{Jones2015a} have presented a relatively detailed description of the
methodology including a comparison with two other stellar evolution codes.

\subsection{Core-collapse supernova simulations}
\label{sec:CCSNsims}

The CCSN models used for this work are a suite of parameterized simulations by
\citet{fryer18} which used the 1D Lagrangian hydrodynamics supernova code
described in \citet{herant94} and \citet{fryer99} with initial conditions
provided by the collapsing KEPLER models.  The parameterization of the 1D
explosions is designed to mimic the 3D convection engine first demonstrated by
\cite{herant94}. Indeed, except for low-mass progenitors, 1D models typically do
not self-consistently produce supernova explosions owing to the lack of
convective energy transport. Uncertainties in the multi-dimensional simulations
of the convective engine allow for a range of results and we use parameterized
models to give us a flavor of the possible explosion conditions. Energy is
injected above the proto-neutron star over different spatial extents to mimic
the size of the convective region and at different durations to mimic the
timescale of the engine \citep[for details see][]{fryer18}. In total there are
63 CCSN simulations of the 3 progenitors that we consider for this work.  We
note that a shortcoming of our approach is that neutrino interactions are
omitted, which are known to substantially modify the electron fraction of the
material above the proto-neutron star in the NSE region
\citep[e.g.][]{Qian1996a,Froehlich2006a,Rampp2000a} and can also modify the
composition of the ejecta via $\nu$-spallation processes. This should not be a
serious concern for the present work, because the production/destruction of
\fe~is not influenced by neutrino interactions, however \citet{Timmes1995a} have
shown the the \al~yield can be increased by up to about 50 per cent when
$\nu$-spallation reactions are included in simulations.

\begin{figure*}
	\centering
	\caption{
		Total mass of $^{60}$Fe in the pre-supernova models above the proto-compact
		remnant (cut), and total ejected mass of~$^{60}$Fe (yield) for all of the simulations
		with the standard $^{59}$Fe$(n,\gamma)^{60}$Fe reaction rate from \textsc{Non-smoker}.
	}
	\label{fig:yields_standard}
	\includegraphics[]{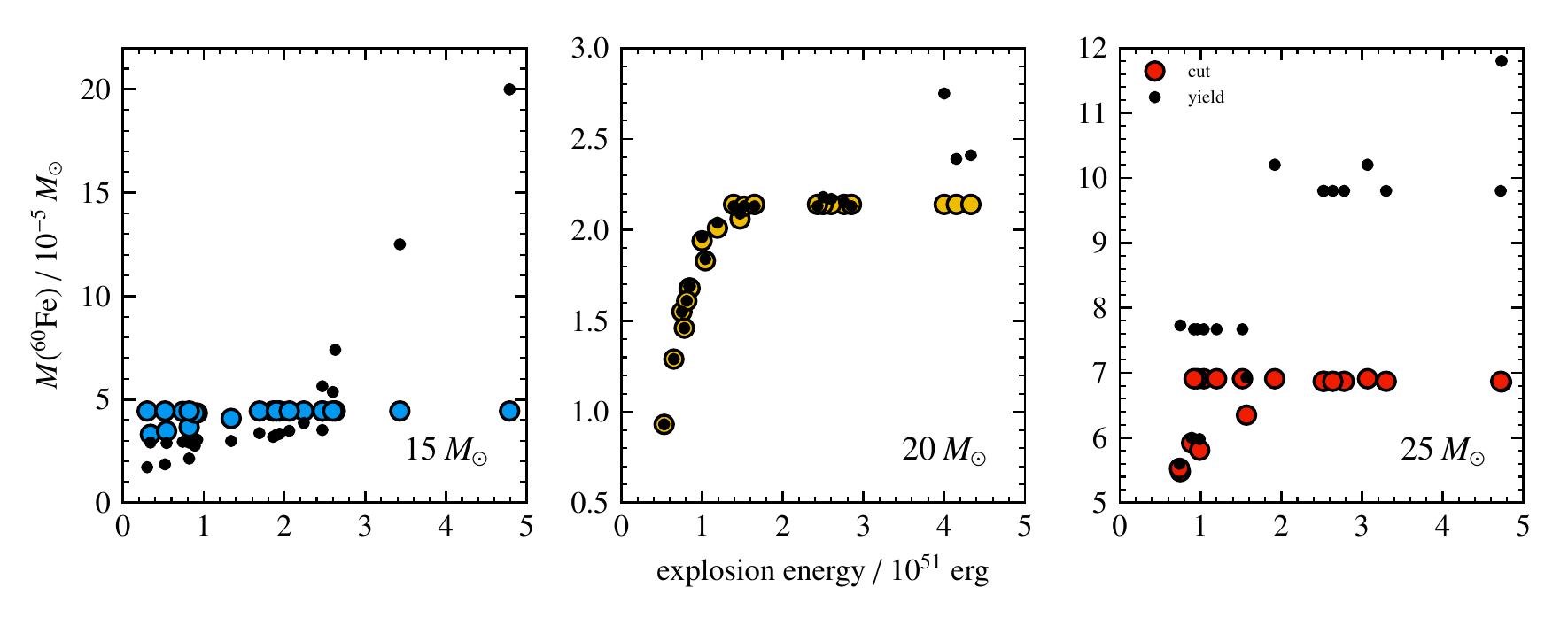}
\end{figure*}

\subsection{Post-processing nucleosynthesis code}
\label{sec:nuc_code}
The nucleosynthesis calculations were performed using a derivative of the NuGrid
nuclear reaction network code \citep[see, e.g.,][]{Pignatari2016a,Ritter2017a}.

The SE calculations were post-processed using the code MPPNP\footnote{Multi-zone
Post-Processing Network -- Parallel}, which evolves the composition forwards in
time for every computational grid cell over one stellar evolution time step and
subsequently performs a mixing step over the whole model with a time-implicit
diffusion solve de-coupled from the network integration using a diffusion
coefficient $D(r)$ from the stellar evolution code. The reaction network for the
evolution up to the pre-supernova stage included a pool of about $1100$ isotopes
and about $14000$ reactions.

For the supernova simulations, we used the code TPPNP\footnote{Tracer particle
Post-Processing Network -- Parallel}, which is simply a different program for
handling the I/O for tracer particles (as opposed to spherically symmetric
stellar models) and capitalising on the embarassingly parallel nature of
post-processing Lagrangian particles between which we assume there to be no
mixing. Since our CCSN simulations were performed with a 1D Lagrangian
hydrodynamics code (see Section~\ref{sec:CCSNsims}) without adaptive mesh
refinement, the evolution of the composition for each grid cell can be
considered to be mutually exclusive from that in all the other grid cells.  The
reaction network for the explosions was substantially larger than for the
stellar evolution models (in fact, probably larger than necessary), with
approximately 5000 isotopes and 67000 reactions.

Both programs MPPNP and TPPNP use the same underlying physics and solver
libraries, to which several additions were made during the course of this work
that we feel are worthy of
mention here. They are:
\begin{itemize}
	\item a semi-implicit extrapolation time integrator (so-called Bader-Deuflhard or generalised
		Bulirsch-St\"oer method; \citealp{Bader1983a,Deuflhard1983a}, but see also
		\citealp{Timmes1999a}),
	\item a Cash-Karp Runge-Kutta integration for the coupling of weak reactions with the NSE
		state,
	\item calculation of reverse reaction rates using the principle of detailed
		balance\footnote{see the appendix of \citet{Calder2007a} for a concise
	formulation including plasma Coulomb corrections},
	\item integration of SuperLU sparse matrix solver library for solving the linear system
		\citep{superlu99,superlu_ug99,li05}, and
	\item implementation of electron screening corrections from \citet{Chugunov2007a}.
\end{itemize}
A discussion of these few additions is deferred to Appendix~\ref{sec:appendix}.

Reaction rates from JINA Reaclib \citep{Cyburt2010}, KaDoNiS
\citep{Dillmann2006a}, NACRE \citep{AnguloNACRE1999} and NON-SMOKER
\citep{Rauscher2000a}, as well as from \citet{FFNweak1985},
\citet{Takahashi1987a}, \citet{Goriely1999a}, \citet{Langanke2000},
\citet{Iliadis2001a} and \citet{ODA94} were used.  Of particular relevance for
this work is that the \feng~reaction rate used was the JINA Reaclib fit to the
reaction rate from \citet{Rauscher2000a}. We used the KaDoNiS reaction rates for
the $^{58}$Fe and $^{60}$Fe  $(n,\gamma)$ reactions, taking the v0.3 values
where available. For the $^{22}\mathrm{Ne}(\alpha,n)^{25}\mathrm{Mg}$ reaction
we use the recommended rate from \citet{Jaeger2001a}. The
$^{59}\mathrm{Fe}\rightarrow{^{59}\mathrm{Co}}+\mathrm{e}^-+\bar{\nu}$ reaction
rate is from \citet{Langanke2000}.

\section{The \feng~cross section}
\label{sec:fe59ng}

In astrophysical scenarios, the primary reaction mechanism to produce $^{60}$Fe
is through \feng{}.  Unfortunately, this reaction is very difficult to measure
directly.  $^{59}$Fe has a short half-life of 44.5 days and emits gamma-rays
with energies in excess of 1~MeV.  The gold-standard for neutron capture
measurements is the differential time-of-flight (TOF) measurement technique.
This requires a sample of the isotope of interest to be produced and placed
inside of a detector array.  Past studies \citep{CoR07} showed that current and
planned neutron experimental facilities have 2-4 orders of magnitude too few
neutrons to perform a TOF measurement on $^{59}$Fe.  An indirect
Coulomb-dissociation measurement by \citet{UAA14} places constraints on the
gamma-ray strength function used in the statistical cross-section calculation,
however, this measurement was only sensitive to $E1$~components.  There is
growing evidence that the $M1$~component, not accessible in the Uberseder
measurement, play an important role in neutron capture cross sections
\citep{LaG10,Mumpower2017a}.  Finally, neutron capture cross sections on
isotopes near shell closures, such as $^{59}$Fe, are notoriously difficult to
predict because individual neutron resonances can dominate the nuclear reaction
rate.   As a result, we have chosen to perform these studies with a
cross-section ranging up and down a factor of ten from the standard
\textsc{nonsmoker} rates.  This is consistent with standard variations for
neutron capture for unstable isotopes \citep{MMS12,Mumpower2016a}.

\begin{figure}
	\centering
	\caption{An overview of where $^{60}$Fe is produced and where it is destroyed in our
	CCSN simulations with the standard $^{59}$Fe$(n,\gamma)^{60}$Fe reaction rate from
	NON-SMOKER \citep{Rauscher2000a}. The thick line shows the pre-supernova mass fraction
	profile inside the star where $^{60}$Fe is most abundant. The thin grey lines are the
	$^{60}$Fe mass fraction profile for all computed explosion models after
	the CCSN shock has passed and the shock-heated
	matter has substantially cooled.}
	\label{fig:ccsn_fe60}
	\includegraphics[width=\linewidth]{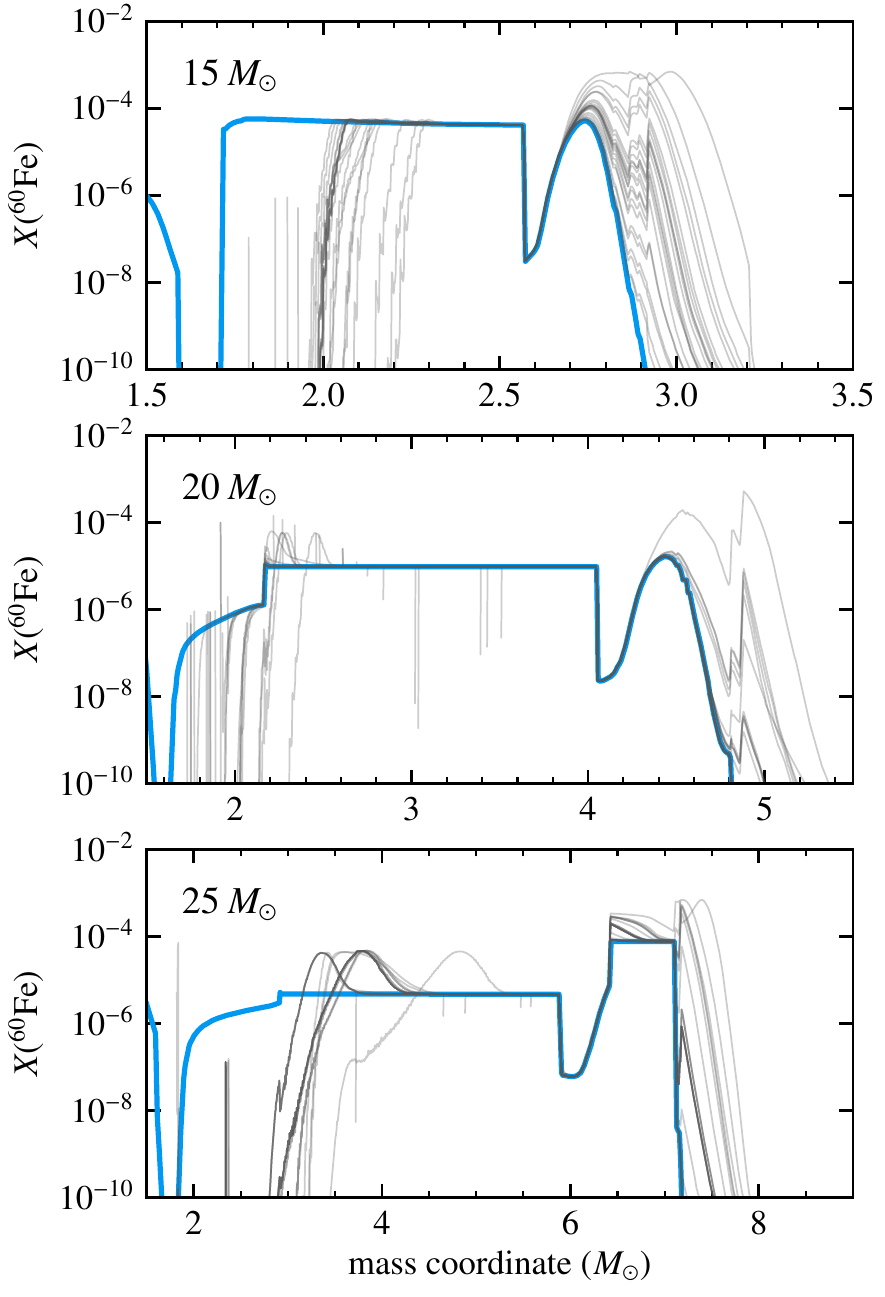}
\end{figure}

\section{Nucleosynthesis results}

In this section, we present the results of the nucleosynthesis calculations as
they pertain to the production of radioactive \fe. We briefly review the
production of \fe~in massive stars via the weak s-process, following which we
present an analysis of the contribution of shock heating in the CCSN to the
synthesis of \fe. Lastly, we assess the impact of varying the \feng~cross
section on the amount of \fe~produced.

\subsection{\fe~in massive stars}

The stellar models initially contain no \fe, owing to our adoption of the Solar
distribution of the metals \citep{GrevesseNoels1993} and \fe~having a
comparatively short half life.  \fe~is predominantly a product of the weak
s-process and is produced by successive neutron captures beginning from
$^{56}\mathrm{Fe}$ seed nuclei from the gas cloud that collapsed to form the
star. The $^{22}\mathrm{Ne}(\alpha,n)^{25}\mathrm{Mg}$ reaction is the source of
the neutrons.  \fe~is therefore considered to be a secondary isotope. The
$(n,\gamma)$ reaction sequence takes place very slowly in the convective
He-burning core towards the end of core He-burning, but much faster in the
C-burning and Ne-burning shells, and in the hotter, deeper portion of the
He-burning shell later in the star's life.  The bulk of the \fe~produced via the
weak s-process is made in the He-, C-, and and Ne-burning shells where the
temperatures are high enough for the
$^{22}\mathrm{Ne}(\alpha,n)^{25}\mathrm{Mg}$ reaction to be sufficiently
activated that the neutron density is high enough for the \feng~reaction to
compete with the $\beta$-decay of $^{59}\mathrm{Fe}$. This requires neutron
densities of $n_n\gtrsim10^{10}$~cm$^{-3}$, as is illustrated in
Figure~\ref{fig:rates_sproc} \citep[see also][]{Limongi2006a}.

At temperatures in excess of about 2~GK, \fe~is destroyed by $(\gamma,n)$,
$(n,\gamma)$ and $(p,n)$ reactions and therefore the \fe~production is most
vigorous in the Ne-burning shells, which approach the maximum temperature for
its production and above which it would be destroyed.  Indeed, O burning
proceeds Ne burning and usually engulfs the entire portion of the star in which
Ne burning operated, raising the temperature above the \fe~destruction
threshold.  Figure~\ref{fig:fe60-kip} shows a map of the mass fraction of \fe~in
the core of the star throughout its entire life (top panel) along with the
neutron density (bottom panel). The cross-hatched regions are convectively
unstable. The production of \fe~is clearly correlated with higher neutron
densities. It is also produced in the iron core during Si-burning and persists
in the central region of the star until the star explodes.  This can be seen in
the top panel of Figure~\ref{fig:fe60-kip}, onwards from
$\log_{10}(t_\mathrm{collapse}-t)\approx-2$ in the central 1.5~\msun~of the
star.  However, this material will likely not escape the gravitational potential
in the collapsing core and will instead become part of the compact remnant,
therefore not contributing to the final yield at all. In fact, this is the case
in all of the CCSN simulations presented in this work (see
Section~\ref{sec:CCSNresults}).

\begin{figure}
	\centering
	\caption{Ratio of initial to final \fe~abundances in the supernova
		simulations for all tracer particles as a function of their peak
		temperature when they are shocked.  There are two distinct
	peaks in production in the He and C shells, and destruction above $\sim3$~GK. Details of the
	relevant processes are given in the text.}
	\label{fig:tpeak_fe60}
	\includegraphics[width=\linewidth]{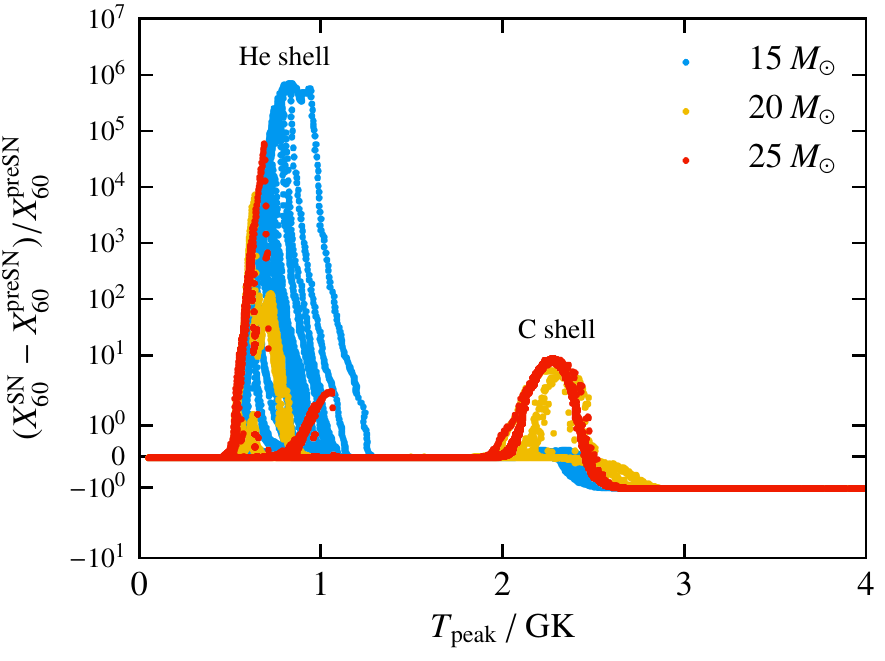}
\end{figure}

\begin{figure*}
	\centering
	\caption{Approximate time scales (s) for various nuclear reaction rates of importance
	for the production or destruction of \fe~during explosive burning.}
	\label{fig:rates}
	\includegraphics[width=\textwidth]{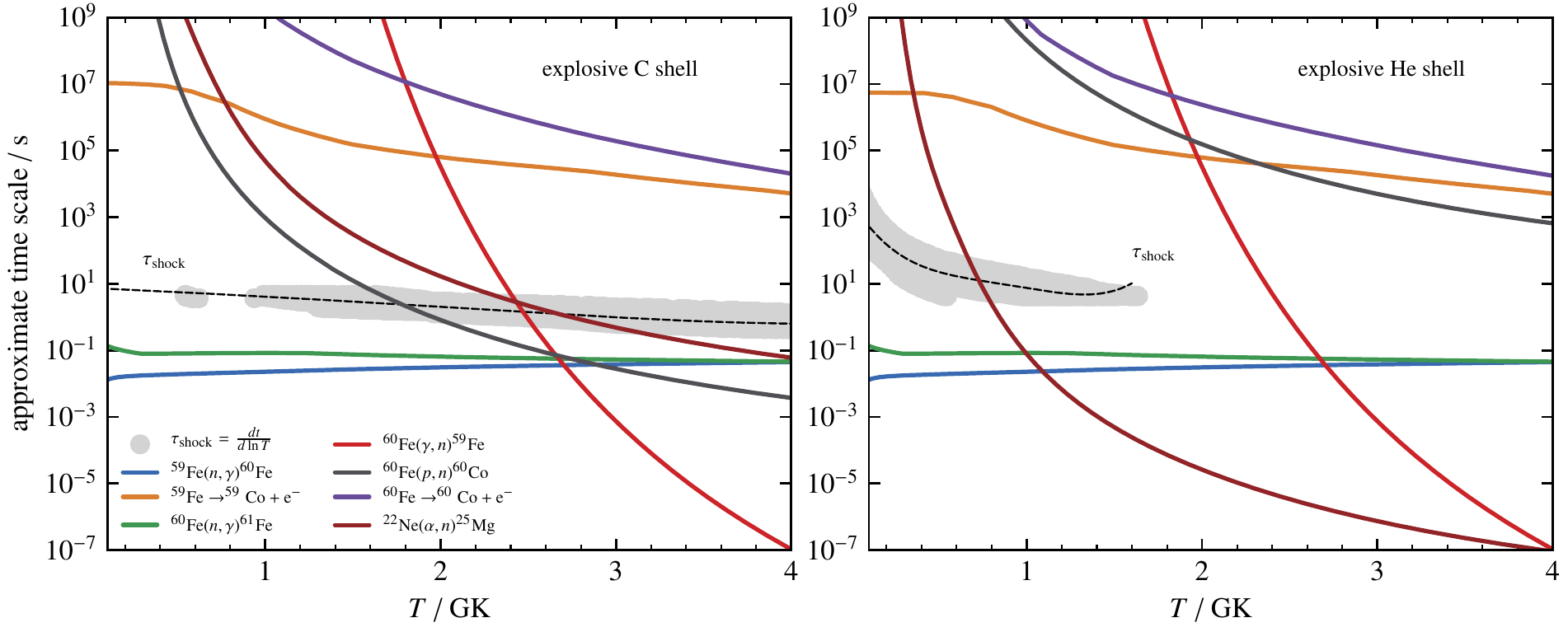}
\end{figure*}

The evolution of \fe~in a massive star culminates at the pre-supernova stage
where the bulk of \fe~is contained in the Fe core, the Si-burning shell,
C-burning shell and in the deeper -- and often convectively stable -- portion of
the He-burning shell (see Figure~\ref{fig:presnfe60}).  The amount of \fe~in the
C-burning and He-burning shells is shown to be sensitive to the uncertain
\feng~reaction cross section but we defer the discussion of this sensitivity to
Section~\ref{sec:rate-impact}.

\subsection{\fe~in CCSNe}
\label{sec:CCSNresults}
There are two key factors affecting the final yield of \fe~from CCSNe once the
amount in the star at the pre-supernova stage is known. These are (i) which
material becomes gravitationally unbound and which remains bound to become the
proto-neutron star (i.e. where is the mass cut) and (ii) heating in the
post-bounce shock wave and cooling during the expansion following the passage of
the shock.

In order to separate out these two effects, in Figure~\ref{fig:yields_standard}
we show the total mass of \fe~in the star at the pre-supernova stage minus what
goes into the compact remnant (``cut", filled circles) and the final yield,
which includes the effect of shock heating on the unbound material (``yield",
black dots). These quantities are plotted as a function of asymptotic kinetic
explosion energy.  While a substantial amount of \fe~is always lost to the
compact remnant, there is a significant contribution from the shock heating in
most of the 15 and 25~\msun~models, but not so much in the 20~\msun~models. In
general the net effect of the shock-heating on the 15~\msun~models is to destroy
\fe~rather than create it, with the exception of the 5 most energetic
explosions.  Conversely, the general net effect of the shock heating in the
25~\msun~models is to produce more \fe, with the largest impact again for the
most energetic explosions. Now that we have an idea of the net effects of the
s-process, mass cut and the shock heating on the final ejected mass of \fe, we
will present a more detailed view of what is happening to the \fe~in the
explosion.

The final abundance profiles of \fe~for all of our simulations is shown in
Figure~\ref{fig:ccsn_fe60} together with the pre-supernova abundance profile.
This figure serves as a reference point for the following discussion.

An overview of the \fe~shock nucleosynthesis is given in
Figure~\ref{fig:tpeak_fe60}, which shows the ratio of final to initial \fe~in
the supernova as a function of peak temperature. Two distinct peaks are seen and
correspond to the conditions met during explosive He and C shell burning.
Additionally there is a deficit in \fe~at high peak temperatures. The details of
the processes taking place are given in the remainder of this section.

Firstly, the shock wave compresses and heats the inner portion of the star, and
sometimes also the lower section of the C shell into NSE. The electron fractions
$Y_\mathrm{e}$ are close to but slightly below 0.5, favoring nuclei with
$N\approx Z$. Much lower electron fractions would be needed in order for \fe~to
be abundant in this scenario: $Y_\mathrm{e}\approx\frac{26}{60} = 0.43$.
Therefore, in the deepest parts of the supernova ejecta virtually all of the
\fe~is destroyed by the readjustment of the composition to its statistical
equilibrium. The electron fraction does not change appreciably during the
passage of the shock in this region because the time scales of the weak rates at
those densities are much longer than the post-shock expansion time scale of the
ejecta. Therefore, it is generally not possible to obtain a $Y_\mathrm{e}$ low
enough to favour \fe~production in these deeper layers of the star during the
explosion.  Indeed, even near the proto-neutron star surface where electron
capture can neutronize the ejecta, neutrino interactions prevent the electron
fraction from becoming this low
\citep[e.g.][]{Qian1996a,Froehlich2006a,Rampp2000a}.

\begin{figure}
\centering
\caption{Neutron density (solid) and proton density (dashed; cm$^{-3}$) in the
C and He shells during the passage of the CCSN shock wave. Neutrons are
released via the reaction \nean. $t=0$ corresponds to the time at core bounce.}
\label{fig:ndens}
\includegraphics[width=\columnwidth]{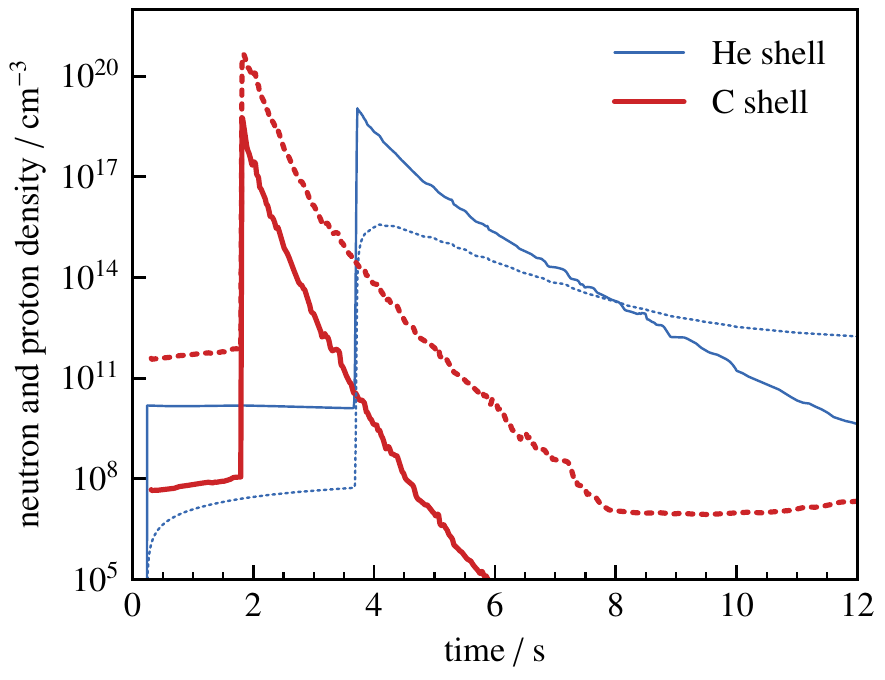}
\end{figure}

As the shock propagates through the C shell and the peak temperature in the
shocked material decreases, the material is no longer heated as far as NSE but
\fe~is still destroyed by $(\gamma,n)$, $(n,\gamma)$ and $(p,n)$ where the peak
temperature exceeds approximately 3~GK.  This is illustrated in
Figure~\ref{fig:rates}, in which the left panel shows time scales of several key
reactions taking place during explosive C shell burning, comparing them to the
post-shock expansion time scale $\tau_\mathrm{shock}$, which is the e-folding
time of the post-shock temperature \citep{fryer18}. Above 3~GK, the
$^{60}\mathrm{Fe}(\gamma,n)$, $^{60}\mathrm{Fe}(p,n)$ and
$^{60}\mathrm{Fe}(n,\gamma)$ reactions are operating on time scales faster than
or comparable to the \feng~reaction that is creating \fe. Below about 2~GK, the
fastest reaction is \feng, however the neutron source reaction \nean~becomes
much slower than the post-shock expansion time scale, so although the
thermodynamic conditions and the composition are optimal for \fe~production, the
whole process operates on too short a time scale to be relevant during the
explosion. Between 2 and 3~GK, however, there is a sweet spot where \feng~is the
fastest reaction, and the destructive reactions have a time scale similar to or
less than $\tau_\mathrm{shock}$.  Between 2~GK and 3~GK the \nean~reaction still
has a time scale that is comparable to $\tau_\mathrm{shock}$, and therefore
neutrons will be released, achieving neutron densities of
$\sim10^{18}$~cm$^{-3}$ (see Figure~\ref{fig:ndens}) which are sufficient to
overcome the enhancement in $\beta$-decay rates of $^{59}$Fe . The increase in
time scale of the \nean~reaction below about 2~GK is the main reason why there
is no \fe~created in the outer portion of the C shell (see
Figure~\ref{fig:ccsn_fe60}).

The neutron densities reached in explosive He shell burning when the shock
reaches the He shell are very similar to those attained in explosive C shell
burning (see Figure~\ref{fig:ndens}).  While the peak temperature of the shock
is lower ($\lesssim1$~GK) the $^{22}\mathrm{Ne}$ abundance is much higher:
$X_{22}=0.02$ compared to $4\times10^{-4}$ in the C shell. The He abundance is
also much higher in the He shell, as one might exepect.  The time scale of
neutron release via \nean~is again faster than or comparable to the post-shock
expansion time scale (Figure~\ref{fig:rates}, right panel) because the proton
densities are so much lower than in the C shell (Figure~\ref{fig:ndens}), there
is no significant destruction of \fe~by $(p,n)$ reactions. Moreover, because the
peak temperature is lower, there is also no discernible destruction of \fe~by
$(\gamma,n)$ reactions either.

\begin{figure*}
	\centering
	\caption{Total ejected mass of~$^{60}$Fe as a function of explosion energy for
	all of the simulations including two variations (increased and decreased by a
	factor of 10) in the $^{59}$Fe$(n,\gamma)^{60}$Fe reaction rate from
	\textsc{Non-smoker}. Where the reaction rate was modified, the reverse reaction
	rate was also modified according to the principle of detailed balance. The
	modified reaction rates were used for both the pre-SN evolution (weak
s-process) and the explosion post-processing calculations.}
\label{fig:ccsn_sens}
\includegraphics[width=\textwidth]{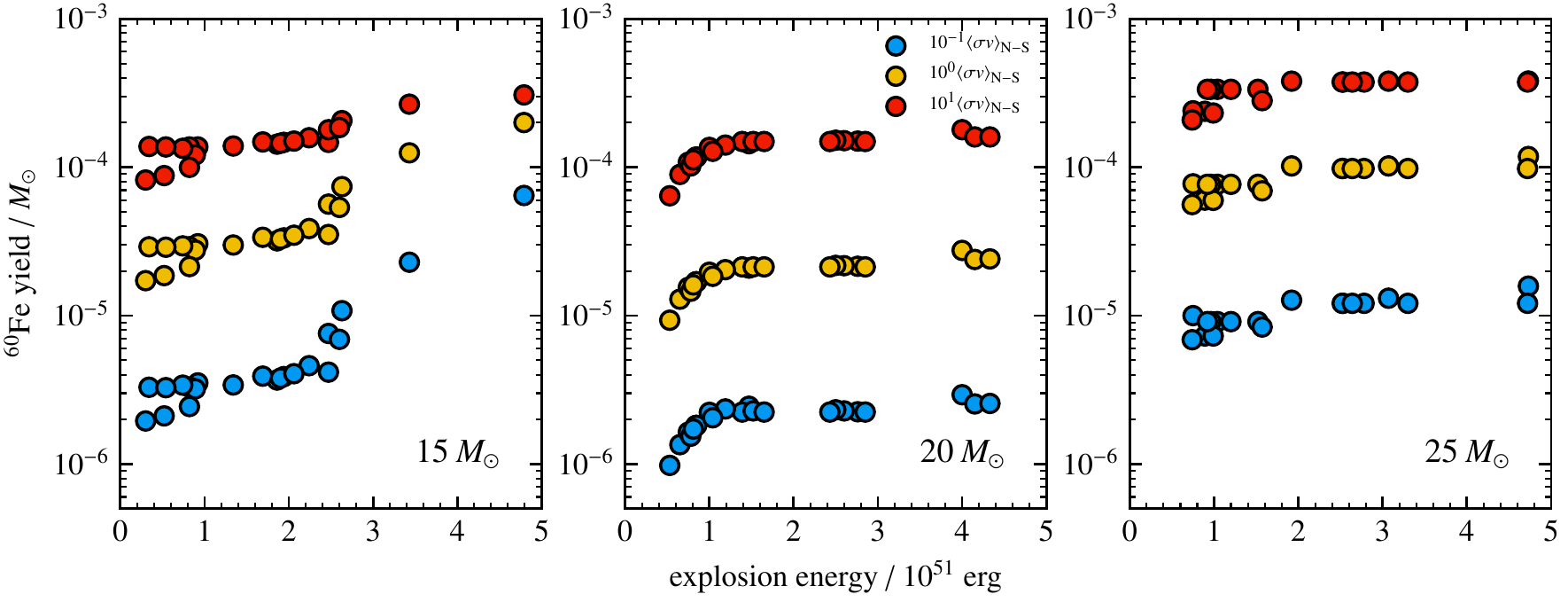}
\end{figure*}

Again, the total final \fe~yield is a monotonic increasing function of the
\feng~reaction rate.  The impact of modifying the reaction rate from NON-SMOKER
through both the stellar evolution and the explosion is shown in
Figure~\ref{fig:ccsn_sens}.  A discussion of the impact of the \feng~cross
section on the production of \fe~is presented in Section~\ref{sec:rate-impact}.

\subsection{Impact of the \feng~cross section and total uncertainty}
\label{sec:rate-impact}

In Figure~\ref{fig:ccsn_sens} we have shown that the total ejected mass of
\fe~is sensitive to and is a monotonic increasing function of the \feng~cross
section. The yield appears to scale almost linearly with the cross section,
suggesting that if the uncertainty in the cross section spans two orders of
magnitude, then the uncertainty in the \fe~yield also spans approximately two
orders of magnitude. In this short section we will examine the sensitivity of
the mass of \fe~in a massive star to the \feng~cross section at the
pre-supernova stage and in the supernova ejecta in a little more detail.

\begin{figure*}
	\centering
	\caption{\emph{Top panels:} Ratio of \fe~mass when using the enhanced cross section
		compared with the standard cross section, and of the \fe~mass when using the
		standard cross section compared with the depressed cross section. Masses of
		\fe~are given at the pre-supernova stage (preSN; top left), minus the mass
		going into the neutron star (cut; top center) and ejected in the explosion
		(yield; top right).
		\emph{Bottom panel:} The same ratios are shown but for the time-evolution of
		\fe~and $^{59}$Fe during a one-zone convective C shell burning example problem.
		See the text for an explanation of the impact of the \feng~reaction cross section
		at the various evolutionary stages.
	}
	\label{fig:rate_sens}
	\includegraphics[width=\textwidth]{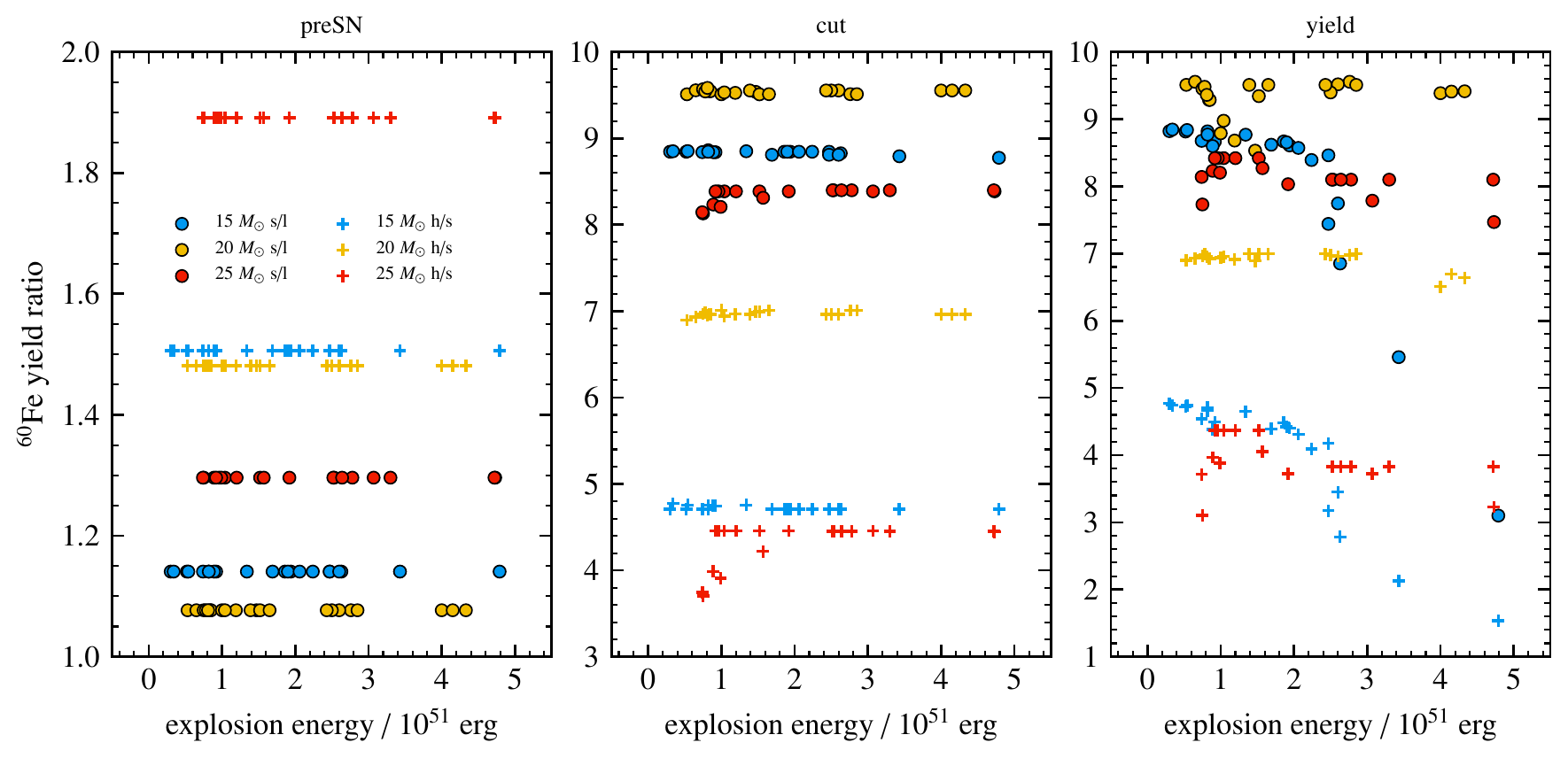} \\
	\includegraphics[width=.5\textwidth]{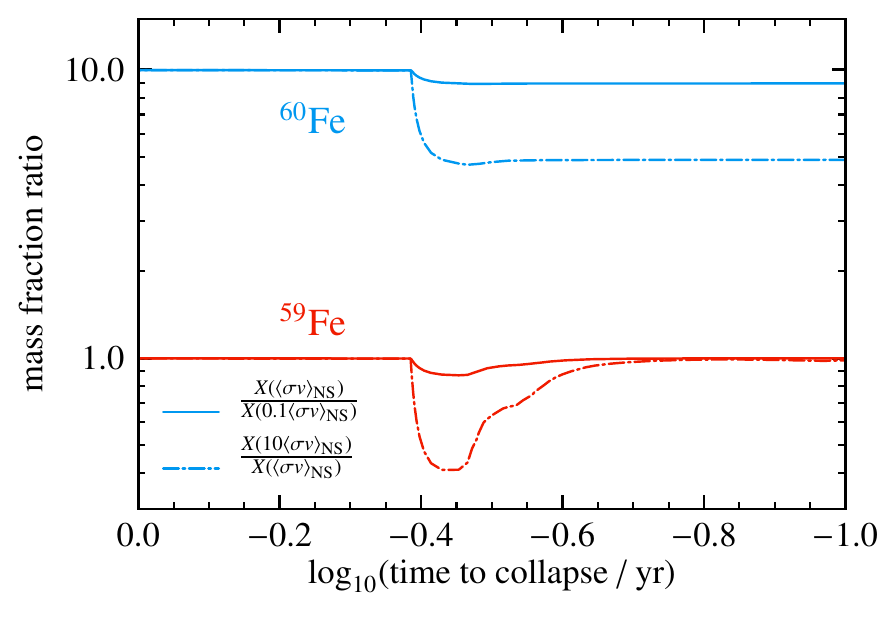}
\end{figure*}

All of the stellar evolution and supernova nucleosynthesis post-processing
calculations were performed three times: with the \feng~cross section from
NON-SMOKER \citep{Rauscher2000a} and with cross sections ten times lower and ten
times higher (s, l and h; standard, low, high, respectively). To get an insight
into the impact of varying the \feng~cross section on the production of \fe, we
look at the masses of \fe~in the simulations with different cross sections
relative to one another. More specifically, we have looked at the ratios
$M(\fe)_s/M(\fe)_l$ and $M(\fe)_h/M(\fe)_s$, where the subscript denotes the
cross section that was used. Three different \fe~masses were used in order to
separate out the different processes affecting the production of \fe: the total
mass of \fe~in the star at the presupernova stage \emph{including} the material
that will form the neutron star (preSN), the total mass of \fe~in the star at
the presupenova stage \emph{excluding} the material that will form the neutron
star (cut), and the total ejected mass of \fe~in the supernova explosion after
shock heating (yield).  This information is presented in the top three panels of
Figure~\ref{fig:rate_sens}.

Increasing or depressing the cross section by a factor of 10 results in at most
a factor of 2 change in the mass of \fe~in the star at the presupernova stage.
This is because a large fraction has been produced in the iron core where the
material is either in NSE or close to NSE. This has been illustrated in
Figures~\ref{fig:fe60-kip} and \ref{fig:presnfe60}. The \fe~abundance for the
NSE material is insensitive to the cross sections being used and instead is only
a function of the masses, chemical potentials and partition functions of the
isotopes in the NSE solver, therefore the same \fe~abundance is obtained in the
iron core regardless of \feng~cross section. There is still some sensitivity to
the cross section, though, which is most apparent in the 25~\msun~models because
of the large amount of \fe~in the He shell at the pre-supernova stage (see
Figure~\ref{fig:presnfe60}).

If the material that will become the proto-neutron star is excluded from the
total pre-supernova mass of \fe~(cut, middle panel of
Figure~\ref{fig:rate_sens}), the models with the fiducial \feng~cross section
produce a factor of $8-10$ times more \fe~than those with the ten times
depressed cross section, i.e.~there is an almost linear relationship between the
\fe~produced and the \feng~cross section. The models with a ten times enhanced
cross section on the other hand produce only a factor of $4-5$ or at most 7
times more \fe~than those with the fiducial cross section. The reason for this
is simply that the seed isotope $^{59}$Fe is depleted enough to have an impact
on the production on \fe. This is illustrated for a simple s-process test
problem in the bottom panel of Figure~\ref{fig:rate_sens}, where the majority of
\fe~is made at roughly 0.4 years before core collapse
($\log_{10}(\mathrm{time~to~collapse/yr}) \approx -0.4$). The dot-dashed lines
show that there is a maximum of a factor of two less $^{59}$Fe while the burning
takes place for the case with the enhanced cross section compared to the
fiducial cross section. Had the $^{59}$Fe abundance not been affected, the
transformation rate of $^{59}$Fe in to \fe~would have been approximately twice
as large and instead of the cross section impact being a factor of $4-5$ or 7,
it would be much closer to 10. This is fairly intuitive because other than
running out of the $^{59}$Fe seed there are essentially no other factors other
than the abundance of neutrons and the \feng~cross section influencing
\fe~production at this stage for a given model. Indeed, the depletion of
$^{59}$Fe is also a factor when comparing the models with fiducial cross section
relative to the depressed cross section, though to a lesser extent. In the test
problem we have used (bottom panel), there is approximately 1.15 times less
$^{59}$Fe in the fiducial model compared with the low cross section model (solid
lines). This is approximately the factor required to correct the spead around
enhancement factors $8-10$ up to factors of 10.

Lastly, after the supernova shock has passed we are left with the yield. A
similar trend can be seen for the yields (right panel) as we have described for
the ``cut", however the yields have more apparent noise in them and this noise
is exacerbated for higher explosion energy. The reason for the additional noise
is that in shock nucleosynthesis the burning takes place at much higher
temperatures than during the star's evolution where many different reaction
channels can contribute to the abundances of $^{59}$Fe and \fe, as has already
been discussed in Section~\ref{sec:CCSNresults} and illustrated in
Figure~\ref{fig:rates}. The interplay or competition between the now-several
reaction branchings complicates the situation, resulting in a larger scatter
that depends quite sensitively on the peak temperature and density in the shock
as it reaches the C and He shells.  At larger explosion energies, peak
temperatures will be even higher for the relevant parts of the star and even
more reaction channels will be opened.

As we have shown in Figure~\ref{fig:yields_standard} (black points), if we
consider that the majority of SNe will have explosion energies below
$2\times10^{51}$~erg, the total \fe~yield exhibits a factor of $\sim2-3$
sensitivity to the explosion energy and how the explosion ensues. Combining this
with the range $\sim 0.1-7$ in the reaction rate the total uncertainty in the
\fe~yield from a CCSN covers approximately two orders of magnitude.

\begin{figure*}
	\centering
	\caption{\emph{Left panel:} Maximum observable distance of the ($\gamma$-ray) decay
		lines from the $^{60}$Fe$\rightarrow^{60}$Co$\rightarrow^{60}$Ni decay chain,
		calculated using the $^{60}$Fe yields from our CCSN simulations assuming a detector
		sensitivity of $5\times10^{-7}$~MeV~cm$^{-2}$~s$^{-1}$ for different ealizations of
		the \feng~cross section. \emph{Right panel:} ejected mass of \fe~for all of our
		simulations, with lower limits on the detection of \fe~in a remnant with a
		given distance from Earth and a given detector sensitivity (horizontal lines).
		Progenitor mass at the ZAMS is indicated by color:
		15~\msun~(blue), 20~\msun~(yellow), 25~\msun~(red).
	}
	\label{fig:generic-detect}
	\includegraphics[width=\textwidth]{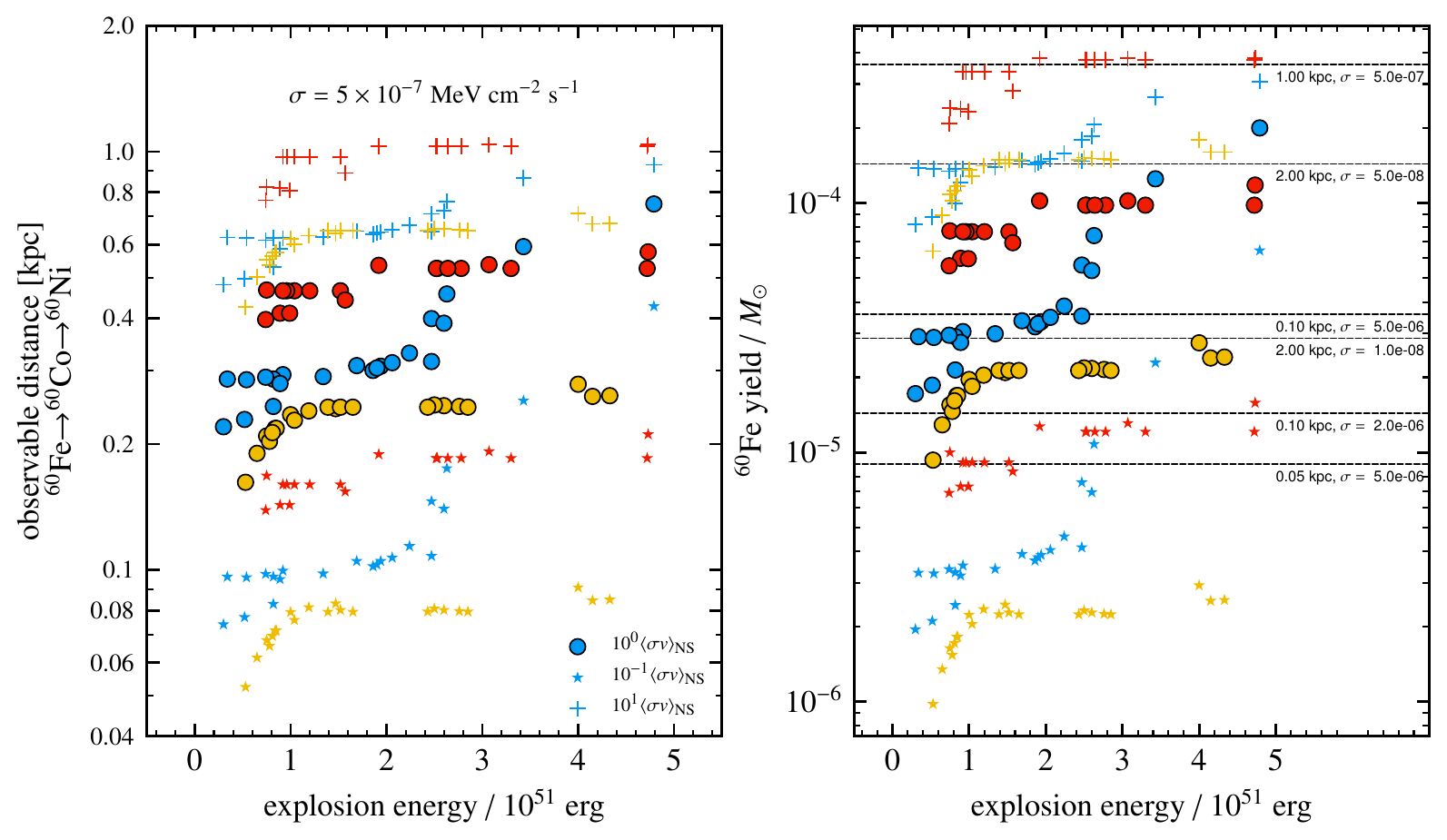}
\end{figure*}

\begin{figure}
	\centering
	\caption{Gamma-ray fluxes at Earth as a function of the remnant's age for a range of
		\fe~ejecta masses assuming the SNRs are point sources either 1~kpc or
		0.5~kpc from Earth.
		}
		\label{fig:gamma_flux}
		\includegraphics[width=\linewidth]{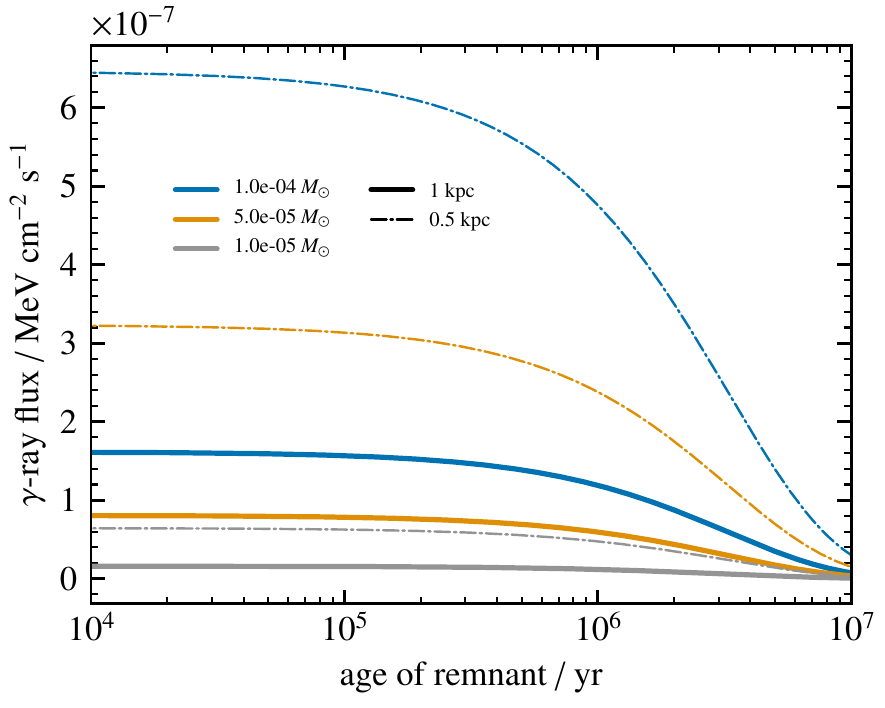}
\end{figure}

\section{Prospects for $\gamma$-ray telescopes}

\subsection{Measuring \fe~in SNRs}

In the past, comparisons of $^{60}$Fe yields have been limited to the total
abundance of this radioactive isotope in the interstellar medium.  \fe~is
inferred from the detection of gamma-lines from the decay of its short-lived
daughter \co. With a CCSN rate of approximately two events
per century \citep{Diehl2006a} and a half-life of $2.6$~Myr, the observed
diffuse $^{60}$Fe abundance pattern is composed of approximately 50,000 SNe.
However, with next generation gamma-ray telescopes, it may be possible to
observe \fe~from individual SNRs if they are sufficiently close to Earth. In
this section we examine the prospects for observing \fe~decay lines in
individual SNRs based on our simulation results.

For a given mass of $^{60}$Fe ejected from a supernova, the radiant power of the
decaying
$^{60}$Fe is approximately
\begin{equation}
	\Phi \approx \dfrac{M_{60}N_\mathrm{A}}{60}\lambda_{60}E_\gamma
	\quad(\mathrm{MeV~s}^{-1}),
\end{equation}
where $M_{60}$ (g) is the ejected mass of $^{60}$Fe, $\lambda_{60}=\ln(2) /
\tau_{1/2}(^{60}\mathrm{Fe})$ is the decay rate of $^{60}$Fe (s$^{-1}$),
$E_\gamma$ is the decay energy (MeV) and $N_\mathrm{A}$ is Avogadro's number.
There are two relevant decay lines at \mbox{1.17} and \mbox{1.33}~MeV ($1.06
\times 10^{-2}$~\AA~and $9.32\times 10^{-3}$ \AA, respectively).

The flux of the radiation at a distance $r$~cm from the source is then of course
\begin{equation}
	F(r) = \dfrac{\Phi}{4\pi r^2} \quad(\mathrm{MeV~cm}^{-2}\mathrm{~s}^{-1}),
\end{equation}
from which we can determine the maximum distance that the $^{60}$Co decay lines
would be detectable from a supernova remnant as a function of the ejected mass
of $^{60}$Fe, given the detector line sensitivity at the relevant energy
$\sigma$~(MeV~cm$^{-2}$~s$^{-1}$). The expression for this is
\begin{equation}
	r_\mathrm{max} = \left(\dfrac{M_{60}N_\mathrm{A}\lambda_{60}E_\gamma}{240\pi\sigma}\right)^\frac{1}{2}.
\end{equation}
The maximum observable distance of each of our simulated CCSNe is shown in the
left panel of Figure~\ref{fig:generic-detect} for a detector line sensitivity of
$\sigma=5\times10^{-7}$~MeV~cm$^{-2}$~s$^{-1}$. In the right panel of
Figure~\ref{fig:generic-detect} the total \fe~yield for each of the simulations
is plotted, and horizontal lines are drawn that indicate the lower limit of the
\fe~ejecta mass from a supernova that would be detectable as a point source for
several combinations of detector line sensitivity and distance of the SNR from
Earth. The gamma-ray fluxes at Earth for three \fe~ejecta masses are shown in
Figure~\ref{fig:gamma_flux} assuming that the SNRs are point sources at either
1~kpc or 0.5~kpc from Earth.

The number of detectable remnants depends of course upon the number of nearby
SNRs. A simple estimate can be obtained by assuming supernovae are uniformly
distributed in the galactic disk.  Most of the Milky Way stars lie in a thin
disk with a scale height below 0.6~kpc.  Assuming a uniform distribution of SNe
out to 15~kpc and a Galactic SN rate $R_\mathrm{SN}$, we would
expect the number of observable supernova remnants within $r$, distance to Earth
(kpc), to be:
\begin{equation}
	N_{\rm SNR}(r)  =  \tau_{1/2}R_{\rm SN} \left(\frac{r}{15}\right)^2.   
\end{equation}
With a supernova rate $R_{\rm SN}\approx0.02$~yr$^{-1}$, we expect about 230~SNe
within 1~kpc.  Even if we only observe out to 0.5\,kpc, we expect to see roughly
60 supernova remnants.  

\begin{figure}
	\centering
	\caption{Observed distribution of SNRs as a function of distance from the
		Earth.  The solid curve uses the entire sample from the
		SNRcat\citep{ferrand12}.  The dotted line removes any pulsar-only determined
	remnants from the catalogue.}
	\label{fig:SNRdist}
	\includegraphics[width=\columnwidth,trim=0 100 0 100, clip]{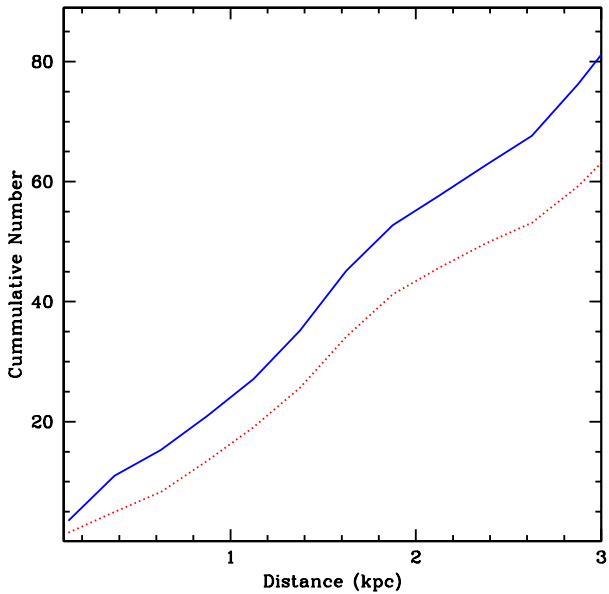}
\end{figure}

So how does this estimate compare to the known supernova remnants? The supernova
remnants observed in high energy emission are compiled in the SNR catalogue,
SNRcat\footnote{\href{http://www.physics.umanitoba.ca/snr/SNRcat/}{http://www.physics.umanitoba.ca/snr/SNRcat/}}
\citep{ferrand12}.  If we restrict our sample to CCSN remnants with known
distances and assume that the probability of the remnant is flat within the
distance errors in the catalogue, we can estimate the SNR distribution as a
function of the distance from the Earth (Figure~\ref{fig:SNRdist}).  The bulk of
the remnants in our sample have a pulsar wind nebula and hence are known to be
core-collapse. However, we include all remnants that are not known to be
thermonuclear supernovae and therefore a few of the systems we have selected,
perhaps 10--20\%, may be thermonuclear supernova remnants and not core-collapse.
If we include all remnants (whether seen as ejecta or compact remnants), we find
a distribution that increases linearly with radius (in the thin disk
approximation, we would expect a complete sample to increase as radius squared).
However, this sample includes some compact remnants observed through the pulsar
or pulsar wind (i.e.  pulsar wind nebulae).  If we remove these systems, the
number of remnants within 1\,kpc is closer to 10-15 remnants (5 within
0.5\,kpc).

The differences between remnant observations and our simple supernova rate
estimates are, in part, due to the fact that SNRs are typically observed while
young and still hot from the passage of the reverse shock.   SNRs evolve through
a series of phases:  free streaming, Sedov-Taylor, and snowplow.  In the
free-streaming phase, the velocity remains roughly constant.  In the adiabatic
Sedov phase, the velocity of the shock is:
\begin{equation}
v_{\rm shock} \approx (E_{\rm SN}/\rho_{\rm ISM})^{2/5} t^{-3/5}
\end{equation}
where $E_{\rm SN}$ is the supernova energy, $\rho_{\rm ISM}$ is the interstellar medium density and 
$t$ is the time.  When cooling 
alters the flow (around 20,000\,y), 
momentum conservation dictates 
the velocity of the ejecta:
\begin{equation}
	v_{\rm shock}=m_{\rm Sedov} v_{\rm Sedov}/(m_{\rm ejecta}+m_{\rm swept up})
\end{equation}
where $m_{\rm Sedov}$ is the mass at the end of the Sedov phase, $v_{\rm Sedov}$
is the velocity at the end of the Sedov phase and $m_{\rm swept up} = 4/3 \pi
r_{\rm SNR}^3 \rho_{\rm ISM}$ with $r_{\rm SNR}$ the radius of the supernova
remnant shock and $\rho_{\rm ISM}$ is the ISM density.  We include all three
phases in our evolution.  As the shock moves out through the snowplow phase, it
cools radiatively (cooling timescales are roughly $10^5$~yr).  But even after it
cools and is no longer visible, it continues to expand until its velocity
decelerates to the sound speed.  Typically, remnants are no longer visible at
the end of the cooling phase where the velocities are around
50~km~s$^{-1}$.

\begin{figure}
	\centering
	\caption{Angular size for a remnant at 1\,kpc as a function of time for a range of
		supernova explosion and interstellar medium properties.
		The angular resolution of AMEGO is $3^\circ$ and it will
			observe these remnants as point sources up to 100,000
		years.
		For nearby systems, especially with the LOX angular resolution,
		it may be possible to map out the $^{60}$Fe. }
	\label{fig:SNRsize}
	\includegraphics[width=\columnwidth,trim=0 10 0 20, clip]{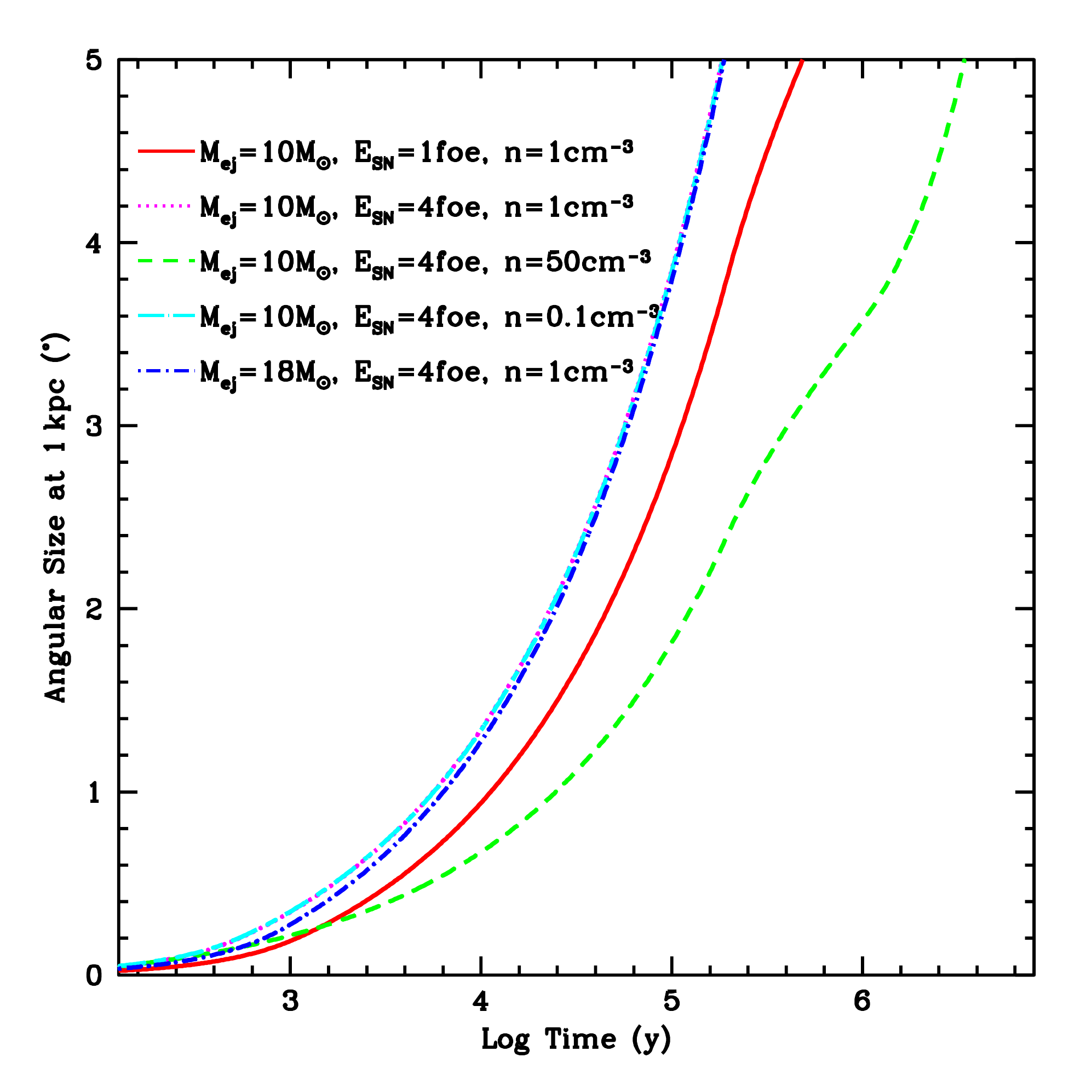}
\end{figure}

However, observations of \fe~are possible to much later times. The $2.6 \times
10^6$~yr lifetime of $^{60}$Fe means that its flux will be roughly constant for
up to about one million years (Figure~\ref{fig:gamma_flux}).  The limitation on
observing these remnants will be the timescale for the ejecta to disperse within
the Galaxy.  For the purposes of this study, we assume this dispersal time
occurs when the ejecta have an angular size larger than the angular resolution
of a typical gamma-ray telescope and that after the SNR velocity decelerates to
the sound speed, we assume that the $^{60}$Fe continues to diffuse into the
Milky Way at the sound speed.  Under these assumptions, we can calculate the
size of the remnant as a function of time.  Figure~\ref{fig:SNRsize} shows the
angular size of SNRs at a distance of 1~kpc for a range of properties of the
explosion and the ISM.  For SNRs relevant to this study, the characteristic age
is 2-3 million years, 20-30 times longer than SNRs can typically be observed in
other EM bands, suggesting that we may see 20-30 times more remnants than the
number predicted by our current sample.  Based on the observed SNR distribution
and the long observing times of \fe~in SNRs, we expect that roughly 200-300 SNRs
within 1~kpc could be detectable, on par with our simple rate estimate.  For
remnants older than 100,000 years, the remnant size will be bigger than the
$3^\circ$ angular resolution of proposed telescopes such as AMEGO and hence will
no longer be a point source in the telescope.  This will reduce the detection
probability but without detailed detector simulations it is difficuly to
quantify by how much. At best, our estimate can be considered an
upper limit.

The majority of CCSN remnants should have had progenitors with ZAMS masses less
than 15~\msun. Assuming a Salpeter IMF, roughly half of the CCSN remnants should
have progenitors with ZAMS masses in the range $8\lesssim M_\mathrm{ZAMS}/\msun
\lesssim 12$. This should not drastically affect our prediction for the number
of detections because the \fe~yield from lower mass stars is similar to the
yields we obtain for 15--25~\msun~stars (between roughly $10^{-6}$ and
$10^{-4}$~\msun) and certainly within the considerable uncertainties
\citep[e.g.][]{Woosley1995}. Furthermore, at the low-mass end are ECSNe that may
collapse into neutron stars or explode as thermonuclear supernovae \citep[see,
e.g.][and references therein]{Jones2016a,Jones2018a}. In the former case,
simulations suggest that the \fe~yield will be around $4\times10^{-5}~\msun$
\citep{Wanajo2013a} and in the latter case around $3\times10^{-3}~\msun$
\citep{Jones2018a}.

\begin{figure}
	\centering
	\caption{Average remnant shock velocity  as a function of time for a range of supernova
	explosion and interstellar medium properties.  We assume the $^{60}$Fe is mixed throughout the remnant and has has this same velocity.}
	\label{fig:SNRvel}
	\includegraphics[width=\columnwidth,trim=0 10 0 20, clip]{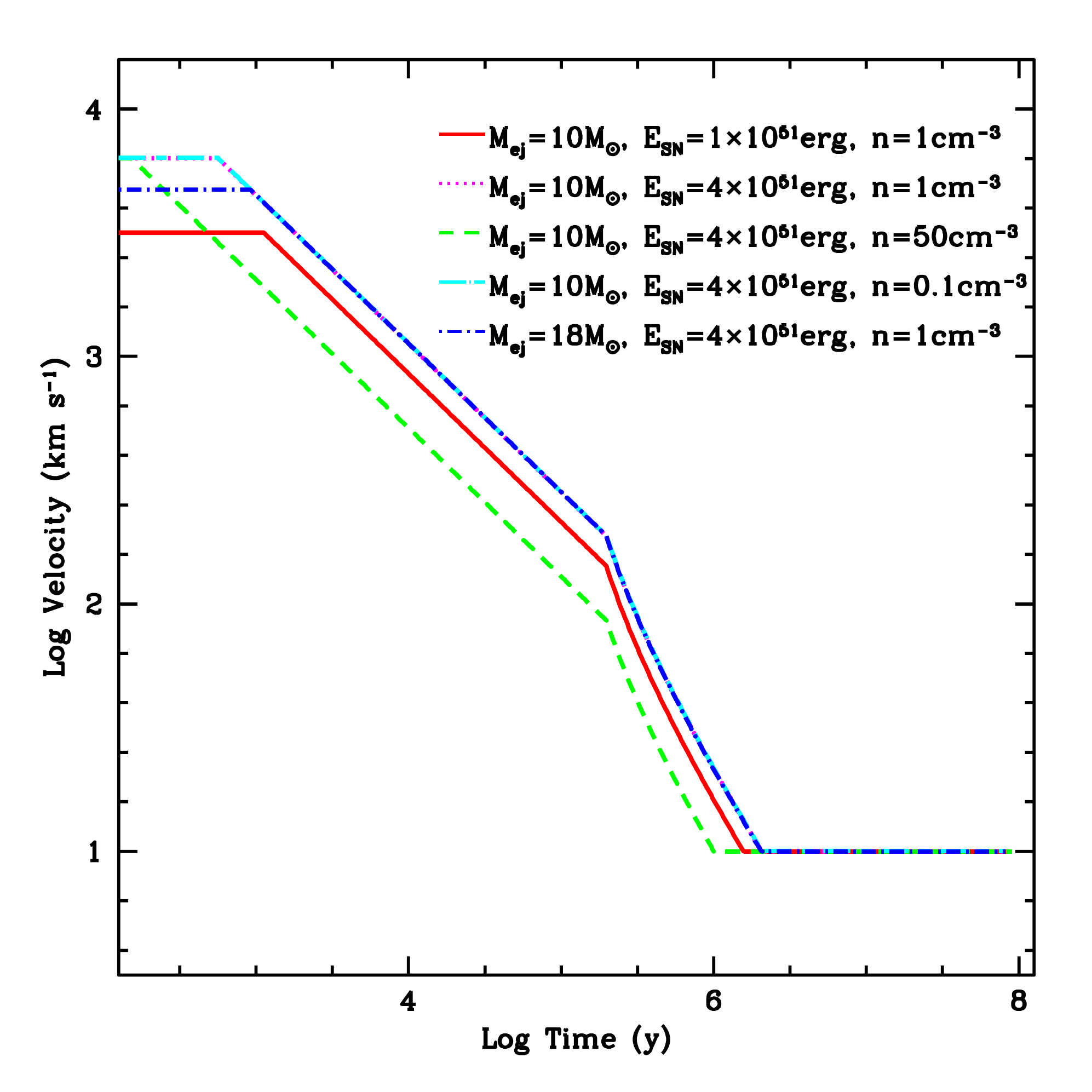}
\end{figure}

The corresponding velocities versus time for the remnant models in
Figure~\ref{fig:SNRsize} are shown in Figure~\ref{fig:SNRvel}.  By 1 million
years, most remnants will decelerate sufficiently to mix with the ISM (expansion
velocity on par with the sound speed of the ISM).  At 100,000~yr, the velocities
can be an order of magnitude above the sound speed and the remnant can be
identified by the Doppler broadening of the emission lines.

For our angular size and remnant velocities, we have assumed densities ranging
from $0.1-50~ {\rm cm^{-3}}$, consistent with what we expect in the ISM.  If the
SN exploded inside of a molecular cloud, the density could be higher, meaning
that at a given time the velocity would be lower and the angular size would be
smaller.  On the other extreme, it has been argued that many supernovae occur in
superbubbles of extremely low-density medium \citep{kretschmer13,krause15}.  In
such cases, the velocities will be higher and angular sizes larger than our
estimates.

\subsection{Future $\gamma$-ray missions}

A number of new missions have been proposed as next generation $\gamma$-ray
satellites that are well-suited to measuring $^{26}$Al and $^{60}$Fe in
supernova remnants:  e.g., All-sky Medium Energy Gamma-ray Observatory
(AMEGO),\footnote{\url{https://asd.gsfc.nasa.gov/amego/}}
e-ASTROGAM,\footnote{\url{http://eastrogam.iaps.inaf.it}} Compton Spectrometer
and Imager (COSI),\footnote{\url{http://cosi.ssl.berkeley.edu}}
Electron-Tracking Compton Camera (ETCC) and Lunar Occultation Explorer (LOX).
Many of these missions remain in design phase, so the exact sensitivities for
these missions remain to be determined.  But a sensitivity of $5 \times
10^{-7}~{\rm MeV~cm^{-2}~s^{-1}}$ is within reach of these next generation
missions.  Even if the signal is detectable, we must be able to distinguish the
signal from a specific remnant from the diffuse emission.  For 100,000~yr old
remnants, the velocity broadening of the decay line can be used to distinguish
the remnant from diffuse emission and germanium detectors such as those used on
COSI would have sufficient spectral resolution to measure this line broadening.

\section{\fe/\al~ratio}

\subsection{Early Solar-System Value}

While it is well established that the early Solar System started out with an
\al/\ex{27}Al ratio of $5.2 \times 10^{-5}$ \citep{jacobsen08}, the initial
abundance of \fe/\ex{56}Fe was disputed until recently. While bulk measurements
of early Solar System phases consistently result in an initial \fe/\ex{56}Fe
ratio of $(1.01 \pm 0.27) \times10^{-8}$ \citep{tang12a,tang15}, in situ
measurements by secondary ion mass spectrometry (SIMS) of individual phases show
higher \fe/\ex{56}Fe ratios of up to $10^{-6}$
\citep{mishra14a,mishra14b,mishra16,telus18}. These SIMS measurements also show
large spread in the determined ratios, indicating heterogeneity in the early
Solar System's \fe/\ex{56}Fe abundance. While the low initial can be explained
as galactic background \citep{tang12a,tang15}, high values of \fe\ are generally
interpreted as a smoking gun for the injection of freshly synthesized supernova
material into the solar nebula just prior to its birth.  Recent work however
\citep{trappitsch18} showed that isotope ratio measurements done by SIMS suffer
from correlated effects. This technique is limited to measure \ex{60}Ni (the
decay product of \fe), \ex{61}Ni, and \ex{62}Ni in meteoritic inclusions.  Due
to the low abundance of \ex{61}Ni, correlated effects in the data evaluation can
show up as an enhancement of \ex{60}Ni if not properly accounted for, which can
be interpreted as a high initial \fe/\ex{56}Fe ratio. The new measurements by
\citet{trappitsch18} found no excess \fe~in the analyzed sample and their
initial \fe/\ex{56}Fe for the Solar system is consistent with the low values
from \citet{tang12a} and \citet{tang15}, as well as with a reevaluation of the
previous measurement of the same sample by \citet{telus18}. \citet{trappitsch18}
thus concluded that it is unlikely that a supernova injection provided the
\al~that was present in the early Solar System, since such an injection would
over predict the abundance of \fe. 

\begin{figure*}
	\centering
	\caption{Predicted \fe/\ex{56}Fe ratio in the early Solar System when contributing the
	correct amount of \al. We used a time delay of $10^5$\,years between the explosion and the
formation of the first Solar System solids.}
\includegraphics[]{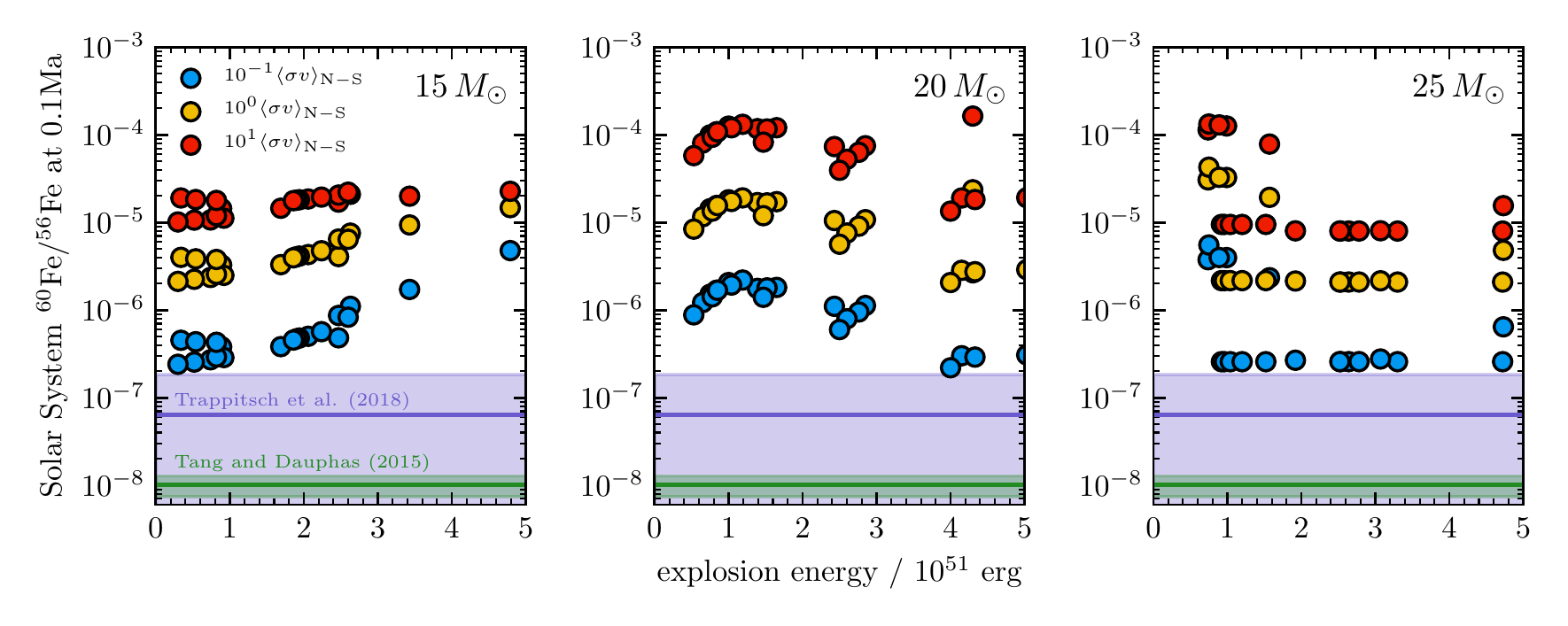}
\label{fig:fe60_solar_system}
\end{figure*}

Using our new yields we can calculate if injecting freshly synthesized CCSN
material into the solar nebula would result in an overabundance of \fe~compared
to the measurements. This calculation is based on the assumption that the same
injection supplied the \al~ abundance in the early Solar System. We can
calculate the mass fraction of the ejecta $x$ that needs to be incorporated in
the Solar System to explain the initial \al/\ex{27}Al ratio. Setting the Si
abundance to $10^6$ atoms in the Solar System, the initial \ex{27}Al$_\tn{ss}$
and \ex{56}Fe$_\tn{ss}$ abundances in the early Solar System (SS) are $8.46
\times 10^4$ and $7.78 \times 10^5$, respectively \citep{lodders09}. Let us
assume that all \al~and \fe~is supplied by the supernova injection. This
assumption is likely true for \al, however, some galactic background can be
expected for \fe~\citep[see, e.g.,][]{lugaro18}. An \fe~addition from galactic
background however would only increase the \fe/\ex{56}Fe level compared to one
calculated here, thus making this calculations a best case scenario. Defining
the composition of a given isotope in the stellar ejecta as \ex{i}X$_\tn{ej}$,
we can write the Solar System's initial \al/\ex{27}Al as
the following mixture:
\begin{equation}
 \frac{^{26}\tn{Al}_\tn{ej} \cdot x}{^{27}\tn{Al}_\tn{ej} \cdot x + (^{27}\tn{Al}_\tn{ss} - ^{27}\tn{Al}_\tn{ej} \cdot x)} = 5.2 \times 10^{-5}
\end{equation}
The amount of \ex{27}Al$_\tn{ej}$ cancels out since the measured amount of
\ex{27}Al$_\tn{ss}$ already includes it. We can thus solve for $x$ and calculate
the mass fraction of the injection as:
\begin{equation}\label{eqn:massfrac_nodecay}
 x = 5.2 \times 10^{-5} \cdot \frac{^{27}\tn{Al}_\tn{ss}}{^{26}\tn{Al}_\tn{ej}}
\end{equation}
If we additionally account for a free decay time $t$ between the time the \al~is
ejected from the supernova and the time it is incorporated into the early Solar
System, equation
\eqref{eqn:massfrac_nodecay} can be rewritten as:
\begin{equation}\label{eqn:massfrac_decay}
 x = 5.2 \times 10^{-5} \cdot \frac{^{27}\tn{Al}_\tn{ss}}{^{26}\tn{Al}_\tn{ej} \exp(-\lambda_{26}t)}
\end{equation}
Here $\lambda_{26}$ is the decay constant of \al. With the calculated mass
fraction we can now calculate the predicted Solar System initial \fe/\ex{56}Fe
content as:
\begin{equation}
 \left(\frac{^{60}\tn{Fe}}{^{56}\tn{Fe}}\right)_\tn{mod} = \frac{\ex{60}\tn{Fe}_\tn{ej} \cdot x \cdot \exp(-\lambda_{60}t)}{^{56}\tn{Fe}_\tn{ss}}
\end{equation}
Here $\lambda_{60}$ is the decay constant of \fe. 

Figure~\ref{fig:fe60_solar_system} shows the comparison of our models with the
measurement limits by \citet{tang15} and \citet{trappitsch18} for bulk and in
situ analyses, respectively.  We calculated the amount of \fe~in the early Solar
System as described above.  A time delay between the SN explosion and the
formation of the first Solar System solids of $10^5$ years is assumed. Using
giant molecular cloud formation simulations, \citet{vasileiadis2013} showed that
some SN injection events into potential solar nebulae can already happen a few
$10^5$ years after explosion. We thus implement this short delay time between SN
explosion and Solar System injection. Longer (and maybe more realistic) delay
times will only raise the initial \fe/\ex{56}Fe with respect to the
\al/\ex{27}Al due to the difference in half-lifes.
Figure~\ref{fig:fe60_solar_system} clearly shows that none of the model results
agree with the low limits of \fe~ that are found in meteorites. Based on our
simulations of 15--25~\msun~stars, we can therefore exclude that a SN injection
from a $15$--$25~\msun$ star (at solar metallicity) triggered the formation of
the Solar System and injected the \al~and \fe.

\begin{figure*}
	\centering
	\caption{Predicted \fe/\al~line flux ratio after $10^6$ years for all of our models.
	The dashed horizontal line is the INTEGRAL/SPI measurement and the shaded blue region
	represents the measurement error \citep{Wang2007a}.}
	\includegraphics[]{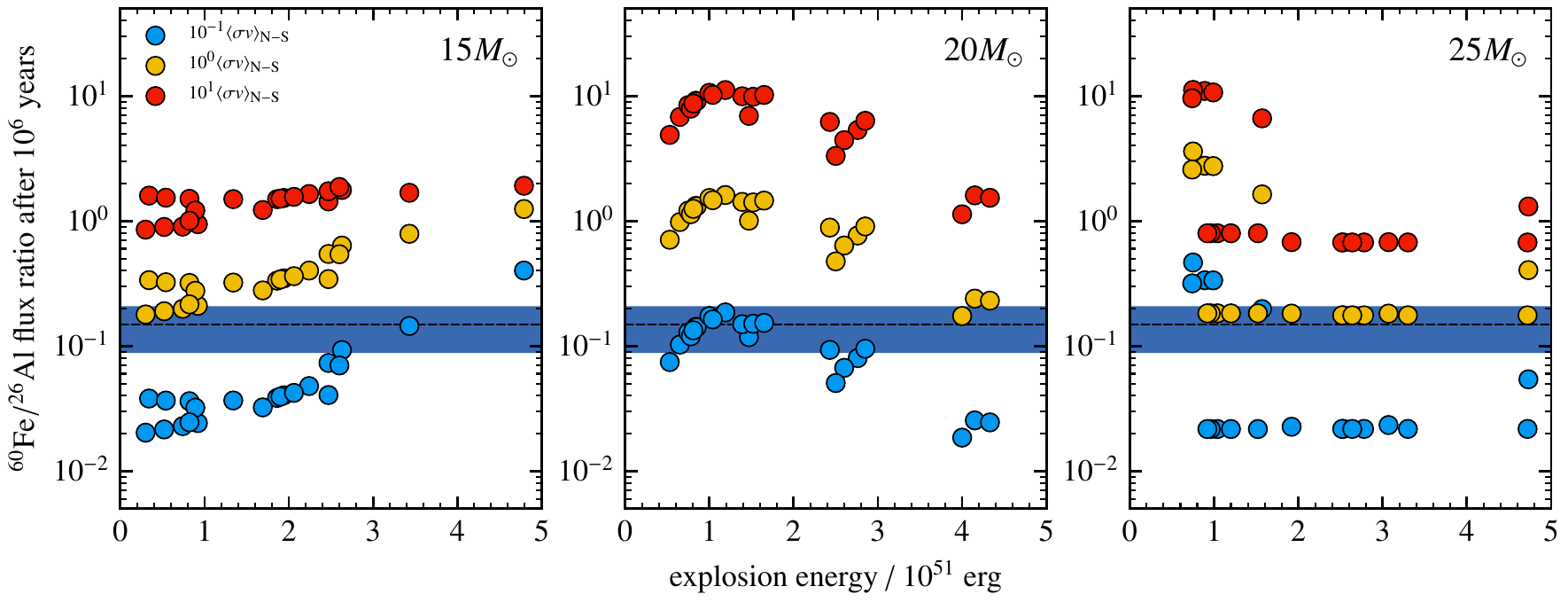}
    \label{fig:flux_ratio}
\end{figure*}

\subsection{Diffuse gamma-ray emission in the ISM}

With the uncertainty we have presented in the \fe~yields of CCSNe arising from a
combination of the respective uncertainties in both the \feng~cross section and
explosion mechanism, it becomes more difficult to say how many supernovae
contributed to the Galactic inventory of \fe~and \al~that can now be observed in
the diffuse ISM, as \citet{Tur2010a} have already shown in the case of
uncertainties in the $3\alpha$ and
$^{12}\mathrm{C}(\alpha,\gamma)^{16}\mathrm{O}$ reaction rates.

In order to use the observed diffuse line flux ratio as a constraint for stellar
evolution and supernova theory, one ideally needs to produce a grid of massive
star models with good coverage of the initial mass space and including
Wolf-Rayet stars, as has been done by \cite{Limongi2006a,Limongi2018a}. This is
not the intention of our study, however we think that it is a valid point to
illustrate the range of \fe/\al~line flux ratios that are possible arising from
different modelling of the energy deposition behind the shock in the CCSN and
the uncertainty in the \feng~cross section.

The line flux ratios for all of our models are shown in
Figure~\ref{fig:flux_ratio} after the ejecta have decayed for $10^6$~yr. The
majority of the models predict line ratios too high compared to the INTEGRAL/SPI
measurement \citep{Wang2007a}, which is plotted with a horizontal dashed line
and a shaded blue region that indicates its error bar.  There are some 15 and
25~\msun~models for which it is possible to obtain the measured ratio when using
the \feng~cross section from \citet{Rauscher2000a}, but the only 20~\msun~models
coming close are those with unrealistically high explosion energies ($\gtrsim
4\times10^{51}$~erg). This does not necessarily rule out this cross section
though, because the diffuse ISM contains the integrated yields from many
supernovae. Figure~\ref{fig:flux_ratio} does suggest, however, that we may be
able to rule out the possibility of the cross section being as large as 10 times
the value from NONSMOKER because models assuming that large a cross section
never get close to the measurement.

The \al~abundance in the ejecta is known to be increased by neutrino
interactions by up to about 50~per cent, which would bring the points in
Figure~\ref{fig:flux_ratio} down a little, but not enough to bring the red
points into agreement with the measured flux ratio.  Accounting for Wolf-Rayet
winds would also boost the amount of \al~in the ISM, however what one needs to
assume about the mass loss rates in order to reproduce the INTEGRAL measurement
will clearly depend upon what the actual \feng~cross section is. Therefore, if
nothing else this could be a useful exercise for constraining Wolf-Rayet star
winds when we finally are able to measure the cross section. While it is true
that there will still be outstanding uncertainties in the \fe~yields from CCSNe
owing to the uncertainty in the explosion mechanism, this appears to be ``only''
on the order of a factor of $2-3$. Progress in the right direction will also
require pinning down the $^{12}\mathrm{C}(\alpha,\gamma)^{16}\mathrm{O}$ cross
section \citep[see][for a recent review]{deBoer2017a}.

\section{X-ray emission}

\citet{Leising2001a} has considered that several long-lived radionuclides that
decay via (atomic) electron capture could be detectable in SNRs via their X-ray
emission following the radiative stabilization of the ion. One particularly
interesting prospect is $^{55}$Mn, from the decay of $^{55}$Fe produced in
SNe~Ia \citep{Seitenzahl2015a}. In this section we consider that
$^{60}$Co~\textsc{ii} (frequently and less precisely denoted \co$^*$ in nuclear
physics terms), produced in the decay of \fe, undergoes internal conversion
resulting in a similar ionization, stabilization and X-ray emission chain of
events. We first model the radiative decay cascade of \co~and go on to examine
the incident X-ray fluxes at Earth and the prospects of detecting them.

\subsection{X-ray spectrum from $^{60}$Co}

In this section we investigate the radiation that can be produced by atomic
transitions in Co by considering the radiative stabilization that occurs when
starting from a $1s$- or $2s$-hole electron configuration of $^{60}$Co. More
specifically, we assume that a neutral $^{60}$Fe atom, described by the electron
ground-state configuration $1s^2 2s^2 2p^6 3s^2 3p^6 3d^6 4s^2$, undergoes beta
decay to produce singly ionized cobalt, i.e. $^{60}$Co~\textsc{ii}, with the same
configuration labeling and an excited nuclear state. The two $4s$ valence
electrons quickly undergo radiative stabilization to the $3d$ subshell to form
the $1s^2 2s^2 2p^6 3s^2 3p^6 3d^8$ ground configuration of
$^{60}$Co~\textsc{ii}.  The nucleus subsequently decays to produce a 58.6~keV
X-ray (2 per cent of the time) or eject a K-shell (81.6 per cent), L-shell (14.2
per cent) or higher-orbital (2.2 percent) electron \citep{BROWNE20131849}, . The
process by which a bound electron is ejected in this context is referred to as
internal conversion.  In the present study, we do not consider second-order
processes that produce two ejected electrons, such as electron shakeoff
\citep{mukoyama75}, which are expected to be negligible in the relevant spectral
analysis.

\begin{figure*}
	\centering
	\caption{Calculated line emission probability, see Eq.~(\ref{emiss_prob}),
	resulting from the cascade network that starts from the
	$1s_{1/2}^1 3d_{3/2}^4 3d_{5/2}^4$ configuration in $^{60}$Co~\textsc{iii}.
	Left panel: full energy range.
	Right panel: zoom-in on lower energy range from 0--100~eV.
	The photon energies have been binned at 1~eV resolution, with all lines
	within a bin being summed and the resulting curve displayed in histogram
	format.
    }
    \includegraphics[clip=true,width=.49\textwidth]{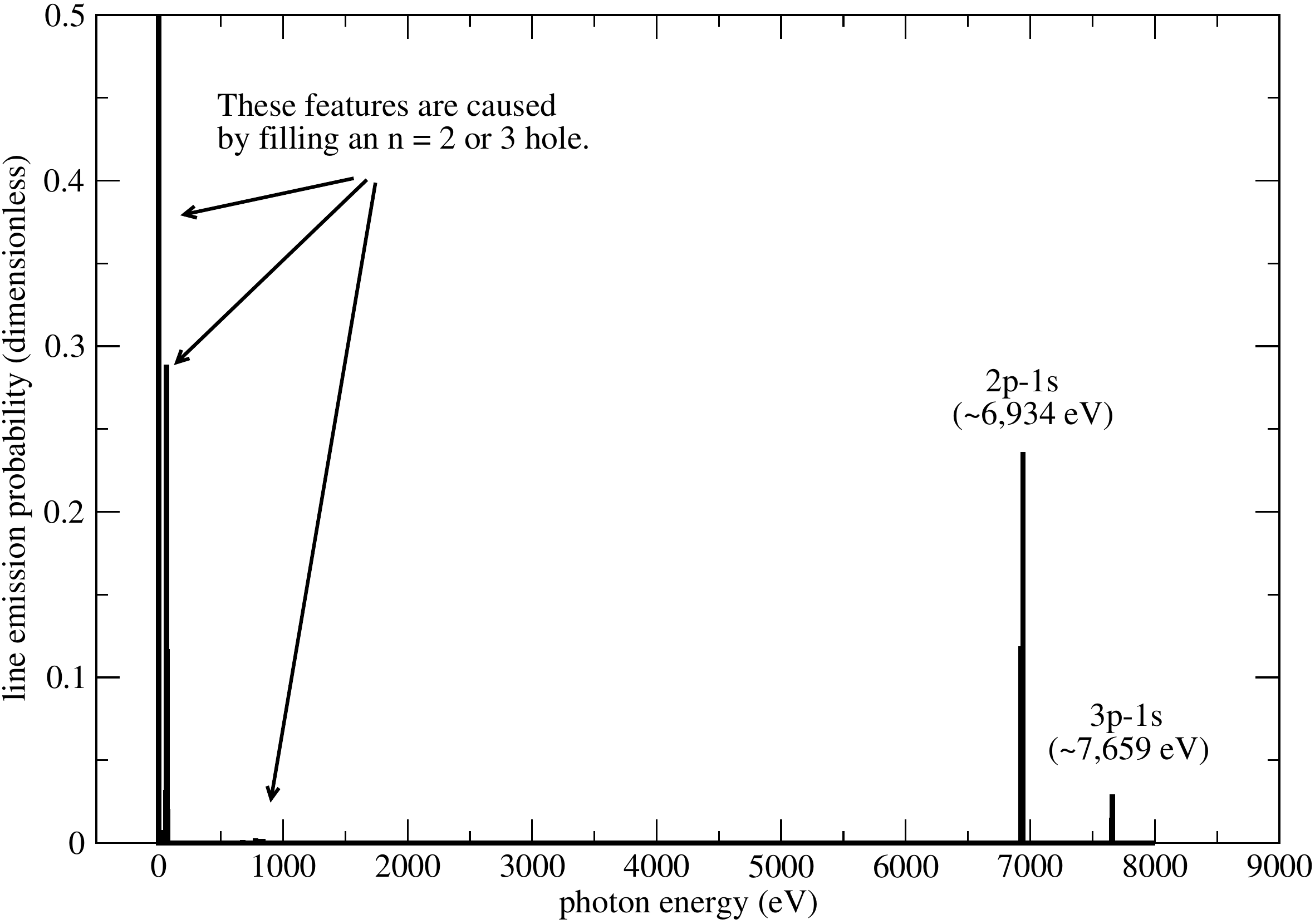}
    \includegraphics[clip=true,width=.49\textwidth]{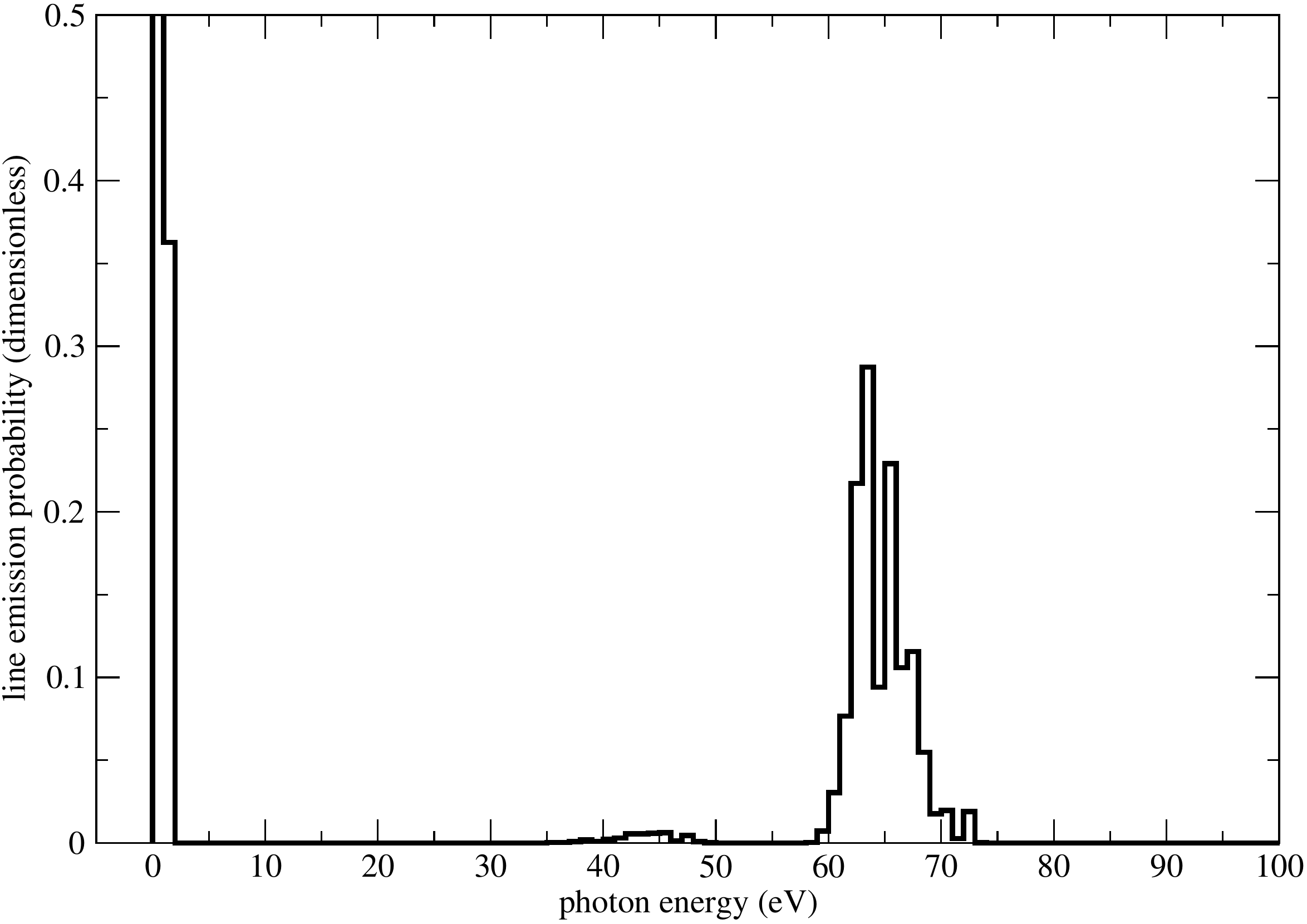}
	\label{fig:co_emis}
\end{figure*}

\begin{figure}[h]
	\centering
	\caption{
		Calculated line emission probability, see Eq.~(\ref{emiss_prob}), resulting
		from the cascade network that starts from the $1s_{1/2}^1 3d_{3/2}^4
		3d_{5/2}^4$ configuration in $^{60}$Co~\textsc{iii}. This figure provides a
		zoom-in of the high-energy region of the left panel in
		Figure~\ref{fig:co_emis}, i.e. 6800--7800~eV, in order to clearly display the
		fine-structure splitting of the $2p-1s$ and $3p-1s$ features. Again, the photon
		energies have been binned at 1~eV resolution, with all lines within a bin being
		summed and the resulting curve displayed in histogram format.
	}
	\includegraphics[clip=true,width=0.49\textwidth]{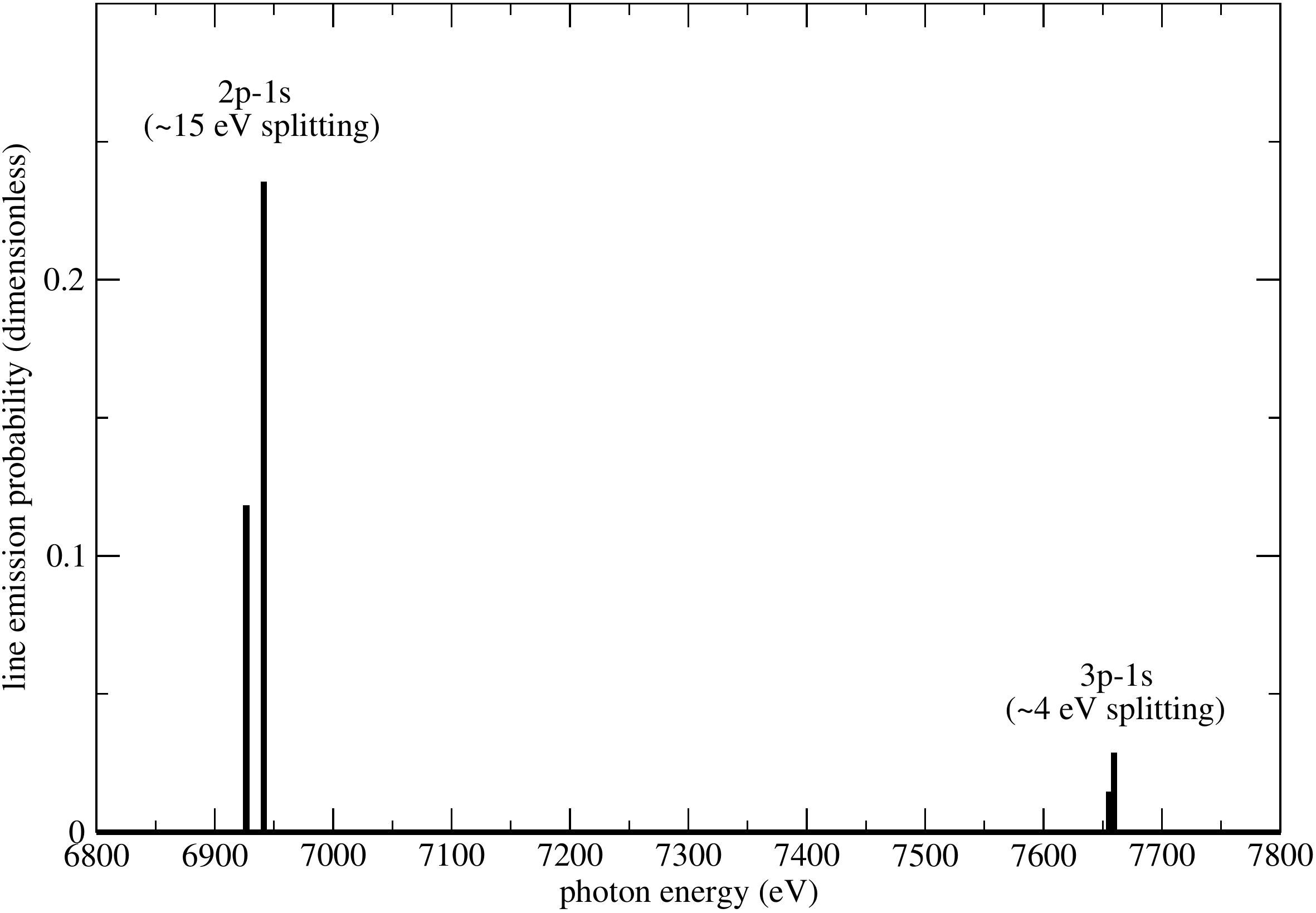}
	\label{fig:co_emis_zoom_highE}
\end{figure}

\begin{figure}
	\centering
	\caption{
		Calculated line emission probability, see Eq.~(\ref{emiss_prob}), resulting
		from the cascade network that starts from the $2s_{1/2}^1 3d_{3/2}^4
		3d_{5/2}^4$ configuration in $^{60}$Co~\textsc{iii}. The top panel displays the
		entire energy range of interest. The middle and bottom panels are zoom-ins of
		the low- and high-energy ranges given by 0--200~eV and 600-900~eV,
		respectively.  The photon energies have been binned at 1~eV resolution, with
		all lines within a bin being summed and the resulting curve displayed in
		histogram format.
	}
	\includegraphics[width=0.45\textwidth]{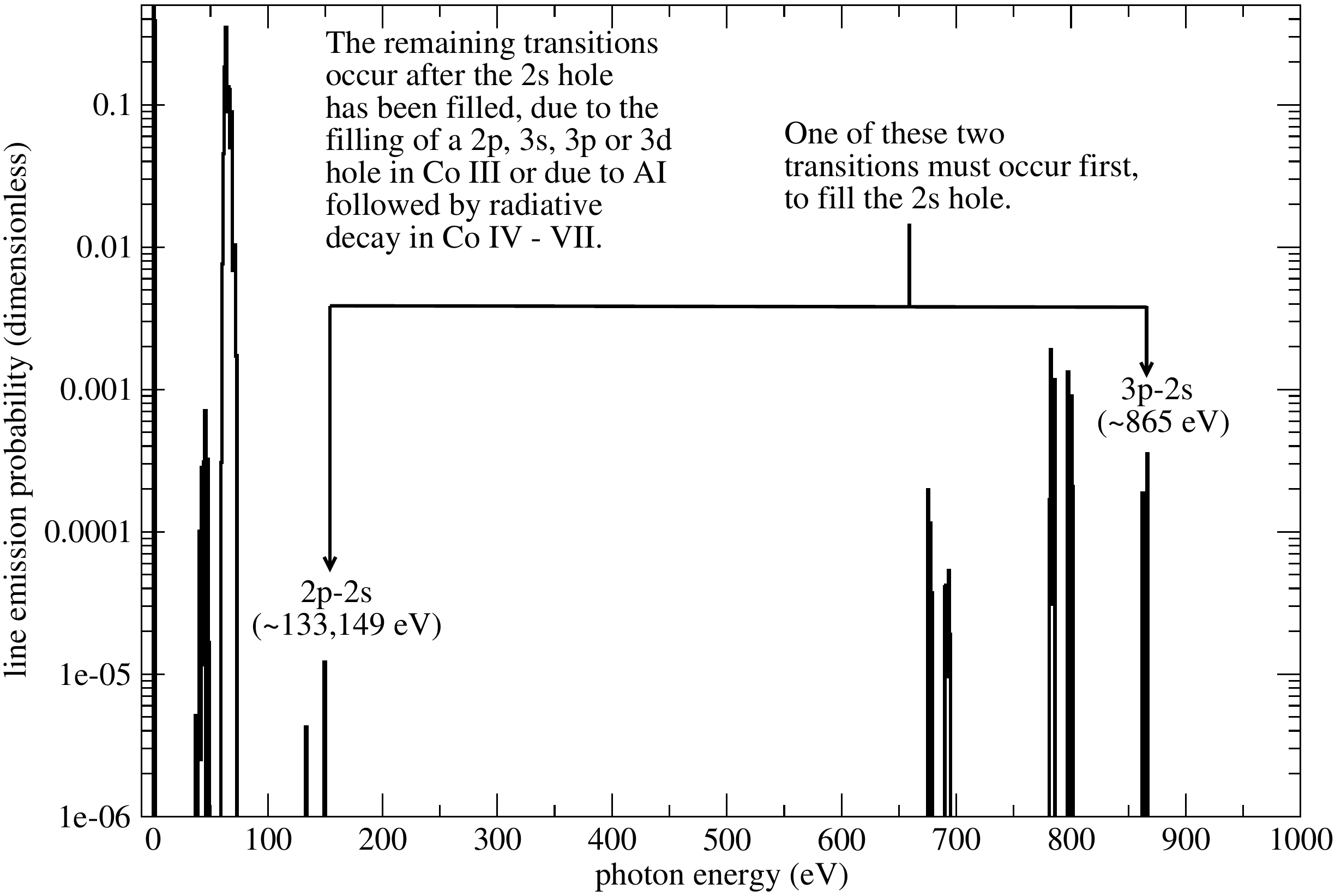} \\
	\includegraphics[width=0.45\textwidth]{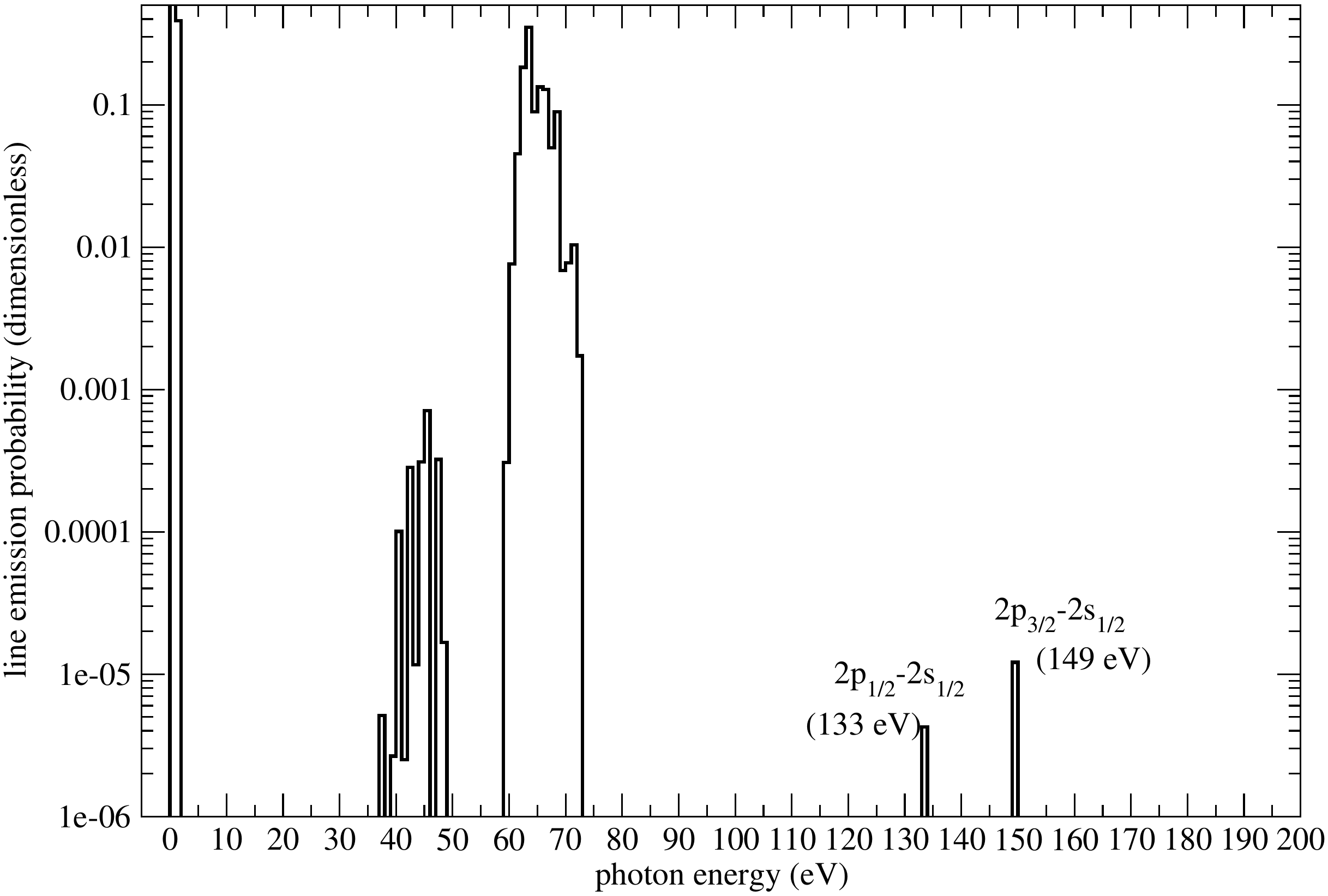} \\
	\includegraphics[width=0.45\textwidth]{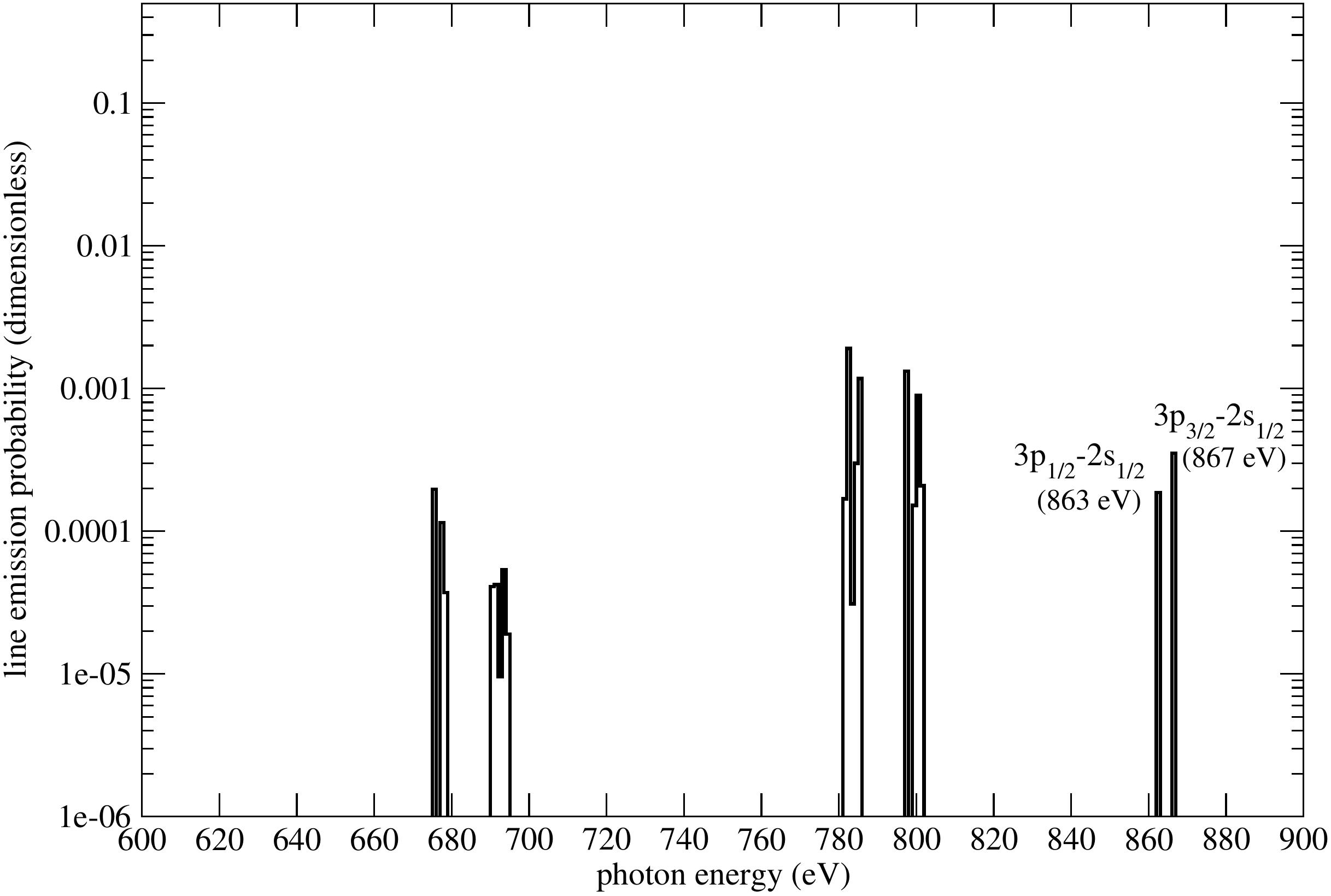}
	\label{fig:co_emis_2s}
\end{figure}

\subsubsection{K-shell electron emission}

For the K-shell option, the result is a doubly ionized cobalt ion,
$^{60}$Co~\textsc{iii}, with an electron configuration given by $1s^1 2s^2 2p^6
3s^2 3p^6 3d^8$, which can be written in the standard shorthand notation $1s^1
3d^8$ to list only the open and valence subshells.  We shall use this type of
shorthand notation for the remainder of this discussion.

To be more precise, since the electron wavefunctions used in this work were
obtained by solving the Dirac equation, the non-relativistic subshell notation,
$nl$, should be replaced with its relativistic counterpart, $nl_j$. The
subscript $j=l\pm1/2$ represents the two possible values of the total angular
momentum of the subshell that result from coupling the orbital angular momentum,
$l$, to the spin angular momentum, 1/2.  The two different $j$ values result in
subshells that have different energies, which is referred to as spin-orbit
splitting. In the relativistic notation, the starting configuration is actually
$1s_{1/2}^1 3d_{3/2}^4 3d_{5/2}^4$, where the eight $3d$ electrons are
distributed in the lowest possible energy permutation among the $3d_{3/2}$ and
$3d_{5/2}$ subshells. In the remaining discussion, we continue to use the
non-relativistic notation for convenience, except when necessary to consider the
spin-orbit splitting of the spectral emission features.

In the low-density environment of these SNRs, ion and electron collisional
processes are negligible and the line emission spectrum results from the cascade
of bound electrons via spontaneous radiative decay. In this example, the $1s$
hole is filled first, followed by the filling of any subsequent holes until the
stable ground configuration of $^{60}$Co~\textsc{iii}, $1s^2 3d^7$, is obtained.
There are many such paths to achieve this stabilization process, which form a
cascade network.  For example, the following symbolic
expression
\begin{equation}
1s^1 3d^8 \rightarrow 1s^2 2p^5 3d^8 + h\nu_1 \rightarrow 1s^2 2p^6 3d^7 + h\nu_2
\nonumber
\end{equation}
describes a 2-step cascade in which a $2p$ electron fills the $1s$ hole,
accompanied by the emission of a photon with energy $h\nu_1$, followed by a $3d$
electron filling the $2p$ hole, accompanied by the emission of a photon with
energy $h\nu_2$.

The rate at which a particular atomic state, $j$, will undergo a spontaneous
radiative decay to a state of lower energy, $k$, is given by the spontaneous
decay rate (or Einstein $A$ coefficient) denoted by $A^{\rm r}_{jk}$.  The
probability that an atom (or ion) in state $j$ will undergo such a transition to
state $k$ is determined by a quantity called the branching ratio. In the present
case, the branching ratio takes into account all of the possible radiative
transitions from state $j$ within its own ion stage, $i$, as well as all of the
possible spontaneous (but radiationless) decays from state $j$ to the next
adjacent ion stage, $i+1$, via the process of autoionization, also known as the
Auger process.

In the autoionization (AI) process, a bound electron drops to a lower subshell,
without the emission of a photon, while providing sufficient energy to ionize
another bound electron.
For example, the following symbolic expression
\begin{equation}
1s^1 3d^8 \rightarrow 1s^2 3d^6 + {\rm e}^-
\nonumber
\end{equation}
describes an AI transition in which a $3d$ electron fills the $1s$ hole and
another $3d$ electron is concurrently ionized, producing the displayed electron
configuration of triply ionized cobalt, $^{60}$Co~\textsc{iv}.  The AI rate for
a transition from state $j$ in ion stage $i$ to state $k'$ in ion stage $i+1$ is
denoted by $A^{\rm a}_{jk'}$. Note that, in general, the resulting state $k'$
may also radiatively decay and/or undergo the AI process.  Thus, the cascade
network resulting from a simple $1s$-hole configuration can produce a complex
emission spectrum with lines that arise from many ion stages.  The AI cascades
proceed from one ion stage to the next until there are no more AI options due to
energy conservation and quantum selection rules, in which case those remaining
states will simply radiatively decay (if allowed) in that final ion stage.

Taking into account the processes of spontaneous radiative decay and AI, the
branching ratio for radiative decay from state $j$ to $k$
can be written as
\begin{equation}
B^{\rm r}_{jk} = { A^{\rm r}_{jk} \over {\sum\limits_{k'} A^{\rm a}_{jk'}} +
{\sum\limits_{l} A^{\rm r}_{jl}} }\,,
\label{bratio_rad}
\end{equation}
where index $k'$ includes all states to which $j$ can autoionize and index $l$
includes all states that are accessible for radiative decay from $j$, including
state $k$.  In a similar manner, the branching ratio for AI from state $j$ to
$k'$ is given by
\begin{equation}
B^{\rm a}_{jk'} = { A^{\rm a}_{jk'} \over {\sum\limits_{k'} A^{\rm a}_{jk'}} +
{\sum\limits_{l} A^{\rm r}_{jl}} }\,,
\label{bratio_ai}
\end{equation}
For additional details about branching ratios, see Section~7 in
\citet{sampson_physrep}.

When considering the entire cascade network, the relative strength of the
emission line associated with an arbitrary radiative transition from state $m$
to $n$ is proportional to the probability to go from the starting point in the
network, i.e. the $1s_{1/2}^1 3d_{3/2}^4 3d_{5/2}^4$ configuration in the
present example, to state $m$. If we label this starting configuration with
index ``1", then the probability that state $m$ will be reached via a {\it
particular} cascade path, e.g. $1 \rightarrow a \rightarrow b \rightarrow c
\ldots g \rightarrow m$, is
given by the product of corresponding branching ratios, i.e.
\begin{equation}
P^\alpha_{1m} = B^x_{1a} B^x_{ab} B^x_{bc} \ldots B^x_{gm}\,,
\end{equation}
where the superscript $x$ can be either ``r" or ``a" to indicate that a
particular transition can be either radiative decay or AI.  A Greek superscript,
$\alpha$, is used to denote this particular cascade path. There can exist many
such paths that lead to state $m$. These probabilities are summed to obtain the
total probability of reaching state $m$ and then the braching ratio, $B^{\rm
r}_{mn}$, provides the probability that the particular radiative transition from
$m$ to $n$ will occur. Thus the {\it total} probability that a radiative
transition from $m$ to $n$ will occur, when starting from state ``1", is given
by
\begin{equation}
P^{\rm r}_{mn} = B^{\rm r}_{mn} \sum_{\beta} P^\beta_{1m}\,,
\label{emiss_prob}
\end{equation}
where the summation index $\beta$ ranges over all possible paths from the
starting state with index ``1" to state $m$.  A ``line emission probability"
spectrum can be constructed from Eq.~(\ref{emiss_prob}), with the resulting
probability plotted at the appropriate photon energy for each radiative
transition of interest.

In this work, we use the Los Alamos suite of atomic physics codes
\citep{LANL_suite} to generate all of the atomic data necessary to evaluate
Eqs.~(\ref{bratio_rad}), (\ref{bratio_ai}) and (\ref{emiss_prob}).  The
calculated line emission probability spectrum that is produced when starting
from the $1s^1 3d^8$ configuration in $^{60}$Co~\textsc{iii} is presented in
Figure~\ref{fig:co_emis}. This calculation includes lines that span seven ion
stages, ranging from Co~\textsc{iii} to Co~\textsc{ix}.  In order to ensure that
all of the emission probability is displayed in this figure, the photon energies
were binned at a resolution of 1~eV and all lines within a bin were summed, with
the resulting curve displayed in histogram format.

The spectrum was generated in the relativistic configuration-average
approximation in order to limit the size of the calculation, while still
providing a reasonable amount of fine-structure splitting in the predicted
spectrum.  A more complete fine-structure calculation would result in
significantly more line splitting in the lower photon energy range (see right
panel) due to angular momentum coupling and configuration interaction. As
presently displayed, significant emission is predicted to occur in a relatively
broad feature centered at $\sim$65~eV. Also, the two bins between 0--2~eV show
large emission probability. These low-energy photons are typically produced in
so-called ``$\Delta n=0$'' transitions that arise from electrons decaying
between the $3d_{3/2}$ and $3d_{5/2}$ orbitals in the lowest-lying
configurations of a given ion stage, i.e. $1s_{1/2}^2 3d_{3/2}^x 3d_{5/2}^y
\rightarrow 1s_{1/2}^2 3d_{3/2}^{x+1} 3d_{5/2}^{y-1} + h\nu$.  These are
non-electric-dipole allowed transitions, typically magnetic dipole (M1)
transitions, that can exhibit significant fine-structure splitting.

On the other hand, the two strong features labeled $2p-1s$ and $3p-1s$ (see left
panel) that arise from the initial filling of the $1s$ hole in Co~\textsc{iii}
are not expected to undergo significantly more fine-structure splitting than
that produced in the present relativistic configuration-average calculation.
The ratio of these two X-ray features is $\sim$8.48, which results from simply
calculating the ratio of their respective radiative branching ratios, $B^{\rm
r}_{2p-1s}/B^{\rm r}_{3p-1s}$, because these two features result from different
cascade paths that occur at the beginning of the network. The relatively simple
manner in which these features are produced suggests that this ratio can be
accurately calculated and is therefore a good candidate for detecting the
presence of $^{60}$Co if the two X-ray features can be observed.

A closer inspection of the left panel in Figure~\ref{fig:co_emis} indicates that
there is some spin-orbit splitting of the $2p-1s$ and $3p-1s$ high-energy
features, as displayed in Figure~\ref{fig:co_emis_zoom_highE}.  As alluded to
above, the use of relativistic configurations splits the $2p$ orbital into its
two relativistic analogs, $2p_{1/2}$ and $2p_{3/2}$, resulting in two
$(2p-1s)$-type decay paths that occur at slightly different energies.
Consequently, the the $2p-1s$ emission feature is composed of two lines that are
separated by $\sim$~15~eV.  The $3p$ orbital is similarly replaced with
relativistic $3p_{1/2}$ and $3p_{3/2}$ orbitals, resulting in two lines that are
separated by $\sim$~4~eV.  Thus, the ratio of the $2p-1s$ and $3p-1s$ features
is more precisely calculated via the relativistic formula $(B^{\rm
r}_{2p_{1/2}-1s} + B^{\rm r}_{2p_{3/2}-1s})/(B^{\rm r}_{3p_{1/2}-1s} + B^{\rm
r}_{3p_{3/2}-1s})$.  Alternatively, if sufficient spectral resolution is
available to measure the amount of spin-orbit splitting displayed in
Figure~\ref{fig:co_emis_zoom_highE}, then the present calculation provides a
reasonable estimate of that effect.

\subsubsection{L-shell electron emission}

Next, we consider the possible emission from the cascade network that proceeds
after a $2s$ electron has been ejected during the internal conversion process.
The analysis and calculations follow in a manner very similar to that provided
in the $1s$-hole discussion above. The starting configuration in this case is
$2s_{1/2}^1 3d_{3/2}^4 3d_{5/2}^4$ in $^{60}$Co~\textsc{iii} and only five ion
stages, Co~\textsc{iii}--\textsc{vii}, are required to encapsulate the entire
cascade network.  The resulting line emission probability is presented in
Figure~\ref{fig:co_emis_2s}.  The top panel of that figure displays the emission
over the complete energy range of interest, while the middle and bottom panels
contain zoom-ins of the low- and high-energy ranges, respectively.

At first glance, the emission probabilities in this case appear to be
significantly smaller than those computed for the $1s$-hole scenario, which is
underscored by the use of a log scale on the y-axis in
Figure~\ref{fig:co_emis_2s}. The initial transitions in this cascade network
fill the $2s$ hole and are denoted by $2p-2s$ and $3p-2s$ in the top panel.
Those two features undergo the same type of spin-orbit splitting as their
$1s$-hole counterparts, as exhibited in the middle and bottom panels.  Their
emission probabilities are significantly suppressed compared to their $1s$-hole
analogs because the AI branching ratio from the initial $2s$-hole configuration
is greater than 99.9 per cent.  However, similar to the $1s$-hole spectrum,
there is a broad feature in Figure~\ref{fig:co_emis_2s} centered at 65~eV with a
peak probability of 0.35, arising from radiative decay in subsequent ion stages.
Also, there exist two high-probability bins below 2~eV that, once again, contain
photons produced by $\Delta n = 0$ transitions.

\subsubsection{X-ray line fluxes}

\begin{figure}
	\centering
	\caption{
		Same as Figure~\ref{fig:gamma_flux} but for the X-ray lines at 6.934~keV and
		7.659~keV.
	}
	\includegraphics[]{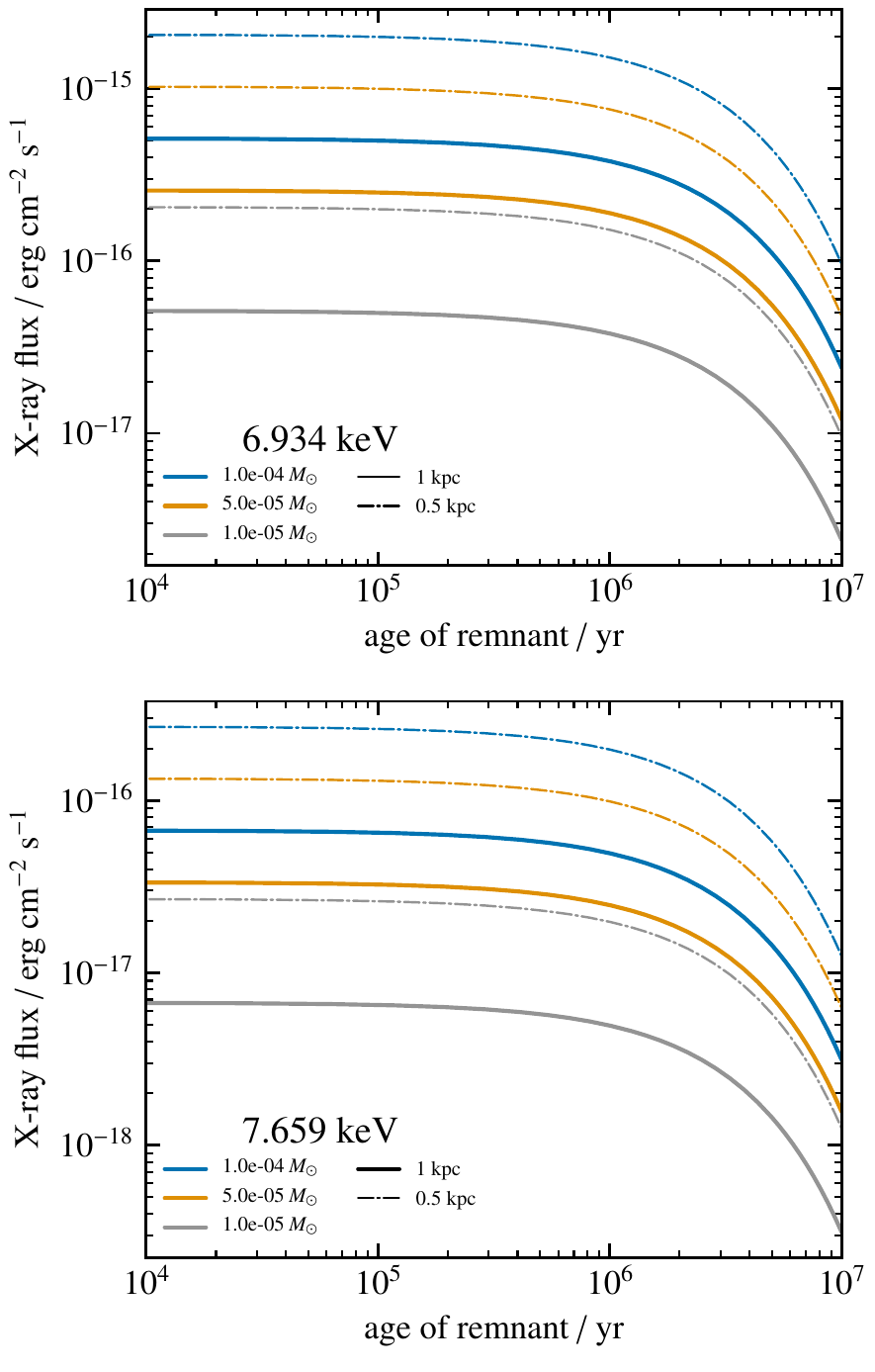}
	\label{fig:xray_flux}
\end{figure}

Recall that the initial decay of the $^{60}$Co~nucleus results in the ionization
of either a K- or L-shell electron 98~per cent of the time via internal
conversion.  Since a K-shell electron is ejected 81.6~per cent of the time, the
subsequent spectrum offers better prospects of detection and we focus on this
case here. For SNRs 0.5~kpc and 1~kpc from Earth we have plotted the X-ray flux
at Earth for a range of \fe~ejecta masses for the 6.934~keV and 7.959~keV lines
in Figure~\ref{fig:xray_flux} assuming that they are a point source.

The lines will be Doppler broadened at early times, becoming narrower at late
times. The line widths for the two emission lines at 6.934 (thick lines) and
7.659~keV (thin lines) are shown as a function of the SNR's age in
Figure~\ref{fig:vbroad}, in which the black horizontal dashed lines represent
the approximate line widths from fine-structure splitting and the other lines
show the velocity Doppler widths. Thermal Doppler broadening is never the
dominant source of line broadening and is negligible here.

\subsection{X-ray detections}

The above-mentioned results put forward a new method for detecting old SNRs in
the X-ray band.  By the time SNRs are $\sim$100 kyr-old, they normally fall
below detectability in the X-ray band since their shock waves have cooled down
to below 10$^5$~K temperatures following the radiative phase. However, as shown
in Fig~19, the 6.934~keV and 7.659~keV lines can potentially probe SNRs in the
$\sim$10$^5$--10$^6$~yr old range, opening a new discovery window that addresses
the missing SNR problem. 

In order to check for their detectability, we estimate the predicted count rates
from these lines with \textit{XMM-Newton}. Currently, this X-ray mission has the
highest sensitivity to low-surface brightness diffuse emission. We find that,
given the flux ranges shown in Figure~19, we need an unreasonably long exposure
time to detect even the brightest case. For example, we estimate that a 1~Ms
observation with \textit{XMM-Newton} will yield only $\lesssim$10 photons from a
10$^{-15}$~erg~cm$^{-2}$~s$^{-1}$ line flux. 

The upcoming mission \textit{eROSITA} \citep{Merloni2012a} will survey SNRs in
the Galaxy, leading to the discovery of many new sources; however its
sensitivity in the hard X-ray band does not exceed that of \textit{XMM-Newton}.
Finally, detectability of these lines will be within the reach of
\textit{ATHENA} \citep{Barcons2017a} given its unprecedented sensitivity (by at
least an order of magnitude improvement over existing missions) combined with
its large field of view (for the WFI instrument) and low instrumental
background. We note however that the field of view is still not large enough to
image degrees-scale nearby SNRs, however we expect the emission to be detected
within small-scale regions of the SNR.

\begin{figure*}
	\centering
	\caption{
		X-ray line widths resulting from velocity broadening of the ejecta af a
		function of remnant age for different supernova ejecta mass, explosion energy
		and ISM density. Thick lines are for the 6.934 keV line and thin lines are for
		the 7.659 keV line. Dashed black horizontal lines represent the approximate
		line widths originating from fine-structure splitting. \emph{Inset:} zoom in of
		the period $10^4-10^6$~yr after the explosion.
	}
	\includegraphics[width=\textwidth]{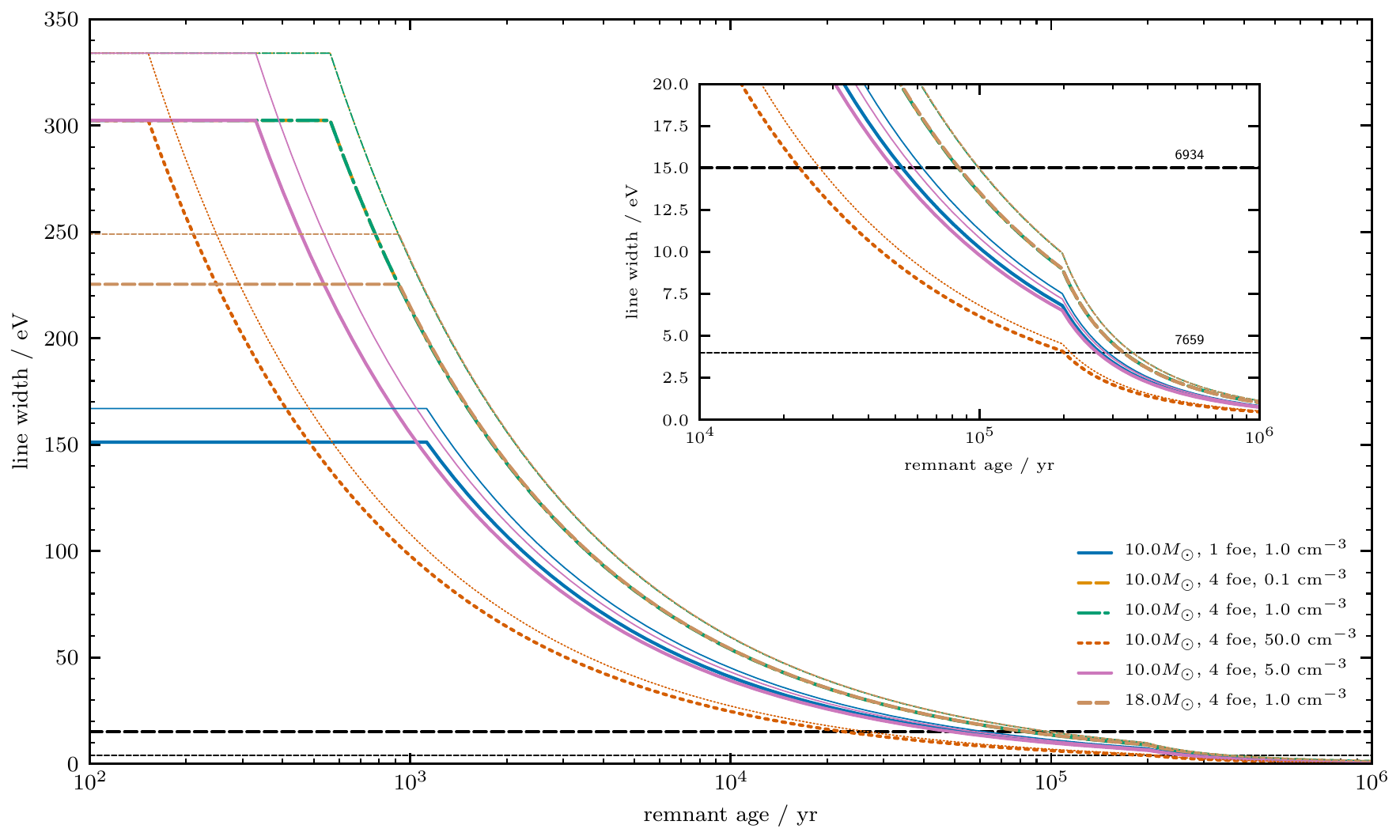}
	\label{fig:vbroad}
\end{figure*}

Finally we note that, as shown in Figure~\ref{fig:vbroad}, the width of the
lines gets narrower as the SNR ages. This, together with the expected large SNR
size (for the oldest, nearby SNRs) and the low line fluxes, will require an
instrument with a large field of view or surveying capabilities, hard X-ray line
sensitivity combined with high-resolution X-ray spectroscopy capabilities.

\section{Summary and conclusions}

We have studied the impact of the uncertainties in the \feng~cross section and
1D CCSN explosion modelling on the yields of \fe~from 15, 20 and
25~\msun~stellar models at Solar metallicity. The total \fe~yield from a CCSN is
sensitive to both the manner in which the energy is deposited in the
parameterised 1D explosion, resulting in a factor 2-3 spread in the \fe~yield
for explosion energies that we consider to be realistic for most CCSNe:
$\lesssim 2\times10^{51}$~erg. The factor of $\sim 10$ uncertainty in the
\feng~cross section results in a larger range of \fe~yields for a given
progenitor: a factor of 0.1--7 relative to models adopting the cross section
from \citet{Rauscher2000a}.

None of the models we have computed produce \fe/\al~yield ratios that are
consistent with a CCSN triggering the formation of the Solar system.  However it
is possible that a lower-mass CCSN progenitor that we have not studied in this
work could still be consistent. The \fe/\al~line flux ratio in the ejecta from
the models is very sensitive to the \feng~cross section, however it is not
possible to reproduce the INTEGRAL/SPI measurement for the diffuse ISM with any
model that assumes the high cross section (ten times larger than NONSMOKER),
although full grids of models would be needed to make a complete and direct
comparison.

Because we can potentially observe SNRs in $^{60}$Fe to late times --
critically, when the SNR is no longer visible in the X-ray bands owing to its
shock heating -- we expect next generation telescopes could discover a large
number (up to 200--300) of nearby SNRs in $^{60}$Fe as well as measure the
amount of \fe~in a handful of known SNRs.  We can use these observations to
better understand the supernova history in the Solar neighborhood. The
dependence of the $^{60}$Fe production on progenitor structure will provide us a
direct probe of the distribution of progenitors and their masses to these
supernova explosions. If we have alternative measurements of the mass and
structure, it is even possible that we can probe the nuclear physics with these
observations.  For very nearby SNRs, the proposed LOX instrument may be able to
map out the $^{60}$Fe, allowing us to truly probe the production of this
isotope.

Next generation $\gamma$-ray telescopes have the potential to detect a single
supernova remnant directly, rather than the diffuse emission from a population
of remnants.  This would allow scientists to truly test their stellar and
explosion models.  But detecting this remnant above the diffuse background
either requires high sensitivity or good energy resolution to separate the
emission of the remnant from the background.

The \co$^*$~produced via the decay of \fe~subsequently decays and via internal
conversion unbinds a K- or L-shell electron. The ensuing radiative decay cascade
produces two prominent hard X-ray lines at 6934 and 7659~eV, with an emission
probability ratio of $\sim8.5$. Detecting these two lines, especially with this
ratio, will be a strong signature of a \co$^*$~and hence \fe~source. After
$10^4-10^5$~yr the dominant source of line broadening will no longer be velocity
of the ejecta but the atomic fine-structure effects, which will be on the order
of 15~eV for the 6934~eV line and 4~eV for the 7659~eV line
(Figure~\ref{fig:vbroad}).

\section*{Acknowledgements}
We thank the anonymous referee for their careful reading of our manuscript and
for their helpful comments and suggestions. SJ acknowledges support from a
Director's Fellowship at Los Alamos National Laboratory and acknowledges support
from the \href{https://www.humboldt-foundation.de}{Alexander von Humboldt
Foundation} and the \href{https://www.klaus-tschira-stiftung.de}{Klaus Tschira
Stiftung} and thanks Alexander Heger for providing stellar evolution models.
SJ, HM, WPE, AC, and MRM acknowledge support by the Laboratory Directed Research
and Development program of Los Alamos National Laboratory under project number
LDRD20160173ER.  This work was supported by the US Department of Energy through
the Los Alamos National Laboratory.  Los Alamos National Laboratory is operated
by Triad National Security, LLC, for the National Nuclear Security
Administration of U.S.  Department of Energy (Contract No.  89233218NCA000001).
Part of this work was performed under the auspices of the U.S. Department of
Energy by Lawrence Livermore National Laboratory under Contract
DE-AC52-07NA27344 and was supported by the Laboratory Directed Research and
Development Program at Lawrence Livermore National Laboratory (17-ERD-001)
LLNL-JRNL-763787. SSH acknowledges support from NSERC.  This research used
resources provided by the Los Alamos National Laboratory Institutional Computing
Program, which is supported by the U.S. Department of Energy National Nuclear
Security Administration under Contract No.  89233218CNA000001.



\begin{thebibliography}{}
\makeatletter
\relax
\def\mn@urlcharsother{\let\do\@makeother \do\$\do\&\do\#\do\^\do\_\do\%\do\~}
\def\mn@doi{\begingroup\mn@urlcharsother \@ifnextchar [ {\mn@doi@}
  {\mn@doi@[]}}
\def\mn@doi@[#1]#2{\def\@tempa{#1}\ifx\@tempa\@empty \href
  {http://dx.doi.org/#2} {doi:#2}\else \href {http://dx.doi.org/#2} {#1}\fi
  \endgroup}
\def\mn@eprint#1#2{\mn@eprint@#1:#2::\@nil}
\def\mn@eprint@arXiv#1{\href {http://arxiv.org/abs/#1} {{\tt arXiv:#1}}}
\def\mn@eprint@dblp#1{\href {http://dblp.uni-trier.de/rec/bibtex/#1.xml}
  {dblp:#1}}
\def\mn@eprint@#1:#2:#3:#4\@nil{\def\@tempa {#1}\def\@tempb {#2}\def\@tempc
  {#3}\ifx \@tempc \@empty \let \@tempc \@tempb \let \@tempb \@tempa \fi \ifx
  \@tempb \@empty \def\@tempb {arXiv}\fi \@ifundefined
  {mn@eprint@\@tempb}{\@tempb:\@tempc}{\expandafter \expandafter \csname
  mn@eprint@\@tempb\endcsname \expandafter{\@tempc}}}

\bibitem[\protect\citeauthoryear{{Angulo} et~al.,}{{Angulo}
  et~al.}{1999}]{AnguloNACRE1999}
{Angulo} C.,  et~al., 1999, \mn@doi [Nuclear Physics A]
  {10.1016/S0375-9474(99)00030-5}, \href
  {http://adsabs.harvard.edu/abs/1999NuPhA.656....3A} {656, 3}

\bibitem[\protect\citeauthoryear{Bader \& Deuflhard}{Bader \&
  Deuflhard}{1983}]{Bader1983a}
Bader G.,  Deuflhard P.,  1983, Numerische Mathematik, 41, 373

\bibitem[\protect\citeauthoryear{{Barcons} et~al.,}{{Barcons}
  et~al.}{2017}]{Barcons2017a}
{Barcons} X.,  et~al., 2017, \mn@doi [Astronomische Nachrichten]
  {10.1002/asna.201713323}, \href
  {http://adsabs.harvard.edu/abs/2017AN....338..153B} {338, 153}

\bibitem[\protect\citeauthoryear{{Bouchet}, {Strong}, {Porter}, {Moskalenko},
  {Jourdain}  \& {Roques}}{{Bouchet} et~al.}{2011}]{Bouchet2011a}
{Bouchet} L.,  {Strong} A.~W.,  {Porter} T.~A.,  {Moskalenko} I.~V.,
  {Jourdain} E.,   {Roques} J.-P.,  2011, \mn@doi [\apj]
  {10.1088/0004-637X/739/1/29}, \href
  {http://adsabs.harvard.edu/abs/2011ApJ...739...29B} {739, 29}

\bibitem[\protect\citeauthoryear{Browne \& Tuli}{Browne \&
  Tuli}{2013}]{BROWNE20131849}
Browne E.,  Tuli J.,  2013, \mn@doi [Nuclear Data Sheets]
  {https://doi.org/10.1016/j.nds.2013.11.002}, 114, 1849

\bibitem[\protect\citeauthoryear{{Calder} et~al.,}{{Calder}
  et~al.}{2007}]{Calder2007a}
{Calder} A.~C.,  et~al., 2007, \mn@doi [\apj] {10.1086/510709}, \href
  {http://adsabs.harvard.edu/abs/2007ApJ...656..313C} {656, 313}

\bibitem[\protect\citeauthoryear{{Chugunov}, {Dewitt}  \&
  {Yakovlev}}{{Chugunov} et~al.}{2007}]{Chugunov2007a}
{Chugunov} A.~I.,  {Dewitt} H.~E.,   {Yakovlev} D.~G.,  2007, \mn@doi [\prd]
  {10.1103/PhysRevD.76.025028}, \href
  {http://adsabs.harvard.edu/abs/2007PhRvD..76b5028C} {76, 025028}

\bibitem[\protect\citeauthoryear{Couture \& Reifarth}{Couture \&
  Reifarth}{2007}]{CoR07}
Couture A.,  Reifarth R.,  2007, Atomic Data and Nuclear Data Tables, 93, 807

\bibitem[\protect\citeauthoryear{{Cyburt} et~al.,}{{Cyburt}
  et~al.}{2010}]{Cyburt2010}
{Cyburt} R.~H.,  et~al., 2010, \mn@doi [\apjs] {10.1088/0067-0049/189/1/240},
  \href {http://adsabs.harvard.edu/abs/2010ApJS..189..240C} {189, 240}

\bibitem[\protect\citeauthoryear{{Davis}, {Jones}  \& {Herwig}}{{Davis}
  et~al.}{2018}]{Davis2018a}
{Davis} A.,  {Jones} S.,   {Herwig} F.,  2018, \mn@doi [\mnras]
  {10.1093/mnras/sty3415}, \href
  {http://adsabs.harvard.edu/abs/2018MNRAS.tmp.3245D} {}

\bibitem[\protect\citeauthoryear{Demmel, Eisenstat, Gilbert, Li  \& Liu}{Demmel
  et~al.}{1999}]{superlu99}
Demmel J.~W.,  Eisenstat S.~C.,  Gilbert J.~R.,  Li X.~S.,   Liu J. W.~H.,
  1999, SIAM J. Matrix Analysis and Applications, 20, 720

\bibitem[\protect\citeauthoryear{Deuflhard}{Deuflhard}{1983}]{Deuflhard1983a}
Deuflhard P.,  1983, \mn@doi [Numerische Mathematik] {10.1007/BF01418332}, 41,
  399

\bibitem[\protect\citeauthoryear{{Diehl}}{{Diehl}}{2013}]{Diehl2013a}
{Diehl} R.,  2013, \mn@doi [Reports on Progress in Physics]
  {10.1088/0034-4885/76/2/026301}, \href
  {http://adsabs.harvard.edu/abs/2013RPPh...76b6301D} {76, 026301}

\bibitem[\protect\citeauthoryear{{Diehl} et~al.,}{{Diehl}
  et~al.}{2006}]{Diehl2006a}
{Diehl} R.,  et~al., 2006, \mn@doi [\nat] {n10.1038/nature04364}, \href
  {http://adsabs.harvard.edu/abs/2006Natur.439...45D} {439, 45}

\bibitem[\protect\citeauthoryear{{Dillmann}, {Heil}, {K{\"a}ppeler}, {Plag},
  {Rauscher}  \& {Thielemann}}{{Dillmann} et~al.}{2006}]{Dillmann2006a}
{Dillmann} I.,  {Heil} M.,  {K{\"a}ppeler} F.,  {Plag} R.,  {Rauscher} T.,
  {Thielemann} F.-K.,  2006, in {Woehr} A.,  {Aprahamian} A.,  eds,  American
  Institute of Physics Conference Series Vol. 819, Capture Gamma-Ray
  Spectroscopy and Related Topics. pp 123--127, \mn@doi{10.1063/1.2187846}

\bibitem[\protect\citeauthoryear{{Doherty}, {Gil-Pons}, {Siess}  \&
  {Lattanzio}}{{Doherty} et~al.}{2017}]{Doherty2017a}
{Doherty} C.~L.,  {Gil-Pons} P.,  {Siess} L.,   {Lattanzio} J.~C.,  2017,
  \mn@doi [\pasa] {10.1017/pasa.2017.52}, \href
  {http://adsabs.harvard.edu/abs/2017PASA...34...56D} {34, e056}

\bibitem[\protect\citeauthoryear{{Edelmann}, {R{\"o}pke}, {Hirschi}, {Georgy}
  \& {Jones}}{{Edelmann} et~al.}{2017}]{Edelmann2017a}
{Edelmann} P.~V.~F.,  {R{\"o}pke} F.~K.,  {Hirschi} R.,  {Georgy} C.,   {Jones}
  S.,  2017, \mn@doi [\aap] {10.1051/0004-6361/201629873}, \href
  {http://adsabs.harvard.edu/abs/2017A%26A...604A..25E} {604, A25}

\bibitem[\protect\citeauthoryear{{Feige}}{{Feige}}{2014}]{Feige2014a}
{Feige} J.,  2014, PhD thesis, {University of Vienna}

\bibitem[\protect\citeauthoryear{Feige et~al.,}{Feige et~al.}{2018}]{feige18}
Feige J.,  et~al., 2018, \mn@doi [Phys. Rev. Lett.]
  {10.1103/PhysRevLett.121.221103}, 121, 221103

\bibitem[\protect\citeauthoryear{{Ferrand} \& {Safi-Harb}}{{Ferrand} \&
  {Safi-Harb}}{2012}]{ferrand12}
{Ferrand} G.,  {Safi-Harb} S.,  2012, \mn@doi [Advances in Space Research]
  {10.1016/j.asr.2012.02.004}, \href
  {http://adsabs.harvard.edu/abs/2012AdSpR..49.1313F} {49, 1313}

\bibitem[\protect\citeauthoryear{{Fields}, {Hochmuth}  \& {Ellis}}{{Fields}
  et~al.}{2005}]{Fields2005a}
{Fields} B.~D.,  {Hochmuth} K.~A.,   {Ellis} J.,  2005, \mn@doi [\apj]
  {10.1086/427797}, \href {http://adsabs.harvard.edu/abs/2005ApJ...621..902F}
  {621, 902}

\bibitem[\protect\citeauthoryear{Fitoussi et~al.,}{Fitoussi
  et~al.}{2008}]{Fitoussi2008a}
Fitoussi C.,  et~al., 2008, \mn@doi [Phys. Rev. Lett.]
  {10.1103/PhysRevLett.101.121101}, 101, 121101

\bibitem[\protect\citeauthoryear{{Fontes} et~al.,}{{Fontes}
  et~al.}{2015}]{LANL_suite}
{Fontes} C.~J.,  et~al., 2015, \mn@doi [Journal of Physics B Atomic Molecular
  Physics] {10.1088/0953-4075/48/14/144014}, \href
  {http://adsabs.harvard.edu/abs/2015JPhB...48n4014F} {48, 144014}

\bibitem[\protect\citeauthoryear{{Fr{\"o}hlich}, {Mart{\'{\i}}nez-Pinedo},
  {Liebend{\"o}rfer}, {Thielemann}, {Bravo}, {Hix}, {Langanke}  \&
  {Zinner}}{{Fr{\"o}hlich} et~al.}{2006}]{Froehlich2006a}
{Fr{\"o}hlich} C.,  {Mart{\'{\i}}nez-Pinedo} G.,  {Liebend{\"o}rfer} M.,
  {Thielemann} F.-K.,  {Bravo} E.,  {Hix} W.~R.,  {Langanke} K.,   {Zinner}
  N.~T.,  2006, \mn@doi [Physical Review Letters]
  {10.1103/PhysRevLett.96.142502}, \href
  {http://adsabs.harvard.edu/abs/2006PhRvL..96n2502F} {96, 142502}

\bibitem[\protect\citeauthoryear{{Fryer}, {Benz}, {Herant}  \&
  {Colgate}}{{Fryer} et~al.}{1999}]{fryer99}
{Fryer} C.,  {Benz} W.,  {Herant} M.,   {Colgate} S.~A.,  1999, \mn@doi [\apj]
  {10.1086/307119}, \href {http://adsabs.harvard.edu/abs/1999ApJ...516..892F}
  {516, 892}

\bibitem[\protect\citeauthoryear{{Fryer}, {Andrews}, {Even}, {Heger}  \&
  {Safi-Harb}}{{Fryer} et~al.}{2018}]{fryer18}
{Fryer} C.~L.,  {Andrews} S.,  {Even} W.,  {Heger} A.,   {Safi-Harb} S.,  2018,
  \mn@doi [\apj] {10.3847/1538-4357/aaaf6f}, \href
  {http://adsabs.harvard.edu/abs/2018ApJ...856...63F} {856, 63}

\bibitem[\protect\citeauthoryear{{Fuchs}, {Breitschwerdt}, {de Avillez},
  {Dettbarn}  \& {Flynn}}{{Fuchs} et~al.}{2006}]{Fuchs2006a}
{Fuchs} B.,  {Breitschwerdt} D.,  {de Avillez} M.~A.,  {Dettbarn} C.,   {Flynn}
  C.,  2006, \mn@doi [\mnras] {10.1111/j.1365-2966.2006.11044.x}, \href
  {http://adsabs.harvard.edu/abs/2006MNRAS.373..993F} {373, 993}

\bibitem[\protect\citeauthoryear{{Fuller}, {Fowler}  \& {Newman}}{{Fuller}
  et~al.}{1985}]{FFNweak1985}
{Fuller} G.~M.,  {Fowler} W.~A.,   {Newman} M.~J.,  1985, \mn@doi [\apj]
  {10.1086/163208}, \href {http://adsabs.harvard.edu/abs/1985ApJ...293....1F}
  {293, 1}

\bibitem[\protect\citeauthoryear{{Fynbo} et~al.,}{{Fynbo}
  et~al.}{2005}]{Fynbo_3a_2005}
{Fynbo} H.~O.~U.,  et~al., 2005, \nat, \href
  {http://adsabs.harvard.edu/abs/2005Natur.433..136F} {433, 136}

\bibitem[\protect\citeauthoryear{{Goriely}}{{Goriely}}{1999}]{Goriely1999a}
{Goriely} S.,  1999, \aap, \href
  {http://adsabs.harvard.edu/abs/1999A%26A...342..881G} {342, 881}

\bibitem[\protect\citeauthoryear{{Green}}{{Green}}{2017}]{Green2017}
{Green} D.~A.,  2017, VizieR Online Data Catalog, \href
  {https://ui.adsabs.harvard.edu/\#abs/2017yCat.7278....0G} {p. VII/278}

\bibitem[\protect\citeauthoryear{{Grevesse} \& {Noels}}{{Grevesse} \&
  {Noels}}{1993}]{GrevesseNoels1993}
{Grevesse} N.,  {Noels} A.,  1993, in {Prantzos} N.,  {Vangioni-Flam} E.,
  {Casse} M.,  eds, Origin and Evolution of the Elements. pp 15--25

\bibitem[\protect\citeauthoryear{{Herant}, {Benz}, {Hix}, {Fryer}  \&
  {Colgate}}{{Herant} et~al.}{1994}]{herant94}
{Herant} M.,  {Benz} W.,  {Hix} W.~R.,  {Fryer} C.~L.,   {Colgate} S.~A.,
  1994, \mn@doi [\apj] {10.1086/174817}, \href
  {http://adsabs.harvard.edu/abs/1994ApJ...435..339H} {435, 339}

\bibitem[\protect\citeauthoryear{{Iliadis}, {D'Auria}, {Starrfield}, {Thompson}
   \& {Wiescher}}{{Iliadis} et~al.}{2001}]{Iliadis2001a}
{Iliadis} C.,  {D'Auria} J.~M.,  {Starrfield} S.,  {Thompson} W.~J.,
  {Wiescher} M.,  2001, \mn@doi [\apjs] {10.1086/320364}, \href
  {http://adsabs.harvard.edu/abs/2001ApJS..134..151I} {134, 151}

\bibitem[\protect\citeauthoryear{{Imbriani} et~al.,}{{Imbriani}
  et~al.}{2004}]{Imbriani2004}
{Imbriani} G.,  et~al., 2004, \mn@doi [\aap] {10.1051/0004-6361:20040981},
  \href {http://adsabs.harvard.edu/abs/2004A%26A...420..625I} {420, 625}

\bibitem[\protect\citeauthoryear{{Jacobsen}, {Yin}, {Moynier}, {Amelin},
  {Krot}, {Nagashima}, {Hutcheon}  \& {Palme}}{{Jacobsen}
  et~al.}{2008}]{jacobsen08}
{Jacobsen} B.,  {Yin} Q.-z.,  {Moynier} F.,  {Amelin} Y.,  {Krot} A.~N.,
  {Nagashima} K.,  {Hutcheon} I.~D.,   {Palme} H.,  2008, \mn@doi [E\&PSL]
  {10.1016/j.epsl.2008.05.003}, \href
  {http://adsabs.harvard.edu/abs/2008E%26PSL.272..353J} {272, 353}

\bibitem[\protect\citeauthoryear{{Jaeger}, {Kunz}, {Mayer}, {Hammer}, {Staudt},
  {Kratz}  \& {Pfeiffer}}{{Jaeger} et~al.}{2001}]{Jaeger2001a}
{Jaeger} M.,  {Kunz} R.,  {Mayer} A.,  {Hammer} J.~W.,  {Staudt} G.,  {Kratz}
  K.~L.,   {Pfeiffer} B.,  2001, \mn@doi [Physical Review Letters]
  {10.1103/PhysRevLett.87.202501}, \href
  {http://adsabs.harvard.edu/abs/2001PhRvL..87t2501J} {87, 202501}

\bibitem[\protect\citeauthoryear{{Jones}, {Hirschi}, {Pignatari}, {Heger},
  {Georgy}, {Nishimura}, {Fryer}  \& {Herwig}}{{Jones}
  et~al.}{2015}]{Jones2015a}
{Jones} S.,  {Hirschi} R.,  {Pignatari} M.,  {Heger} A.,  {Georgy} C.,
  {Nishimura} N.,  {Fryer} C.,   {Herwig} F.,  2015, \mn@doi [\mnras]
  {10.1093/mnras/stu2657}, \href
  {http://adsabs.harvard.edu/abs/2015MNRAS.447.3115J} {447, 3115}

\bibitem[\protect\citeauthoryear{{Jones}, {R{\"o}pke}, {Pakmor}, {Seitenzahl},
  {Ohlmann}  \& {Edelmann}}{{Jones} et~al.}{2016}]{Jones2016a}
{Jones} S.,  {R{\"o}pke} F.~K.,  {Pakmor} R.,  {Seitenzahl} I.~R.,  {Ohlmann}
  S.~T.,   {Edelmann} P.~V.~F.,  2016, \mn@doi [\aap]
  {10.1051/0004-6361/201628321}, \href
  {http://adsabs.harvard.edu/abs/2016A%26A...593A..72J} {593, A72}

\bibitem[\protect\citeauthoryear{{Jones}, {Andrassy}, {Sandalski}, {Davis},
  {Woodward}  \& {Herwig}}{{Jones} et~al.}{2017}]{Jones2017a}
{Jones} S.,  {Andrassy} R.,  {Sandalski} S.,  {Davis} A.,  {Woodward} P.,
  {Herwig} F.,  2017, \mn@doi [\mnras] {10.1093/mnras/stw2783}, \href
  {http://adsabs.harvard.edu/abs/2017MNRAS.465.2991J} {465, 2991}

\bibitem[\protect\citeauthoryear{{Jones} et~al.,}{{Jones}
  et~al.}{2018}]{Jones2018a}
{Jones} S.,  et~al., 2018, arXiv e-prints, \href
  {http://adsabs.harvard.edu/abs/2018arXiv181208230J} {}

\bibitem[\protect\citeauthoryear{{Knie}, {Korschinek}, {Faestermann},
  {Wallner}, {Scholten}  \& {Hillebrandt}}{{Knie} et~al.}{1999}]{Knie1999a}
{Knie} K.,  {Korschinek} G.,  {Faestermann} T.,  {Wallner} C.,  {Scholten} J.,
   {Hillebrandt} W.,  1999, \mn@doi [Physical Review Letters]
  {10.1103/PhysRevLett.83.18}, \href
  {http://adsabs.harvard.edu/abs/1999PhRvL..83...18K} {83, 18}

\bibitem[\protect\citeauthoryear{{Knie}, {Korschinek}, {Faestermann}, {Dorfi},
  {Rugel}  \& {Wallner}}{{Knie} et~al.}{2004}]{Knie2004a}
{Knie} K.,  {Korschinek} G.,  {Faestermann} T.,  {Dorfi} E.~A.,  {Rugel} G.,
  {Wallner} A.,  2004, \mn@doi [Physical Review Letters]
  {10.1103/PhysRevLett.93.171103}, \href
  {http://adsabs.harvard.edu/abs/2004PhRvL..93q1103K} {93, 171103}

\bibitem[\protect\citeauthoryear{{Krause} et~al.,}{{Krause}
  et~al.}{2015}]{krause15}
{Krause} M.~G.~H.,  et~al., 2015, \mn@doi [\aap] {10.1051/0004-6361/201525847},
  \href {http://adsabs.harvard.edu/abs/2015A%26A...578A.113K} {578, A113}

\bibitem[\protect\citeauthoryear{{Kretschmer}, {Diehl}, {Krause}, {Burkert},
  {Fierlinger}, {Gerhard}, {Greiner}  \& {Wang}}{{Kretschmer}
  et~al.}{2013}]{kretschmer13}
{Kretschmer} K.,  {Diehl} R.,  {Krause} M.,  {Burkert} A.,  {Fierlinger} K.,
  {Gerhard} O.,  {Greiner} J.,   {Wang} W.,  2013, \mn@doi [\aap]
  {10.1051/0004-6361/201322563}, \href
  {http://adsabs.harvard.edu/abs/2013A%26A...559A..99K} {559, A99}

\bibitem[\protect\citeauthoryear{{Kunz}, {Fey}, {Jaeger}, {Mayer}, {Hammer},
  {Staudt}, {Harissopulos}  \& {Paradellis}}{{Kunz} et~al.}{2002}]{Kunz2002}
{Kunz} R.,  {Fey} M.,  {Jaeger} M.,  {Mayer} A.,  {Hammer} J.~W.,  {Staudt} G.,
   {Harissopulos} S.,   {Paradellis} T.,  2002, \mn@doi [\apj]
  {10.1086/338384}, \href {http://adsabs.harvard.edu/abs/2002ApJ...567..643K}
  {567, 643}

\bibitem[\protect\citeauthoryear{{Langanke} \&
  {Mart{\'{\i}}nez-Pinedo}}{{Langanke} \&
  {Mart{\'{\i}}nez-Pinedo}}{2000}]{Langanke2000}
{Langanke} K.,  {Mart{\'{\i}}nez-Pinedo} G.,  2000, \mn@doi [Nuclear Physics A]
  {10.1016/S0375-9474(00)00131-7}, \href
  {http://adsabs.harvard.edu/abs/2000NuPhA.673..481L} {673, 481}

\bibitem[\protect\citeauthoryear{{Langer}}{{Langer}}{1989}]{Langer1989a}
{Langer} N.,  1989, \aap, \href
  {http://adsabs.harvard.edu/abs/1989A%26A...220..135L} {220, 135}

\bibitem[\protect\citeauthoryear{{Larsen} \& {Goriely}}{{Larsen} \&
  {Goriely}}{2010}]{LaG10}
{Larsen} A.~C.,  {Goriely} S.,  2010, \mn@doi [\prc]
  {10.1103/PhysRevC.82.014318}, \href
  {http://adsabs.harvard.edu/abs/2010PhRvC..82a4318L} {82, 014318}

\bibitem[\protect\citeauthoryear{{Leising}}{{Leising}}{2001}]{Leising2001a}
{Leising} M.~D.,  2001, \mn@doi [\apj] {10.1086/323776}, \href
  {http://adsabs.harvard.edu/abs/2001ApJ...563..185L} {563, 185}

\bibitem[\protect\citeauthoryear{Li}{Li}{2005}]{li05}
Li X.~S.,  2005, 31, 302

\bibitem[\protect\citeauthoryear{Li, Demmel, Gilbert, iL. Grigori, Shao  \&
  Yamazaki}{Li et~al.}{1999}]{superlu_ug99}
Li X.,  Demmel J.,  Gilbert J.,  iL. Grigori Shao M.,   Yamazaki I.,  1999,
  Technical Report LBNL-44289, {SuperLU Users' Guide}.
Lawrence Berkeley National Laboratory

\bibitem[\protect\citeauthoryear{{Limongi} \& {Chieffi}}{{Limongi} \&
  {Chieffi}}{2003}]{Limongi2003a}
{Limongi} M.,  {Chieffi} A.,  2003, \mn@doi [\apj] {10.1086/375703}, \href
  {http://adsabs.harvard.edu/abs/2003ApJ...592..404L} {592, 404}

\bibitem[\protect\citeauthoryear{{Limongi} \& {Chieffi}}{{Limongi} \&
  {Chieffi}}{2006}]{Limongi2006a}
{Limongi} M.,  {Chieffi} A.,  2006, \mn@doi [\apj] {10.1086/505164}, \href
  {http://adsabs.harvard.edu/abs/2006ApJ...647..483L} {647, 483}

\bibitem[\protect\citeauthoryear{{Limongi} \& {Chieffi}}{{Limongi} \&
  {Chieffi}}{2018}]{Limongi2018a}
{Limongi} M.,  {Chieffi} A.,  2018, \mn@doi [\apjs] {10.3847/1538-4365/aacb24},
  \href {http://adsabs.harvard.edu/abs/2018ApJS..237...13L} {237, 13}

\bibitem[\protect\citeauthoryear{{Lippuner} \& {Roberts}}{{Lippuner} \&
  {Roberts}}{2017}]{Lippuner2017a}
{Lippuner} J.,  {Roberts} L.~F.,  2017, \mn@doi [\apjs]
  {10.3847/1538-4365/aa94cb}, \href
  {http://adsabs.harvard.edu/abs/2017ApJS..233...18L} {233, 18}

\bibitem[\protect\citeauthoryear{{Lodders}, {Palme}  \& {Gail}}{{Lodders}
  et~al.}{2009}]{lodders09}
{Lodders} K.,  {Palme} H.,   {Gail} H.-P.,  2009, {Abundances of the Elements
  in the Solar System}.
Springer-Verlag Berlin Heidelberg, pp 560--630 (\mn@eprint {arXiv}
  {0901.1149}), \mn@doi{10.1007/978-3-540-88055-4_34}

\bibitem[\protect\citeauthoryear{{Longland}, {Martin}  \&
  {Jos{\'e}}}{{Longland} et~al.}{2014}]{Longland2014a}
{Longland} R.,  {Martin} D.,   {Jos{\'e}} J.,  2014, \mn@doi [\aap]
  {10.1051/0004-6361/201321958}, \href
  {http://adsabs.harvard.edu/abs/2014A%26A...563A..67L} {563, A67}

\bibitem[\protect\citeauthoryear{{Lugaro}, {Doherty}, {Karakas}, {Maddison},
  {Liffman}, {Garc{\'{\i}}a-Hern{\'a}ndez}, {Siess}  \& {Lattanzio}}{{Lugaro}
  et~al.}{2012}]{Lugaro2012a}
{Lugaro} M.,  {Doherty} C.~L.,  {Karakas} A.~I.,  {Maddison} S.~T.,  {Liffman}
  K.,  {Garc{\'{\i}}a-Hern{\'a}ndez} D.~A.,  {Siess} L.,   {Lattanzio} J.~C.,
  2012, \mn@doi [Meteoritics and Planetary Science]
  {10.1111/j.1945-5100.2012.01411.x}, \href
  {http://adsabs.harvard.edu/abs/2012M%26PS...47.1998L} {47, 1998}

\bibitem[\protect\citeauthoryear{Lugaro, Ott  \& Kereszturi}{Lugaro
  et~al.}{2018}]{lugaro18}
Lugaro M.,  Ott U.,   Kereszturi .,  2018, \mn@doi [Progress in Particle and
  Nuclear Physics] {https://doi.org/10.1016/j.ppnp.2018.05.002}, 102, 1

\bibitem[\protect\citeauthoryear{{Mahoney}, {Ling}, {Jacobson}  \&
  {Lingenfelter}}{{Mahoney} et~al.}{1982}]{Mahoney1982a}
{Mahoney} W.~A.,  {Ling} J.~C.,  {Jacobson} A.~S.,   {Lingenfelter} R.~E.,
  1982, \mn@doi [\apj] {10.1086/160469}, \href
  {http://adsabs.harvard.edu/abs/1982ApJ...262..742M} {262, 742}

\bibitem[\protect\citeauthoryear{{Meakin} \& {Arnett}}{{Meakin} \&
  {Arnett}}{2007}]{Meakin2007}
{Meakin} C.~A.,  {Arnett} D.,  2007, \mn@doi [\apj] {10.1086/520318}, \href
  {http://adsabs.harvard.edu/abs/2007ApJ...667..448M} {667, 448}

\bibitem[\protect\citeauthoryear{{Merloni} et~al.,}{{Merloni}
  et~al.}{2012}]{Merloni2012a}
{Merloni} A.,  et~al., 2012, preprint, \href
  {http://adsabs.harvard.edu/abs/2012arXiv1209.3114M} {} (\mn@eprint {arXiv}
  {1209.3114})

\bibitem[\protect\citeauthoryear{{Mishra} \& {Chaussidon}}{{Mishra} \&
  {Chaussidon}}{2014}]{mishra14b}
{Mishra} R.~K.,  {Chaussidon} M.,  2014, \mn@doi [E\&PSL]
  {10.1016/j.epsl.2014.04.032}, \href
  {http://adsabs.harvard.edu/abs/2014E%26PSL.398...90M} {398, 90}

\bibitem[\protect\citeauthoryear{{Mishra} \& {Goswami}}{{Mishra} \&
  {Goswami}}{2014}]{mishra14a}
{Mishra} R.~K.,  {Goswami} J.~N.,  2014, \mn@doi [\gca]
  {10.1016/j.gca.2014.01.011}, \href
  {http://adsabs.harvard.edu/abs/2014GeCoA.132..440M} {132, 440}

\bibitem[\protect\citeauthoryear{{Mishra}, {Marhas}  \& {Sameer}}{{Mishra}
  et~al.}{2016}]{mishra16}
{Mishra} R.~K.,  {Marhas} K.~K.,   {Sameer} 2016, \mn@doi [E\&PSL]
  {10.1016/j.epsl.2015.12.007}, \href
  {http://adsabs.harvard.edu/abs/2016E%26PSL.436...71M} {436, 71}

\bibitem[\protect\citeauthoryear{Mukoyama \& Shimizu}{Mukoyama \&
  Shimizu}{1975}]{mukoyama75}
Mukoyama T.,  Shimizu S.,  1975, \mn@doi [Phys. Rev. C]
  {10.1103/PhysRevC.11.1353}, 11, 1353

\bibitem[\protect\citeauthoryear{{Mumpower}, {McLaughlin}  \&
  {Surman}}{{Mumpower} et~al.}{2012}]{MMS12}
{Mumpower} M.~R.,  {McLaughlin} G.~C.,   {Surman} R.,  2012, \mn@doi [\prc]
  {10.1103/PhysRevC.86.035803}, \href
  {http://adsabs.harvard.edu/abs/2012PhRvC..86c5803M} {86, 035803}

\bibitem[\protect\citeauthoryear{{Mumpower}, {Surman}, {McLaughlin}  \&
  {Aprahamian}}{{Mumpower} et~al.}{2016}]{Mumpower2016a}
{Mumpower} M.~R.,  {Surman} R.,  {McLaughlin} G.~C.,   {Aprahamian} A.,  2016,
  \mn@doi [Progress in Particle and Nuclear Physics]
  {10.1016/j.ppnp.2015.09.001}, \href
  {http://adsabs.harvard.edu/abs/2016PrPNP..86...86M} {86, 86}

\bibitem[\protect\citeauthoryear{{Mumpower}, {Kawano}, {Ullmann}, {Krti{\v
  c}ka}  \& {Sprouse}}{{Mumpower} et~al.}{2017}]{Mumpower2017a}
{Mumpower} M.~R.,  {Kawano} T.,  {Ullmann} J.~L.,  {Krti{\v c}ka} M.,
  {Sprouse} T.~M.,  2017, \mn@doi [\prc] {10.1103/PhysRevC.96.024612}, \href
  {http://adsabs.harvard.edu/abs/2017PhRvC..96b4612M} {96, 024612}

\bibitem[\protect\citeauthoryear{{Nugis} \& {Lamers}}{{Nugis} \&
  {Lamers}}{2000}]{Nugis2000}
{Nugis} T.,  {Lamers} H.~J.~G.~L.~M.,  2000, \aap, \href
  {http://adsabs.harvard.edu/abs/2000A%26A...360..227N} {360, 227}

\bibitem[\protect\citeauthoryear{{Oda}, {Hino}, {Muto}, {Takahara}  \&
  {Sato}}{{Oda} et~al.}{1994}]{ODA94}
{Oda} T.,  {Hino} M.,  {Muto} K.,  {Takahara} M.,   {Sato} K.,  1994, \mn@doi
  [Atomic Data and Nuclear Data Tables] {10.1006/adnd.1994.1007}, \href
  {http://adsabs.harvard.edu/abs/1994ADNDT..56..231O} {56, 231}

\bibitem[\protect\citeauthoryear{{Palacios}, {Meynet}, {Vuissoz},
  {Kn{\"o}dlseder}, {Schaerer}, {Cervi{\~n}o}  \& {Mowlavi}}{{Palacios}
  et~al.}{2005}]{Palacios2005a}
{Palacios} A.,  {Meynet} G.,  {Vuissoz} C.,  {Kn{\"o}dlseder} J.,  {Schaerer}
  D.,  {Cervi{\~n}o} M.,   {Mowlavi} N.,  2005, \mn@doi [\aap]
  {10.1051/0004-6361:20041757}, \href
  {http://adsabs.harvard.edu/abs/2005A%26A...429..613P} {429, 613}

\bibitem[\protect\citeauthoryear{{Pignatari} et~al.,}{{Pignatari}
  et~al.}{2016}]{Pignatari2016a}
{Pignatari} M.,  et~al., 2016, \mn@doi [\apjs] {10.3847/0067-0049/225/2/24},
  \href {http://adsabs.harvard.edu/abs/2016ApJS..225...24P} {225, 24}

\bibitem[\protect\citeauthoryear{{Prantzos}}{{Prantzos}}{2004}]{Prantzos2004a}
{Prantzos} N.,  2004, \mn@doi [\aap] {10.1051/0004-6361:20035766}, \href
  {http://adsabs.harvard.edu/abs/2004A%26A...420.1033P} {420, 1033}

\bibitem[\protect\citeauthoryear{{Qian} \& {Woosley}}{{Qian} \&
  {Woosley}}{1996}]{Qian1996a}
{Qian} Y.-Z.,  {Woosley} S.~E.,  1996, \mn@doi [\apj] {10.1086/177973}, \href
  {http://adsabs.harvard.edu/abs/1996ApJ...471..331Q} {471, 331}

\bibitem[\protect\citeauthoryear{{Rampp} \& {Janka}}{{Rampp} \&
  {Janka}}{2000}]{Rampp2000a}
{Rampp} M.,  {Janka} H.-T.,  2000, \mn@doi [\apjl] {10.1086/312837}, \href
  {http://adsabs.harvard.edu/abs/2000ApJ...539L..33R} {539, L33}

\bibitem[\protect\citeauthoryear{{Rauscher} \& {Thielemann}}{{Rauscher} \&
  {Thielemann}}{2000}]{Rauscher2000a}
{Rauscher} T.,  {Thielemann} F.-K.,  2000, \mn@doi [Atomic Data and Nuclear
  Data Tables] {10.1006/adnd.2000.0834}, \href
  {http://adsabs.harvard.edu/abs/2000ADNDT..75....1R} {75, 1}

\bibitem[\protect\citeauthoryear{{Rauscher}, {Heger}, {Hoffman}  \&
  {Woosley}}{{Rauscher} et~al.}{2002}]{Rauscher2002}
{Rauscher} T.,  {Heger} A.,  {Hoffman} R.~D.,   {Woosley} S.~E.,  2002, \mn@doi
  [\apj] {10.1086/341728}, \href
  {http://adsabs.harvard.edu/abs/2002ApJ...576..323R} {576, 323}

\bibitem[\protect\citeauthoryear{{Renzo}, {Ott}, {Shore}  \& {de Mink}}{{Renzo}
  et~al.}{2017}]{Renzo2017a}
{Renzo} M.,  {Ott} C.~D.,  {Shore} S.~N.,   {de Mink} S.~E.,  2017, \mn@doi
  [\aap] {10.1051/0004-6361/201730698}, \href
  {http://adsabs.harvard.edu/abs/2017A%26A...603A.118R} {603, A118}

\bibitem[\protect\citeauthoryear{{Ritter}, {Herwig}, {Jones}, {Pignatari},
  {Fryer}  \& {Hirschi}}{{Ritter} et~al.}{2017}]{Ritter2017a}
{Ritter} C.,  {Herwig} F.,  {Jones} S.,  {Pignatari} M.,  {Fryer} C.,
  {Hirschi} R.,  2017, preprint, \href
  {http://adsabs.harvard.edu/abs/2017arXiv170908677R} {} (\mn@eprint {arXiv}
  {1709.08677})

\bibitem[\protect\citeauthoryear{{Ritter}, {Andrassy}, {C{\^o}t{\'e}},
  {Herwig}, {Woodward}, {Pignatari}  \& {Jones}}{{Ritter}
  et~al.}{2018}]{Ritter2018a}
{Ritter} C.,  {Andrassy} R.,  {C{\^o}t{\'e}} B.,  {Herwig} F.,  {Woodward}
  P.~R.,  {Pignatari} M.,   {Jones} S.,  2018, \mn@doi [\mnras]
  {10.1093/mnrasl/slx126}, \href
  {http://adsabs.harvard.edu/abs/2018MNRAS.474L...1R} {474, L1}

\bibitem[\protect\citeauthoryear{{Sampson}, {Zhang}  \& {Fontes}}{{Sampson}
  et~al.}{2009}]{sampson_physrep}
{Sampson} D.~H.,  {Zhang} H.~L.,   {Fontes} C.~J.,  2009, \mn@doi [\physrep]
  {10.1016/j.physrep.2009.04.002}, \href
  {http://adsabs.harvard.edu/abs/2009PhR...477..111S} {477, 111}

\bibitem[\protect\citeauthoryear{{Schulreich}, {Breitschwerdt}, {Feige}  \&
  {Dettbarn}}{{Schulreich} et~al.}{2017}]{Schulreich2017a}
{Schulreich} M.~M.,  {Breitschwerdt} D.,  {Feige} J.,   {Dettbarn} C.,  2017,
  \mn@doi [\aap] {10.1051/0004-6361/201629837}, \href
  {http://adsabs.harvard.edu/abs/2017A%26A...604A..81S} {604, A81}

\bibitem[\protect\citeauthoryear{{Seitenzahl} et~al.,}{{Seitenzahl}
  et~al.}{2015}]{Seitenzahl2015a}
{Seitenzahl} I.~R.,  et~al., 2015, \mn@doi [\mnras] {10.1093/mnras/stu2537},
  \href {http://adsabs.harvard.edu/abs/2015MNRAS.447.1484S} {447, 1484}

\bibitem[\protect\citeauthoryear{{Smith}}{{Smith}}{2004}]{Smith2004a}
{Smith} D.~M.,  2004, in {Schoenfelder} V.,  {Lichti} G.,   {Winkler} C.,  eds,
   ESA Special Publication Vol. 552, 5th INTEGRAL Workshop on the INTEGRAL
  Universe. p.~45 (\mn@eprint {} {astro-ph/0404594})

\bibitem[\protect\citeauthoryear{{Takahashi} \& {Yokoi}}{{Takahashi} \&
  {Yokoi}}{1987}]{Takahashi1987a}
{Takahashi} K.,  {Yokoi} K.,  1987, \mn@doi [Atomic Data and Nuclear Data
  Tables] {10.1016/0092-640X(87)90010-6}, \href
  {http://adsabs.harvard.edu/abs/1987ADNDT..36..375T} {36, 375}

\bibitem[\protect\citeauthoryear{{Tammann}, {Loeffler}  \&
  {Schroeder}}{{Tammann} et~al.}{1994}]{Tammann1994}
{Tammann} G.~A.,  {Loeffler} W.,   {Schroeder} A.,  1994, \mn@doi [The
  Astrophysical Journal Supplement Series] {10.1086/192002}, \href
  {https://ui.adsabs.harvard.edu/\#abs/1994ApJS...92..487T} {92, 487}

\bibitem[\protect\citeauthoryear{{Tang} \& {Dauphas}}{{Tang} \&
  {Dauphas}}{2012}]{tang12a}
{Tang} H.,  {Dauphas} N.,  2012, \mn@doi [E\&PSL] {10.1016/j.epsl.2012.10.011},
  \href {http://adsabs.harvard.edu/abs/2012E%26PSL.359..248T} {359, 248}

\bibitem[\protect\citeauthoryear{{Tang} \& {Dauphas}}{{Tang} \&
  {Dauphas}}{2015}]{tang15}
{Tang} H.,  {Dauphas} N.,  2015, \mn@doi [\apj] {10.1088/0004-637X/802/1/22},
  \href {http://adsabs.harvard.edu/abs/2015ApJ...802...22T} {802, 22}

\bibitem[\protect\citeauthoryear{{Telus}, Huss, Nagashima, Ogliore  \&
  Tachibana}{{Telus} et~al.}{2018}]{telus18}
{Telus} M.,  Huss G.~R.,  Nagashima K.,  Ogliore R.~C.,   Tachibana S.,  2018,
  \gca, 221, 342

\bibitem[\protect\citeauthoryear{{Timmes}}{{Timmes}}{1999}]{Timmes1999a}
{Timmes} F.~X.,  1999, \mn@doi [\apjs] {10.1086/313257}, \href
  {http://adsabs.harvard.edu/abs/1999ApJS..124..241T} {124, 241}

\bibitem[\protect\citeauthoryear{{Timmes}, {Woosley}, {Hartmann}, {Hoffman},
  {Weaver}  \& {Matteucci}}{{Timmes} et~al.}{1995}]{Timmes1995a}
{Timmes} F.~X.,  {Woosley} S.~E.,  {Hartmann} D.~H.,  {Hoffman} R.~D.,
  {Weaver} T.~A.,   {Matteucci} F.,  1995, \mn@doi [\apj] {10.1086/176046},
  \href {http://adsabs.harvard.edu/abs/1995ApJ...449..204T} {449, 204}

\bibitem[\protect\citeauthoryear{Trappitsch et~al.,}{Trappitsch
  et~al.}{2018}]{trappitsch18}
Trappitsch R.,  et~al., 2018, The Astrophysical Journal Letters, in review

\bibitem[\protect\citeauthoryear{{Tur}, {Heger}  \& {Austin}}{{Tur}
  et~al.}{2010}]{Tur2010a}
{Tur} C.,  {Heger} A.,   {Austin} S.~M.,  2010, \mn@doi [\apj]
  {10.1088/0004-637X/718/1/357}, \href
  {http://adsabs.harvard.edu/abs/2010ApJ...718..357T} {718, 357}

\bibitem[\protect\citeauthoryear{Uberseder et~al.,}{Uberseder
  et~al.}{2014}]{UAA14}
Uberseder E.,  et~al., 2014, \mn@doi [Phys. Rev. Lett.]
  {10.1103/PhysRevLett.112.211101}, 112, 211101

\bibitem[\protect\citeauthoryear{Vasileiadis, Nordlund  \&
  Bizzarro}{Vasileiadis et~al.}{2013}]{vasileiadis2013}
Vasileiadis A.,  Nordlund {\AA}.,   Bizzarro M.,  2013, The Astrophysical
  Journal Letters, 769, L8

\bibitem[\protect\citeauthoryear{{Wallner} et~al.,}{{Wallner}
  et~al.}{2016}]{Wallner2016a}
{Wallner} A.,  et~al., 2016, \mn@doi [\nat] {10.1038/nature17196}, \href
  {http://adsabs.harvard.edu/abs/2016Natur.532...69W} {532, 69}

\bibitem[\protect\citeauthoryear{{Wanajo}, {Janka}  \& {M{\"u}ller}}{{Wanajo}
  et~al.}{2013}]{Wanajo2013a}
{Wanajo} S.,  {Janka} H.-T.,   {M{\"u}ller} B.,  2013, \mn@doi [\apjl]
  {10.1088/2041-8205/774/1/L6}, \href
  {http://adsabs.harvard.edu/abs/2013ApJ...774L...6W} {774, L6}

\bibitem[\protect\citeauthoryear{{Wanajo}, {M{\"u}ller}, {Janka}  \&
  {Heger}}{{Wanajo} et~al.}{2018}]{Wanajo2018a}
{Wanajo} S.,  {M{\"u}ller} B.,  {Janka} H.-T.,   {Heger} A.,  2018, \mn@doi
  [\apj] {10.3847/1538-4357/aa9d97}, \href
  {http://adsabs.harvard.edu/abs/2018ApJ...852...40W} {852, 40}

\bibitem[\protect\citeauthoryear{{Wang} et~al.,}{{Wang}
  et~al.}{2007}]{Wang2007a}
{Wang} W.,  et~al., 2007, \mn@doi [\aap] {10.1051/0004-6361:20066982}, \href
  {http://adsabs.harvard.edu/abs/2007A%26A...469.1005W} {469, 1005}

\bibitem[\protect\citeauthoryear{{Weaver}, {Zimmerman}  \& {Woosley}}{{Weaver}
  et~al.}{1978}]{WZW78}
{Weaver} T.~A.,  {Zimmerman} G.~B.,   {Woosley} S.~E.,  1978, \mn@doi [\apj]
  {10.1086/156569}, \href {http://adsabs.harvard.edu/abs/1978ApJ...225.1021W}
  {225, 1021}

\bibitem[\protect\citeauthoryear{{Woosley}}{{Woosley}}{1997}]{Woosley1997a}
{Woosley} S.~E.,  1997, \mn@doi [\apj] {10.1086/303650}, \href
  {http://adsabs.harvard.edu/abs/1997ApJ...476..801W} {476, 801}

\bibitem[\protect\citeauthoryear{{Woosley} \& {Heger}}{{Woosley} \&
  {Heger}}{2007a}]{Woosley2007a}
{Woosley} S.~E.,  {Heger} A.,  2007a, \mn@doi [\physrep]
  {10.1016/j.physrep.2007.02.009}, \href
  {http://adsabs.harvard.edu/abs/2007PhR...442..269W} {442, 269}

\bibitem[\protect\citeauthoryear{{Woosley} \& {Heger}}{{Woosley} \&
  {Heger}}{2007b}]{WH07}
{Woosley} S.~E.,  {Heger} A.,  2007b, \mn@doi [\physrep]
  {10.1016/j.physrep.2007.02.009}, \href
  {http://adsabs.harvard.edu/abs/2007PhR...442..269W} {442, 269}

\bibitem[\protect\citeauthoryear{{Woosley} \& {Weaver}}{{Woosley} \&
  {Weaver}}{1995}]{Woosley1995}
{Woosley} S.~E.,  {Weaver} T.~A.,  1995, \mn@doi [\apjs] {10.1086/192237},
  \href {http://adsabs.harvard.edu/abs/1995ApJS..101..181W} {101, 181}

\bibitem[\protect\citeauthoryear{{Woosley}, {Fowler}, {Holmes}  \&
  {Zimmerman}}{{Woosley} et~al.}{1978}]{Woosley1978a}
{Woosley} S.~E.,  {Fowler} W.~A.,  {Holmes} J.~A.,   {Zimmerman} B.~A.,  1978,
  \mn@doi [Atomic Data and Nuclear Data Tables] {10.1016/0092-640X(78)90018-9},
  \href {http://adsabs.harvard.edu/abs/1978ADNDT..22..371W} {22, 371}

\bibitem[\protect\citeauthoryear{{Woosley}, {Heger}, {Rauscher}  \&
  {Hoffman}}{{Woosley} et~al.}{2003}]{Woosley2003a}
{Woosley} S.~E.,  {Heger} A.,  {Rauscher} T.,   {Hoffman} R.~D.,  2003, \mn@doi
  [Nuclear Physics A] {10.1016/S0375-9474(03)00673-0}, \href
  {http://adsabs.harvard.edu/abs/2003NuPhA.718....3W} {718, 3}

\bibitem[\protect\citeauthoryear{{deBoer} et~al.,}{{deBoer}
  et~al.}{2017}]{deBoer2017a}
{deBoer} R.~J.,  et~al., 2017, \mn@doi [Reviews of Modern Physics]
  {10.1103/RevModPhys.89.035007}, \href
  {http://adsabs.harvard.edu/abs/2017RvMP...89c5007D} {89, 035007}

\makeatother
\end{thebibliography}
\input{main.bbl}


\begin{appendices}
	\section{Appendices}
\subsection{Nucleosynthesis code details}

\label{sec:appendix}

\begin{figure*}
	\caption{Equilibration of the reaction network at $T=6$~GK and $\rho=10^6$g~cm$^{-3}$
	without weak interactions. Left panel: all reverse reaction rates taken from their respective
	sources, either tables or fits. Right panel: all reverse reaction rates calculated from the
	principle of detailed balance. In both cases, the network does equilibrate, but only in the
	case with detailed balance (right panel) does it equilibrate to the NSE solution (horizontal
	dashed lines).}
	\label{fig:nse6}
	\includegraphics[width=\textwidth]{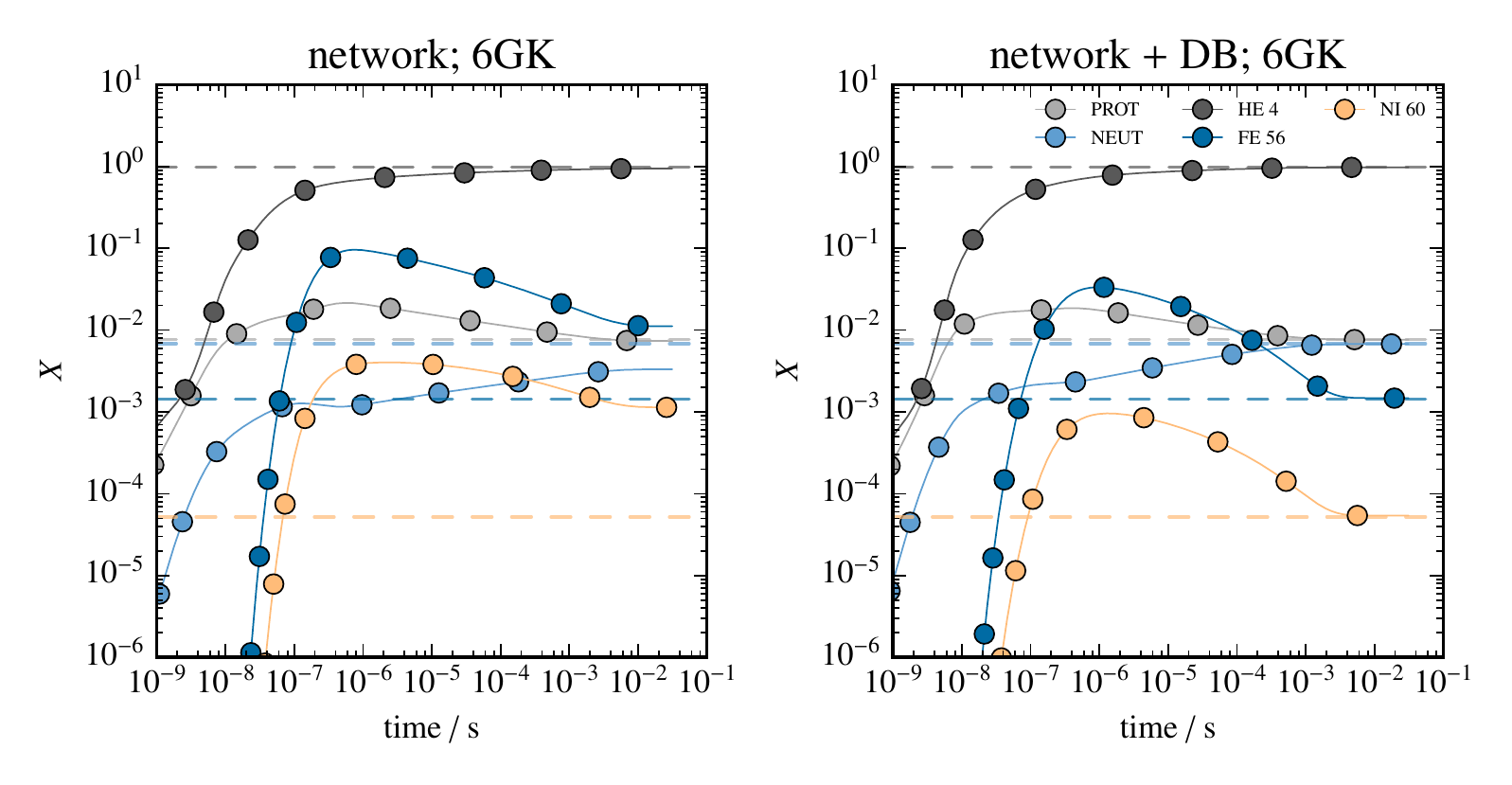}
\end{figure*}

In Section~\ref{sec:nuc_code} we have described the nuclear reaction network
code that was used in the present work. In this Appendix we will briefly discuss
some of the developments that were made for this work.

Firstly, the topic of reverse reaction rates. While we have access to the rates
for the reverse reactions from the various compilations and sources we are
using, it is important that consistency between the forward and reverse rates
and the NSE solver is maintained. Therefore, reverse reaction rates for most but
not all reactions are now computed at run time from the rates of the
corresponding forward reactions using the principle of detailed balance using
the formulation outlined in the Appendix of \citet{Calder2007a}. There is not a
general rule for determining precisely \emph{which} reactions should be
considered a forward reaction and which are reverse reactions. An example of
such a simple consideration would be to assume that all photodisintegration
reactions are reverses, e.g. $(\gamma,p)$, $(\gamma,n)$ and so on.  Typically,
reactions where there is an experimental measurement are considered to be the
forward reaction, and this varies greatly on the experimental technique used. In
our code, we use the JINA Reaclib database to identify which reactions we will
treat as a reverse reaction even if we do not take the reaction rate from JINA
Reaclib.

The impact of using (right panel) or not using (left panel) detailed balance for
the reverse reaction rates is illustrated in Figure~\ref{fig:nse6}, which shows
the mass fractions of a few abundant isotopes resulting from an NSE solve at
temperature 6~GK and density $10^6$~g~cm$^{-3}$ (dashed horizontal lines)
together with the time evolution of the same isotopes from a network
integration. The weak reaction rates are excluded from the network calculation
because we are only interested in the equilibration of the strong rates here.
One can see that in both panels the network reaches an equilibrium at
$\sim3\times10^{-3}$~s, however only in the case where the reverse reaction
rates were computed using detailed balance (right panel) does the network
equilibrate to the result obtained by the NSE solver. In the reaction network
code we assume no energetic feedback from the nuclear reactions because this is
assumed to be captured accurately enough in the stellar evolution or
hydrodynamic code that produced the trajectories. If it were included, one can
see that the point at which the code switched from performing a network
integration to using an NSE solver (typically this switch is made above
temperature of 6~GK), the immediate jump to an inconsistent composition would
result in an instantaneous and potentially large gain or loss of internal energy
in the system, which would affect the ensuing hydrodynamics. While this
particular issue does not affect our present calculations, other properties
could be affected such as $Y_\mathrm{e}$. In any case, it is good to be
consistent. We refer the interested reader to the work by \citet{Lippuner2017a},
who have also discussed and resolved this issue.

The other development that we want to mention here is the implementation of the
semi-implicit extrapolation or Bader-Deuflhard (BD) time integration method,
with which we have chosen to use in this work over the fully implicit
backward-Euler Newton-Raphson (BE-NR) method for the explosion simulations. The
method has also been called the generalised Bulirsch-St\"oer method, and has
been described by \citealp{Bader1983a,Deuflhard1983a}, \citet{Timmes1999a} and
\citet{Longland2014a}. It is a variable-order modified midpoint rule combined
with Richardson extrapolation. We would like to demonstrate its enhanced
accuracy and its excellent balance between accuracy and computational
efficiency.

For both integrators the time step is adaptive. For BE-NR the time step will be
reduced if the Newton Raphson scheme does not converge after 6 iterations and
will increase if the solution converged (increase is only possible if the time
step was decreased during a previous attempt at the integration), according to
\begin{equation}
        \Delta t_\mathrm{new} = \max\left( \min\left( \left[ \frac{ \delta }{ \max(\epsilon,10^{-15}) }\right]^\frac{1}{2},
	\alpha_\mathrm{max}\right), \alpha_\mathrm{min}\right) \Delta t_\mathrm{old},
\end{equation}
where $\delta$ is the numerical tolerance (typically $10^{-3}$), $\epsilon$ is
the numerical error (greatest absolute relative difference in mass fractions
between previous and current iterations), $\alpha_\mathrm{min}=0.2$ and
$\alpha_\mathrm{max}=2$.  The BD integrator uses the time stepping algorithm
outlined by \citet{Deuflhard1983a}. For either integrator we never allow a
larger time step than the local time resolution of the hydrodynamics or stellar
evolution data being post-processed. In principle, though, the BD integrator can
take much larger time steps provided that the evaluations of the right-hand-side
of the network system, $\mathrm{d}\mathbf{Y}/\mathrm{d}t$, at each time level
stage for each order accounts for the time-dependence of temperature and density
in the trajectory being post-processed. This can potentially decrease the
computational cost of the method.

\begin{figure*}
	\caption{A simple hydrogen-burning problem at $T=5.5\times10^7$~K and
		$\rho=10^2$~g~cm$^{-3}$ with various time step sizes. In all cases the network
		would be considered to have `converged'. In order to obtain the same accuracy, the
		backward-Euler method requires substantially more time steps than semi-implicit extrapolation.}
	\label{fig:bd-vs-be}
	\includegraphics[width=\textwidth]{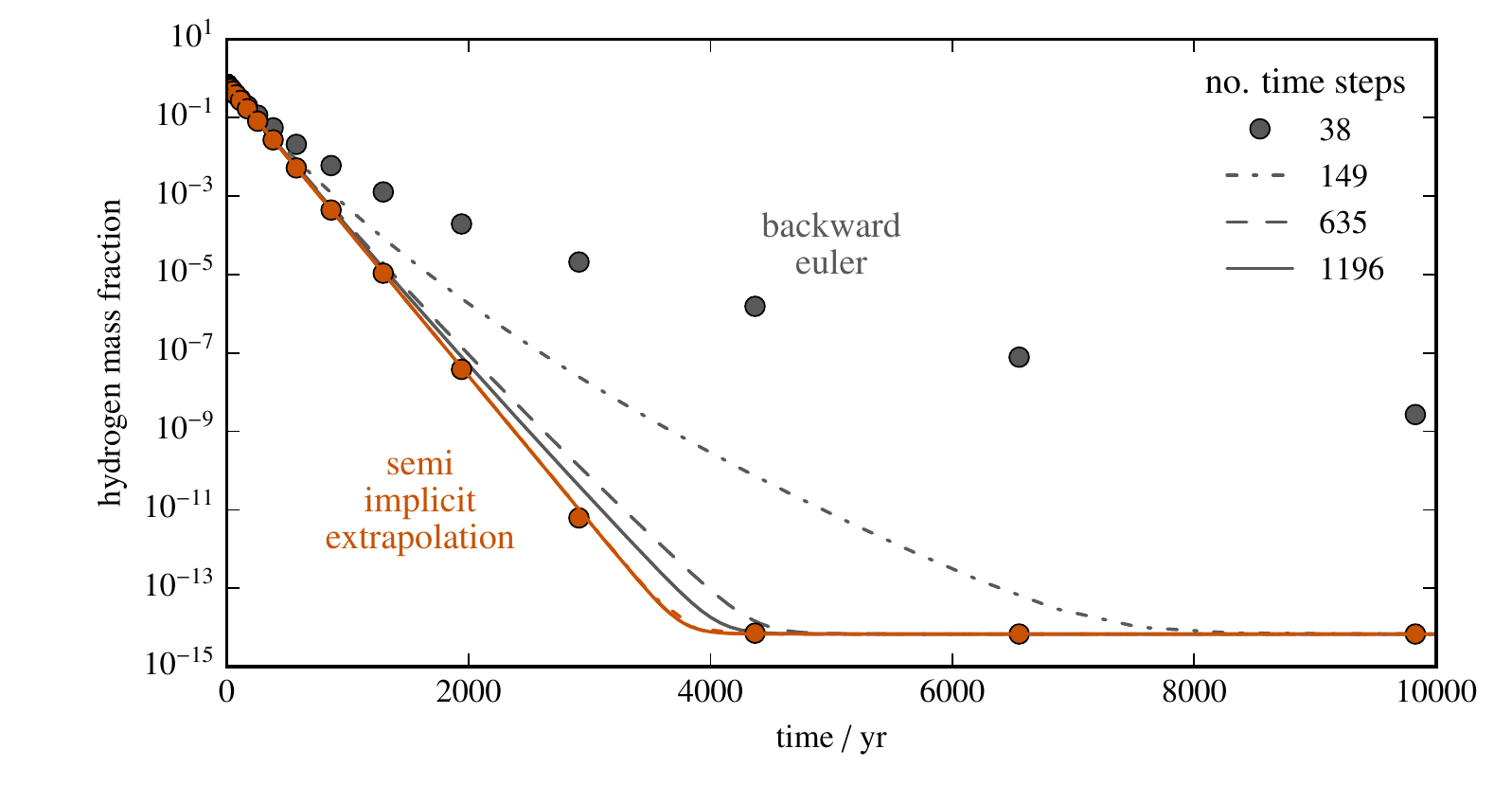}
\end{figure*}

Figure~\ref{fig:bd-vs-be} shows the mass fraction of H as a function of time for
a simple constant temperature and constant density one-zone H burning problem.
Results are given for the backward Euler integrator (grey) and the
Bader-Deuflhard integrator (red). The same global time is covered in all the
simulations but with different numbers of time steps, and in all simulations the
solution is considered to be converged at every time step as far as the
integration method is concerned. One can see that the semi-implicit
extrapolation (Bader-Deuflhard) integrator obtains almost the same solution
regardless of the time step size owing to its adaptive order. The backward Euler
Newton-Raphson integrator, however, converges to a different result with
different time step size although appears to be converging to the
Bader-Deuflhard solutions in the limit of infinitessimally small time steps, as
one would expect.

Now we will compare the computational expense of the methods for a few one-zone
problems. Keep in mind, however, that in all examples, the same global time step
size was used for both integrators, however each integrator was allowed to
adaptively adjust the time step as necessary within the global time step. We
note that in these comparisons, the measure of computational expense is not for
the same accuracy. For example for H burning in Figure~\ref{fig:bd-vs-be} we
could be comparing the computational time for either method to complete the 38
time step case. In order to compare more fairly (e.g. computational time needed
to achieve similar accuracy), one would need to compare the time for 38 time
steps of BD with the time for 1196 steps of BE-NR.

\begin{figure*}
	\caption{Comparison of execution times for three simulations of astrophysical nucleosynthesis
		sites. On the x-axis are different solver options: LEQS (dense, Gaussian elimination),
		NR(PP) (dense, partial pivoting), ACML (dense, AMD core math routines) and SuperLU
		\citep[sparse, partial pivoting;][]{li05}. The semi-implicit extrapolation
		\citep{Bader1983a,Deuflhard1983a} integration method was only ran with SuperLU; it is
		typically 30-300\% more costly than backward-Euler, which seems an appropriate
		trade-off for the superior accuracy \citep[see][for a more thorough comparison of time
		integration methods and matrix inversion software for reaction networks]{Timmes1999a}.}
		\label{fig:solver-times}
		\includegraphics[width=\textwidth]{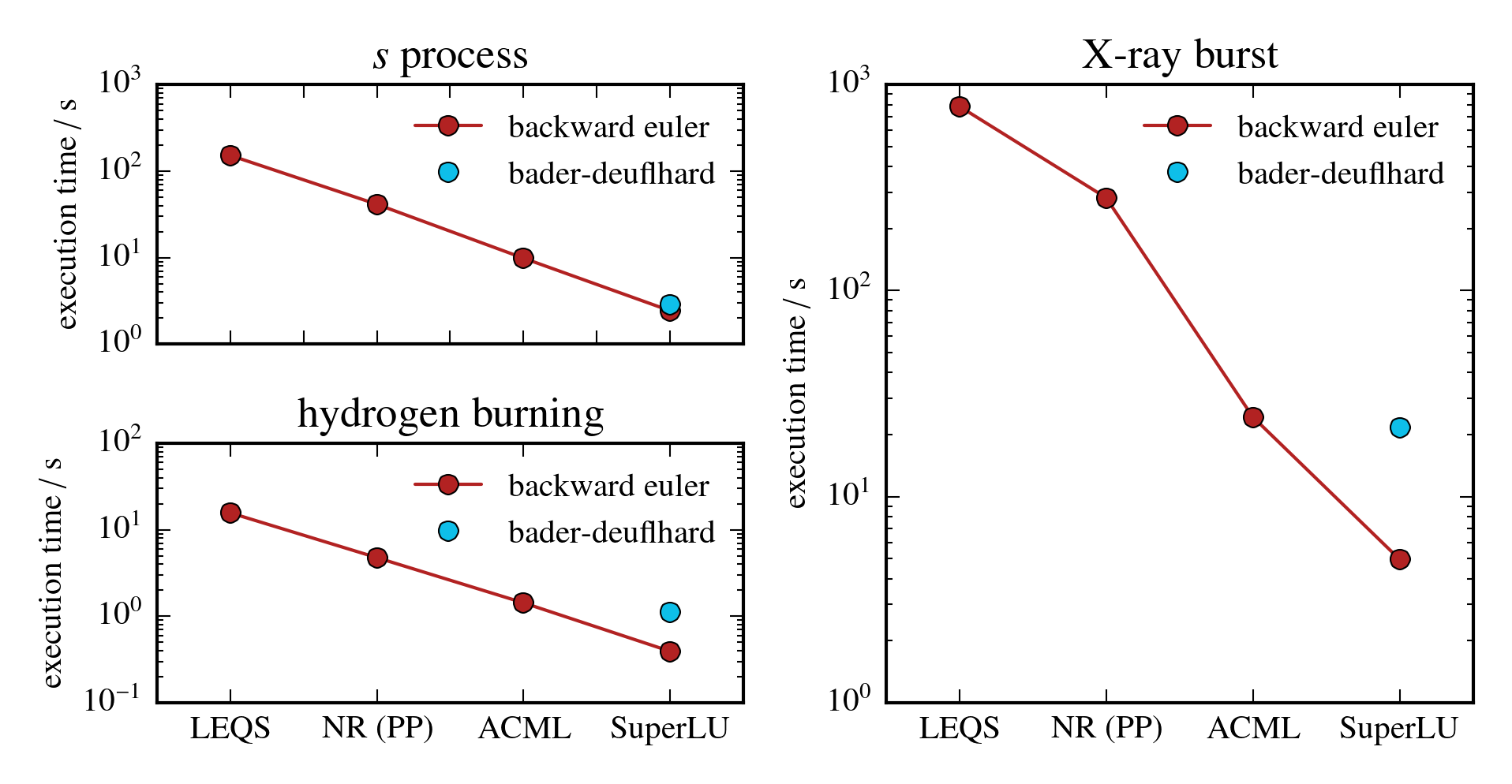}
	\end{figure*}

The execution time for one-zone $s$-process, H-burning X-ray burst problems are
plotted in Figure~\ref{fig:solver-times} as a function of matrix inversion
software or technique, in the order (left to right) in which they were
implemented into the code. Most recently, we have added the SuperLU sparse
matrix package which is currently the fastest available in the code and renders
the code approximately 6 times faster than when using the ACML dense matrix
routines for a $\sim5000$-isotope problem (i.e.~X-ray burst) and approximately 3
times faster for a $\sim1000$-isotope problem (e.g.~H burning). The BD
integrator is only programmed to operate with the SuperLU solver, but in that
case integrating a complete trajectory using the BD integrator is typically only
$20-300$~per cent more expensive than using the BE-NR integrator, depending of
course on the problem.

\subsection{Summary of results of all calculations}
A summary of the relevant results from each of the simulations performed is
given in Table~\ref{tab:all_results}: ZAMS mass, remnant mass, explosion energy,
preSN, cut and yield \fe~masses for the three different \feng~cross sections and
the \al~yield.

\begin{table*}
	\centering
	\caption{Results from all stellar evolution/explosion simulations.
	}
	\label{tab:all_results}
    \begin{tabular}{ccc @{\hspace{6\tabcolsep}} ccccccccc @{\hspace{4\tabcolsep}}c}
\midrule
     & & & \multicolumn{3}{c}{$10^{-1}\langle \sigma v \rangle_{\mathrm{N-S}}$} & \multicolumn{3}{c}{$\langle \sigma v \rangle_{\mathrm{N-S}}$} & \multicolumn{3}{c}{$10^{1}\langle \sigma v \rangle_{\mathrm{N-S}}$} & \\
        \cmidrule(lr){4-6} \cmidrule(lr){7-9} \cmidrule(lr){10-12}
             $M_\star$ & $M_\mathrm{rem}$ &  $E_\mathrm{exp}$ & pre-SN & cut & yield & pre-SN & cut & yield & pre-SN & cut & yield & $^{26}\mathrm{Al}$ yield\\
      $[M_\odot]$ & $[M_\odot]$ & $[10^{51}$~erg] & \multicolumn{3}{c}{$[10^{-4}M_\odot]$} & \multicolumn{3}{c}{$[10^{-4}M_\odot]$} & \multicolumn{3}{c}{$[10^{-4}M_\odot]$}  & $[10^{-4}M_\odot]$\\
\midrule
    15 &  1.80 &  1.34 &      2.84 &      0.05 &      0.03 &      3.24 &      0.41 &      0.30 &      4.88 &      1.94 &      1.39 &      0.20  \\
    15 &  1.71 &  0.30 &      2.84 &      0.05 &      0.02 &      3.24 &      0.44 &      0.17 &      4.88 &      2.09 &      0.82 &      0.21  \\
    15 &  1.52 &  2.47 &      2.84 &      0.05 &      0.04 &      3.24 &      0.44 &      0.35 &      4.88 &      2.09 &      1.47 &      0.22  \\
    15 &  1.61 &  1.94 &      2.84 &      0.05 &      0.04 &      3.24 &      0.44 &      0.33 &      4.88 &      2.09 &      1.47 &      0.21  \\
    15 &  1.75 &  0.92 &      2.84 &      0.05 &      0.04 &      3.24 &      0.43 &      0.31 &      4.88 &      2.06 &      1.37 &      0.32  \\
    15 &  1.53 &  2.63 &      2.84 &      0.05 &      0.11 &      3.24 &      0.44 &      0.74 &      4.88 &      2.09 &      2.06 &      0.25  \\
    15 &  1.88 &  0.82 &      2.84 &      0.04 &      0.03 &      3.24 &      0.37 &      0.29 &      4.88 &      1.74 &      1.37 &      0.20  \\
    15 &  1.94 &  0.34 &      2.84 &      0.04 &      0.03 &      3.24 &      0.33 &      0.29 &      4.88 &      1.58 &      1.38 &      0.19  \\
    15 &  1.56 &  2.24 &      2.84 &      0.05 &      0.05 &      3.24 &      0.44 &      0.39 &      4.88 &      2.09 &      1.58 &      0.21  \\
    15 &  1.50 &  4.79 &      2.84 &      0.05 &      0.65 &      3.24 &      0.44 &      2.00 &      4.88 &      2.09 &      3.07 &      0.35  \\
    15 &  1.74 &  0.89 &      2.84 &      0.05 &      0.03 &      3.24 &      0.43 &      0.28 &      4.88 &      2.06 &      1.21 &      0.22  \\
    15 &  1.63 &  1.86 &      2.84 &      0.05 &      0.04 &      3.24 &      0.44 &      0.32 &      4.88 &      2.09 &      1.43 &      0.21  \\
    15 &  1.52 &  1.69 &      2.84 &      0.05 &      0.04 &      3.24 &      0.44 &      0.34 &      4.88 &      2.09 &      1.48 &      0.26  \\
    15 &  1.73 &  0.74 &      2.84 &      0.05 &      0.03 &      3.24 &      0.44 &      0.30 &      4.88 &      2.08 &      1.34 &      0.32  \\
    15 &  1.51 &  3.43 &      2.84 &      0.05 &      0.23 &      3.24 &      0.44 &      1.25 &      4.88 &      2.09 &      2.66 &      0.34  \\
    15 &  1.62 &  1.90 &      2.84 &      0.05 &      0.04 &      3.24 &      0.44 &      0.33 &      4.88 &      2.09 &      1.45 &      0.21  \\
    15 &  1.71 &  0.52 &      2.84 &      0.05 &      0.02 &      3.24 &      0.44 &      0.19 &      4.88 &      2.09 &      0.88 &      0.21  \\
    15 &  1.91 &  0.54 &      2.84 &      0.04 &      0.03 &      3.24 &      0.35 &      0.29 &      4.88 &      1.65 &      1.37 &      0.19  \\
    15 &  1.59 &  2.06 &      2.84 &      0.05 &      0.04 &      3.24 &      0.44 &      0.35 &      4.88 &      2.09 &      1.50 &      0.21  \\
    15 &  1.71 &  0.82 &      2.84 &      0.05 &      0.02 &      3.24 &      0.44 &      0.21 &      4.88 &      2.09 &      1.00 &      0.22  \\
    15 &  1.51 &  2.47 &      2.84 &      0.05 &      0.08 &      3.24 &      0.44 &      0.56 &      4.88 &      2.09 &      1.79 &      0.23  \\
    15 &  1.52 &  2.60 &      2.84 &      0.05 &      0.07 &      3.24 &      0.44 &      0.54 &      4.88 &      2.09 &      1.85 &      0.22  \\
    20 &  1.86 &  4.00 &      2.47 &      0.02 &      0.03 &      2.66 &      0.21 &      0.27 &      3.94 &      1.49 &      1.79 &      0.34  \\
    20 &  2.23 &  1.47 &      2.47 &      0.02 &      0.02 &      2.66 &      0.21 &      0.21 &      3.94 &      1.44 &      1.44 &      0.05  \\
    20 &  1.85 &  4.15 &      2.47 &      0.02 &      0.03 &      2.66 &      0.21 &      0.24 &      3.94 &      1.49 &      1.60 &      0.22  \\
    20 &  1.93 &  1.39 &      2.47 &      0.02 &      0.02 &      2.66 &      0.21 &      0.21 &      3.94 &      1.49 &      1.49 &      0.03  \\
    20 &  1.76 &  2.76 &      2.47 &      0.02 &      0.02 &      2.66 &      0.21 &      0.21 &      3.94 &      1.50 &      1.50 &      0.06  \\
    20 &  3.40 &  0.53 &      2.47 &      0.01 &      0.01 &      2.66 &      0.09 &      0.09 &      3.94 &      0.64 &      0.64 &      0.03  \\
    20 &  3.03 &  0.65 &      2.47 &      0.01 &      0.01 &      2.66 &      0.13 &      0.13 &      3.94 &      0.89 &      0.89 &      0.03  \\
    20 &  2.28 &  1.19 &      2.47 &      0.02 &      0.02 &      2.66 &      0.20 &      0.20 &      3.94 &      1.40 &      1.41 &      0.03  \\
    20 &  1.87 &  4.33 &      2.47 &      0.02 &      0.03 &      2.66 &      0.21 &      0.24 &      3.94 &      1.49 &      1.60 &      0.23  \\
    20 &  1.90 &  2.60 &      2.47 &      0.02 &      0.02 &      2.66 &      0.21 &      0.22 &      3.94 &      1.49 &      1.51 &      0.07  \\
    20 &  1.97 &  1.52 &      2.47 &      0.02 &      0.02 &      2.66 &      0.21 &      0.21 &      3.94 &      1.49 &      1.49 &      0.03  \\
    20 &  2.62 &  0.84 &      2.47 &      0.02 &      0.02 &      2.66 &      0.17 &      0.17 &      3.94 &      1.17 &      1.17 &      0.03  \\
    20 &  1.93 &  2.50 &      2.47 &      0.02 &      0.02 &      2.66 &      0.21 &      0.22 &      3.94 &      1.49 &      1.52 &      0.10  \\
    20 &  2.62 &  0.85 &      2.47 &      0.02 &      0.02 &      2.66 &      0.17 &      0.17 &      3.94 &      1.17 &      1.17 &      0.03  \\
    20 &  1.74 &  2.85 &      2.47 &      0.02 &      0.02 &      2.66 &      0.21 &      0.21 &      3.94 &      1.50 &      1.49 &      0.05  \\
    20 &  2.35 &  1.00 &      2.47 &      0.02 &      0.02 &      2.66 &      0.19 &      0.20 &      3.94 &      1.36 &      1.36 &      0.03  \\
    20 &  2.76 &  0.75 &      2.47 &      0.02 &      0.02 &      2.66 &      0.15 &      0.15 &      3.94 &      1.08 &      1.08 &      0.03  \\
    20 &  1.78 &  1.65 &      2.47 &      0.02 &      0.02 &      2.66 &      0.21 &      0.21 &      3.94 &      1.50 &      1.49 &      0.03  \\
    20 &  1.86 &  2.43 &      2.47 &      0.02 &      0.02 &      2.66 &      0.21 &      0.21 &      3.94 &      1.49 &      1.49 &      0.05  \\
    20 &  2.85 &  0.78 &      2.47 &      0.02 &      0.02 &      2.66 &      0.15 &      0.15 &      3.94 &      1.02 &      1.02 &      0.03  \\
    20 &  2.70 &  0.81 &      2.47 &      0.02 &      0.02 &      2.66 &      0.16 &      0.16 &      3.94 &      1.12 &      1.12 &      0.03  \\
    20 &  2.47 &  1.04 &      2.47 &      0.02 &      0.02 &      2.66 &      0.18 &      0.18 &      3.94 &      1.27 &      1.28 &      0.03  \\
    25 &  1.84 &  1.04 &      2.06 &      0.08 &      0.09 &      2.67 &      0.69 &      0.77 &      5.05 &      3.08 &      3.35 &      0.91  \\
    25 &  1.83 &  3.07 &      2.06 &      0.08 &      0.13 &      2.67 &      0.69 &      1.02 &      5.05 &      3.08 &      3.80 &      1.22  \\
    25 &  2.38 &  4.73 &      2.06 &      0.08 &      0.16 &      2.67 &      0.69 &      1.18 &      5.05 &      3.05 &      3.81 &      0.63  \\
    25 &  4.66 &  0.89 &      2.06 &      0.07 &      0.07 &      2.67 &      0.59 &      0.60 &      5.05 &      2.36 &      2.38 &      0.05  \\
    25 &  1.83 &  1.52 &      2.06 &      0.08 &      0.09 &      2.67 &      0.69 &      0.77 &      5.05 &      3.08 &      3.35 &      0.91  \\
    25 &  1.83 &  0.96 &      2.06 &      0.08 &      0.09 &      2.67 &      0.69 &      0.77 &      5.05 &      3.08 &      3.35 &      0.91  \\
    25 &  3.13 &  1.92 &      2.06 &      0.08 &      0.13 &      2.67 &      0.69 &      1.02 &      5.05 &      3.08 &      3.80 &      1.22  \\
    25 &  2.35 &  2.53 &      2.06 &      0.08 &      0.12 &      2.67 &      0.69 &      0.98 &      5.05 &      3.06 &      3.75 &      1.21  \\
    25 &  2.35 &  4.72 &      2.06 &      0.08 &      0.12 &      2.67 &      0.69 &      0.98 &      5.05 &      3.06 &      3.75 &      1.21  \\
    25 &  2.35 &  2.78 &      2.06 &      0.08 &      0.12 &      2.67 &      0.69 &      0.98 &      5.05 &      3.06 &      3.75 &      1.21  \\
    25 &  5.60 &  0.75 &      2.06 &      0.07 &      0.10 &      2.67 &      0.55 &      0.77 &      5.05 &      2.03 &      2.40 &      0.05  \\
    25 &  2.35 &  3.30 &      2.06 &      0.08 &      0.12 &      2.67 &      0.69 &      0.98 &      5.05 &      3.06 &      3.75 &      1.21  \\
    25 &  4.89 &  0.99 &      2.06 &      0.07 &      0.07 &      2.67 &      0.58 &      0.60 &      5.05 &      2.27 &      2.32 &      0.05  \\
    25 &  3.73 &  1.57 &      2.06 &      0.08 &      0.08 &      2.67 &      0.64 &      0.69 &      5.05 &      2.68 &      2.81 &      0.09  \\
    25 &  1.84 &  1.20 &      2.06 &      0.08 &      0.09 &      2.67 &      0.69 &      0.77 &      5.05 &      3.08 &      3.35 &      0.91  \\
    25 &  5.47 &  0.74 &      2.06 &      0.07 &      0.07 &      2.67 &      0.55 &      0.56 &      5.05 &      2.07 &      2.08 &      0.05  \\
    25 &  2.34 &  2.52 &      2.06 &      0.08 &      0.12 &      2.67 &      0.69 &      0.98 &      5.05 &      3.06 &      3.75 &      1.21  \\
    25 &  1.84 &  0.92 &      2.06 &      0.08 &      0.09 &      2.67 &      0.69 &      0.77 &      5.05 &      3.08 &      3.35 &      0.91  \\
    25 &  2.35 &  2.64 &      2.06 &      0.08 &      0.12 &      2.67 &      0.69 &      0.98 &      5.05 &      3.06 &      3.75 &      1.21  \\
   \hline
	\end{tabular}
\end{table*}

\end{appendices}


\bsp	
\label{lastpage}
\end{document}